\documentclass[11pt,a4paper]{article}

\usepackage{jheppub}
\usepackage{amsmath}
\usepackage{amssymb}
\usepackage{amsfonts}
\usepackage{braket}
\usepackage[normalem]{ulem}
\usepackage{color}
\usepackage{bbold}
\usepackage{ulem}
\usepackage{amsthm}
\usepackage{todonotes}
\usepackage{subcaption}
\usepackage[T1]{fontenc}
\usepackage{url}

\renewcommand{\S}{\widetilde{S}}

\newcommand{\trho}{\tilde{\rho}}
\newcommand{\tf}{\tilde{f}}
\newcommand{\tF}{\widetilde{F}}

\newcommand{\aux}{\text{aux}}
\newcommand{\OO}{\mathcal{O}}

\DeclareMathOperator{\tr}{tr}

\title{A Renyi Quantum Null Energy Condition: Proof for Free Field Theories}

\author[1]{Mudassir Moosa}
\author[2,3]{, Pratik Rath}
\author[2,3]{ and Vincent Paul Su}

\affiliation[1]{Department of Physics, Cornell University, Ithaca, NY, 14853, USA
}
\affiliation[2]{Center for Theoretical Physics and Department of Physics,\\
University of California, Berkeley, CA 94720, U.S.A.}
\affiliation[3]{Lawrence Berkeley National Laboratory, Berkeley, CA 94720, U.S.A.}

\emailAdd{mudassir.moosa@cornell.edu}
\emailAdd{pratik\_rath@berkeley.edu}
\emailAdd{vipasu@berkeley.edu}

\abstract{The Quantum Null Energy Condition (QNEC) is a lower bound on the stress-energy tensor in quantum field theory that has been proved quite generally.
It can equivalently be phrased as a positivity condition on the second null shape derivative of the relative entropy $S_{\text{rel}}(\rho||\sigma)$ of an arbitrary state $\rho$ with respect to the vacuum $\sigma$.
The relative entropy has a natural one-parameter family generalization, the Sandwiched Renyi divergence $S_n(\rho||\sigma)$, which also measures the distinguishability of two states for arbitrary $n\in[1/2,\infty)$.
A Renyi QNEC, a positivity condition on the second null shape derivative of $S_n(\rho||\sigma)$, was conjectured in previous work.
In this work, we study the Renyi QNEC for free and superrenormalizable field theories in spacetime dimension $d>2$ using the technique of null quantization.
In the above setting, we prove the Renyi QNEC in the case $n>1$ for arbitrary states.
We also provide counterexamples to the Renyi QNEC for $n<1$.
}

\begin{document}
\maketitle

\section{Introduction}\label{sec-intro}

In recent years, the fascinating interplay between semiclassical gravity, quantum information and relativistic quantum field theory (QFT) has led to a lot of deep insights.
Ideas motivated from semiclassical gravity often have interesting non-gravitational limits that lead to novel results in QFT \cite{Bousso:2015mna,Bousso:2019dxk}.

The quintessential result that arose from this connection is the Quantum Null Energy Condition (QNEC), which follows from the Quantum Focusing Conjecture \cite{Bousso:2015mna}.
The QNEC is a lower bound on the stress-energy tensor $T_{\mu \nu}$ in a relativistic QFT that takes the form
\begin{align}\label{eq-QNEC}
    \langle T_{vv}(y) \rangle &\geq \frac{\hbar}{2\pi} S_{vv}''(y),
\end{align}
where Minkowski spacetime is written in terms of coordinates $\{u,v,y^i\}$, $u$ and $v$ representing null lightcone coordinates, and $y^i$ representing the transverse directions.
$S_{vv}''(y)$ represents the diagonal part of the second null shape derivative of entropy, i.e.,
\begin{align}\label{eq-shape}
   \frac{\delta^2 S(R)}{\delta V(y) \,\delta V(y')} &= S_{vv}''(y) \delta^{(d-2)}(y-y') + (\text{off-diagonal terms}),
\end{align}
where $S(R)$ is the von Neumann entropy of a subregion $R$, with an entangling surface $\partial R$ given by an arbitrary cut of a null plane, $u=0$ and $v=V(y)$, as seen in Fig.~(\ref{fig:null-quant}).

The individual quantities in Eq.~\eqref{eq-QNEC} are in fact ultraviolet (UV) divergent in general, although the combination that appears in the QNEC is UV finite.
To avoid this issue, one can instead formulate the QNEC directly in terms of the relative entropy $S_{\text{rel}}(\rho_R||\sigma_R)$ defined by
\begin{align}\label{eq-Srel}
    S_{\text{rel}}(\rho_R||\sigma_R)&\equiv \tr (\rho_R \log \rho_R) - \tr(\rho_R \log \sigma_R)\\
    &=\Delta \langle K_R^{\sigma}\rangle - \Delta S(R),
\end{align}
where in our case of interest, $\sigma_R$ will represent the vacuum reduced density matrix for subregion $R$, and $K_R^{\sigma}$ is the associated modular Hamiltonian.
To see this, one can use the fact that $K_R^{\sigma}$ for arbitrary cuts of a null plane can be written as a local integral on the null plane \cite{Casini:2017roe}, i.e.,
\begin{align}\label{eq-vac}
    K_R^{\sigma}&=\int_{V(y)}^{\infty} (v-V(y))T_{vv}(y) \,dv\,d^{d-2}y.
\end{align}
Using this result, it can be shown that the QNEC is equivalent to a purely information theoretic constraint \cite{Leichenauer:2018obf}
\begin{align}\label{eq-QNECequiv}
    \frac{\delta^2 S_{\text{rel}}(\rho_R||\sigma_R)}{\delta V(y)\delta V(y')}\geq 0
\end{align}
where the shape derivative $\frac{\delta}{\delta V(y)}$ involves an infinitesimal deformation of the subregion $R$ along the null plane at the transverse location $y$.
The diagonal part of Eq.~\eqref{eq-QNECequiv}, i.e., the limit $y\to y'$, reduces to Eq.~\eqref{eq-QNEC}, while the off-diagonal part follows from strong subadditivity.
We shall henceforth refer to the quantity in Eq.~\eqref{eq-QNECequiv} as the second relative entropy variation.
The QNEC is thus the statement of the positivity of the diagonal part of the second relative entropy variation.

The QNEC was initially proven in various restricted settings such as for free field theories and holographic theories \cite{Bousso:2015wca,Koeller:2015qmn,Malik:2019dpg}.
The QNEC has since been proven quite generally using the appropriate generalization of quantum information theory to algebraic QFT that goes under the name of Tomita-Takesaki modular theory \cite{Balakrishnan:2017bjg,Ceyhan:2018zfg}.
In general, a lot of progress has been made in recent years by considering information theoretic quantities in QFT like the QNEC, which often carry valuable hints about the underlying structure of the theory \cite{Lashkari:2018nsl,Mezei:2019sla,Balakrishnan:2019gxl,Ecker:2019ocp,Ecker:2020gnw}.
We will take this as inspiration to understand a generalized version of the QNEC motivated from quantum information theory \cite{Lashkari:2018nsl}.

The formulation of the QNEC in Eq.~\eqref{eq-QNECequiv} suggests a natural generalization based on a one-parameter generalization of the relative entropy called the Sandwiched Renyi Divergence $S_n(\rho_R||\sigma_R)$, defined by
\begin{align}\label{srd_density1}
    S_n(\rho_R||\sigma_R) = \frac{1}{n-1} \log \tr  \left(\sigma_R^{\frac{1-n}{2n}} \rho_R \sigma_R^{\frac{1-n}{2n}} \right)^n.
\end{align}
The analogous Renyi generalization of Eq.~\eqref{eq-QNECequiv} then reads
\begin{align}\label{eq-Renyi}
        \frac{\delta^2 S_{n}(\rho_R||\sigma_R)}{\delta V(y)\delta V(y')}\geq 0,
\end{align}
and was proposed in Ref.~\cite{Lashkari:2018nsl} which provided some evidence in favour of it.
In analogy with Eq.~\eqref{eq-QNECequiv}, we will refer to the quantity in Eq.~\eqref{eq-Renyi} as the second SRD variation.
The second SRD variation also involves a diagonal and an off-diagonal part similar to the second relative entropy variation.
In this paper, our focus will be on the diagonal part of Eq.~\eqref{eq-Renyi}, which we henceforth refer to as the Renyi QNEC.

In this paper, we prove the Renyi QNEC in the parameter regime $n>1$ for free and superrenormalizable field theories in spacetime dimensions $d>2$.
Further, we disprove the Renyi QNEC in the regime $n<1$ in a similar setting.

We now provide a brief overview of the paper.
In Sec.~(\ref{sec-SRD}), we review the definition and properties of the Sandwiched Renyi divergence (SRD), the quantity that shows up in the Renyi QNEC.
In particular, we review how SRD is defined in QFTs and discuss a reasonable class of states for which SRD is finite.

In Sec.~(\ref{sec-free-QNEC}), we set up the calculation of the diagonal part of the second SRD variation in free field theories.
The main technique that is used in this calculation is null quantization of free field theories.
This technique is also applicable to superrenormalizable deformations of a free field theory and our results easily extend to this case as well.
We review relevant aspects of this formalism and then utilize it to recast the Renyi QNEC in a suitable form.
Having done this, we show how the free field calculation can be related to a perturbative calculation of SRD which allows us to use existing results in the literature.

Sec.~(\ref{sec-proof})  is our main result where we prove the Renyi QNEC for free field theories.
Having set up the problem, we first prove the Renyi QNEC for the simpler case of integer $n>1$, where the proof follows from reflection positivity.
For the case of general $n$, we explicitly compute the second SRD variation for arbitrary states.
Using this result, we then show that the second SRD variation is positive for $n>1$, thus proving the Renyi QNEC in this case.
We also provide counterexamples for the conjecture in the case $n<1$.

Having completed the proof for free field theories, we then discuss various possible generalizations of our result in Sec.~(\ref{sec-disc}).
First, we consider the Renyi QNEC in interacting theories in $d>2$.
By considering states that are perturbatively close to the vacuum and computing the second SRD variation in a perturbative expansion, we provide evidence that the Renyi QNEC could in fact be saturated, just like the QNEC \cite{Balakrishnan:2019gxl}.
We then discuss issues with proving the positivity of the off-diagonal part of the second SRD variation.
Since our proof only applies to $d>2$, we provide some numerical evidence that the Renyi QNEC could in fact be true even in $d=2$.
Finally, we discuss other possible generalizations of the Renyi QNEC motivated by other distinguishability measures in information theory.

\section{Sandwiched Renyi Divergence}\label{sec-SRD}


The Sandwiched Renyi Divergence (SRD), $S_{n}^{\mathcal{M}}(\Psi||\Phi)$,  is a measure of distinguishability of two quantum states $\ket{\Psi}$ and $\ket{\Phi}$ given an algebra of operators $\mathcal{M}$ \cite{muller2013quantum,Wilde:2014eda}.
SRD is a one-parameter generalization of the relative entropy $S_{\text{rel}}^{\mathcal{M}}(\Psi||\Phi)$, for a parameter $n \in [1/2,\infty)$.
Since the Renyi QNEC is formulated in terms of the SRD, we first review its definition and properties.

In Sec.~(\ref{sec-SRDdensity}), we first discuss the definition of SRD in terms of density matrices.
This definition is not directly applicable to QFT where reduced density matrices do not exist in the continuum limit.
However, we will return to this formulation later.
In Sec.~(\ref{sec-SRDQFT}), we will then review a suitably generalized definition of SRD in quantum field theory (QFT) using techniques from Tomita-Takesaki theory \cite{Berta:2018ecp,Jencova-1,Jencova-2,Lashkari:2018nsl}.
Having defined SRD, our main focus will be constructing a physically reasonable set of states using the Euclidean path integral for which SRD with respect to the vacuum state is finite.

\subsection{Sandwiched Renyi Divergence for Density Matrices}\label{sec-SRDdensity}

A Type-I von Neumann algebra $\mathcal{M}$ induces a decomposition of the Hilbert space $\mathcal{H}=\mathcal{H}_R \otimes \mathcal{H}_{\bar{R}}$ such that $\mathcal{M}=\mathcal{L}(\mathcal{H}_R)$, the algebra of bounded operators on the Hilbert space $\mathcal{H}_A$.\footnote{Here we ignore the possibility of a non-trivial center of the algebra, i.e., operators that belong to both $\mathcal{M}$ and its commutant $\mathcal{M}'$. These have played an important role in other situations, especially in AdS/CFT, e.g., \cite{Harlow:2016vwg,Akers:2018fow,Dong:2018seb}. The discussion we provide here can be easily generalized to include such a possibility.}
A special case of interest is when $\mathcal{M}$ is the algebra of operators associated to a spatial region $R$.
Given this decomposition, one can define the reduced density matrices of $\ket{\Psi}$ and $\ket{\Phi}$, denoted $\rho_R$ and $\sigma_R$ respectively.

The SRD of $\rho_R$ with respect to $\sigma_R$ is then defined as
\begin{align}\label{srd_density}
    S_n(\rho_R||\sigma_R) = \frac{1}{n-1} \log \tr  \left(\sigma_R^{\frac{1-n}{2n}} \rho_R \sigma_R^{\frac{1-n}{2n}} \right)^n,
\end{align}
where we have suppressed the label $\mathcal{M}$ in the SRD since the dependence on the algebra $\mathcal{M}$ is completely captured by the reduced density matrices $\rho_R$ and $\sigma_R$.
For $n>1$, the SRD is defined to be infinite if the support of $\rho_R$ is not contained in the support of $\sigma_R$.
For $n<1$ on the other hand, the SRD is finite as long as $\rho_R$ and $\sigma_R$ are not orthogonal.

The SRD in the range $n\in[\frac{1}{2},\infty)$ has been shown to satisfy all the required properties of a measure of distinguishability of quantum states \cite{muller2013quantum,frank2013monotonicity,beigi2013sandwiched,Wilde:2014eda}.
Principal among these is the data-processing inequality which states that the SRD decreases under any completely-positive trace-preserving (CPTP) map $\mathcal{N}$, i.e.,
\begin{align}\label{eq:DPI}
    S_n(\rho_R||\sigma_R) \geq S_n(\mathcal{N}(\rho_R)||\mathcal{N}(\sigma_R)).
\end{align}
In particular, SRD decreases under tracing out a portion of $\mathcal{H}_R$ as the states are harder to distinguish given access to a smaller algebra of operators.
Apart from this, SRD is also positive for all states and unitarily invariant.
In the limit $n \to 1$, SRD approaches the familiar relative entropy
\begin{align}\label{srel_density}
    S_{\text{rel}}(\rho_R||\sigma_R) = \tr \rho_R \log \rho_R - \tr \rho_R \log \sigma_R,
\end{align}
and thus, serves as a natural one-parameter generalization of the relative entropy.

In this work we will assume that in the presence of a suitable ultraviolet (UV) cutoff $\epsilon$, the Hilbert space of QFT factorizes into the Hilbert spaces of subregions $R$ and its complement $\bar{R}$ as described in this section.
For the purposes of computing relative entropy or SRD with respect to the vacuum, we expect that this is a reasonable assumption for states which have a finite SRD in the limit $\epsilon \to 0$.
In such cases, one could perform computations using density matrices and send the cutoff $\epsilon\to0$ to get a finite low energy answer irrespective of the details of the cutoff prescription.
In particular, computations in later sections will be done using the Euclidean path integral where this is manifestly true when a suitable state is chosen.
Thus, we now focus our attention on determining a class of states in QFT where the SRD with respect to the vacuum is indeed finite.

\subsection{Sandwiched Renyi Divergence in QFT}
\label{sec-SRDQFT}

In QFT, the algebra associated to any spatial subregion of a Cauchy slice is a Type-III von Neumann algebra and the Hilbert space does not factorize into Hilbert spaces of subregions.
This means that the reduced density matrix of a state on a subregion is not well-defined and hence, the definitions of relative entropy and SRD in terms of density matrices, i.e. Eq.~\eqref{srel_density} and Eq.~\eqref{srd_density} respectively, are not applicable.
However, relative entropy and SRD between two states can still be defined using the algebra of operators in a subregion.
We now review these definitions that are based on Tomita-Takesaki theory \cite{tomita1967quasi,takesaki2006tomita} (see also Refs.~\cite{haag2012local,Borchers:2000pv,Hollands:2017dov,Witten:2018zxz}.

Let $\mathcal{M}$ be the algebra of operators associated with some spatial subregion $R$ of a Cauchy slice $\Sigma$.
The commutant algebra $\mathcal{M}'$ is then associated with the complementary
subregion $\bar{R}$.\footnote{This is true under the assumption of Haag duality which does not hold in general. In this paper, we will mostly be interested in Rindler regions for which this property is known to hold \cite{Bisognano:1976za}.}
Consider two arbitrary states $\ket{\Psi}$ and $\ket{\Phi}$ which are cyclic and separating for both $\mathcal{M}$ and $\mathcal{M}'$.\footnote{Many results can be obtained without the assumption of cyclic and separating states. See \cite{Ceyhan:2018zfg,Faulkner:2020iou} for discussions that do not make this assumption.}
The relative Tomita-Takesaki operator $\mathcal{S}_{\Phi|\Psi}$ is defined by its action on the dense set of states:
\begin{align}
    \mathcal{S}_{\Psi|\Phi} \, \OO_{A} \ket{\Psi} \, \equiv \, \OO_{A}^{\dagger}\ket{\Phi} \, ,
\end{align}
for all $\OO_{A} \in \mathcal{M}$.\footnote{Note that we are using the convention in \cite{Witten:2018zxz}, whereas some other literature, e.g., \cite{Lashkari:2018nsl} uses $\mathcal{S}_{\Phi|\Psi}$ to represent the same object.}
The relative modular operator $\Delta_{\Psi|\Phi}$ is a positive-definite, Hermitian operator defined as
\begin{align}
    \Delta_{\Psi|\Phi} \equiv \, \mathcal{S}_{\Psi|\Phi}^{\dagger} \, \mathcal{S}_{\Psi|\Phi} \, ,
\end{align}
where we have kept the $\mathcal{M}$ dependence of $\mathcal{S}_{\Psi|\Phi}$ and $\Delta_{\Psi|\Phi}$ implicit.

With these basic objects in hand, the relative entropy between states $\ket{\Psi}$ and $\ket{\Phi}$ for algebra $\mathcal{M}$ is defined by \cite{Araki:1976}
\begin{align}
    S_{\text{rel}}^{\mathcal{M}}(\Psi||\Phi) \, = \, - \, \bra{\Psi} \, \log \, \Delta_{\Psi|\Phi} \ket{\Psi} \, .\label{eq-RE-TT}
\end{align}
When the algebra $\mathcal{M}$ is a Type-I algebra associated to a subregion $R$, the reduced density matrices for the states $\ket{\Psi}$ and $\ket{\Phi}$, denoted $\rho_R$ and $\sigma_R$ are well defined.
Similarly, the reduced density matrices for the complementary subregion $\bar{R}$ are also well defined, denoted by $\rho_{\bar{R}}$ and $\sigma_{\bar{R}}$.
The relative modular operator $\Delta_{\Psi|\Phi}$ is given by
\begin{align}\label{eq-Delta-mat}
    \Delta_{\Psi|\Phi} = \sigma_R \otimes \rho_{\bar{R}}^{-1}.
\end{align}
Using this, one can show that the definition for the relative entropy in Eq.~\eqref{eq-RE-TT} reduces to the one in Eq.~\eqref{srel_density} whenever the algebra $\mathcal{M}$ is a Type-I algebra \cite{Witten:2018zxz}.
However, the definition in Eq.~\eqref{eq-RE-TT} continues to make sense in QFT and provides a reasonable measure of distinguishability given the algebra $\mathcal{M}$.

Analogously, the sandwiched Renyi divergence (SRD) can be defined for an algebra $\mathcal{M}$ in QFT   \cite{Berta:2018ecp,Jencova-1,Jencova-2,Lashkari:2018nsl}.
This definition makes use of the $p$-norm of unbounded operators defined by Araki and Masuda \cite{Araki:1982}.
In particular, the SRD between states $\ket{\Psi}$ and $\ket{\Phi}$ for algebra $\mathcal{M}$ is
\begin{align}
    S_{n}^{\mathcal{M}}(\Psi||\Phi) \, \equiv \, \frac{n}{n-1} \, \sup_{\ket{\chi}\in \mathcal{H}} \, \log \, \bra{\Psi} \, \left(\Delta_{\chi|\Phi}\right)^{\frac{1-n}{n}} \, \ket{\Psi} \, , \label{eq-SRD-TT-1}
\end{align}
for $n>1$, and
\begin{align}
    S_{n}^{\mathcal{M}}(\Psi||\Phi) \, \equiv \, \frac{n}{n-1} \, \inf_{\ket{\chi}\in \mathcal{H}} \, \log \, \bra{\Psi} \, \left(\Delta_{\chi|\Phi}\right)^{\frac{1-n}{n}} \, \ket{\Psi} \, , \label{eq-SRD-TT-2}
\end{align}
for $1/2 \leq n < 1$.
Using this definition, it can be shown that SRD satisfies the relevant properties of a measure of distinguishability such as positivity and unitary invariance.
Most importantly, it satisfies a data processing inequality analogous to Eq.~\eqref{eq:DPI}, i.e., SRD decreases monotonically upon reducing the algebra $\mathcal{M}$ \cite{Berta:2018ecp,Jencova-1,Jencova-2,Lashkari:2018nsl}.

\begin{figure}
    \centering
    \includegraphics[width=\textwidth]{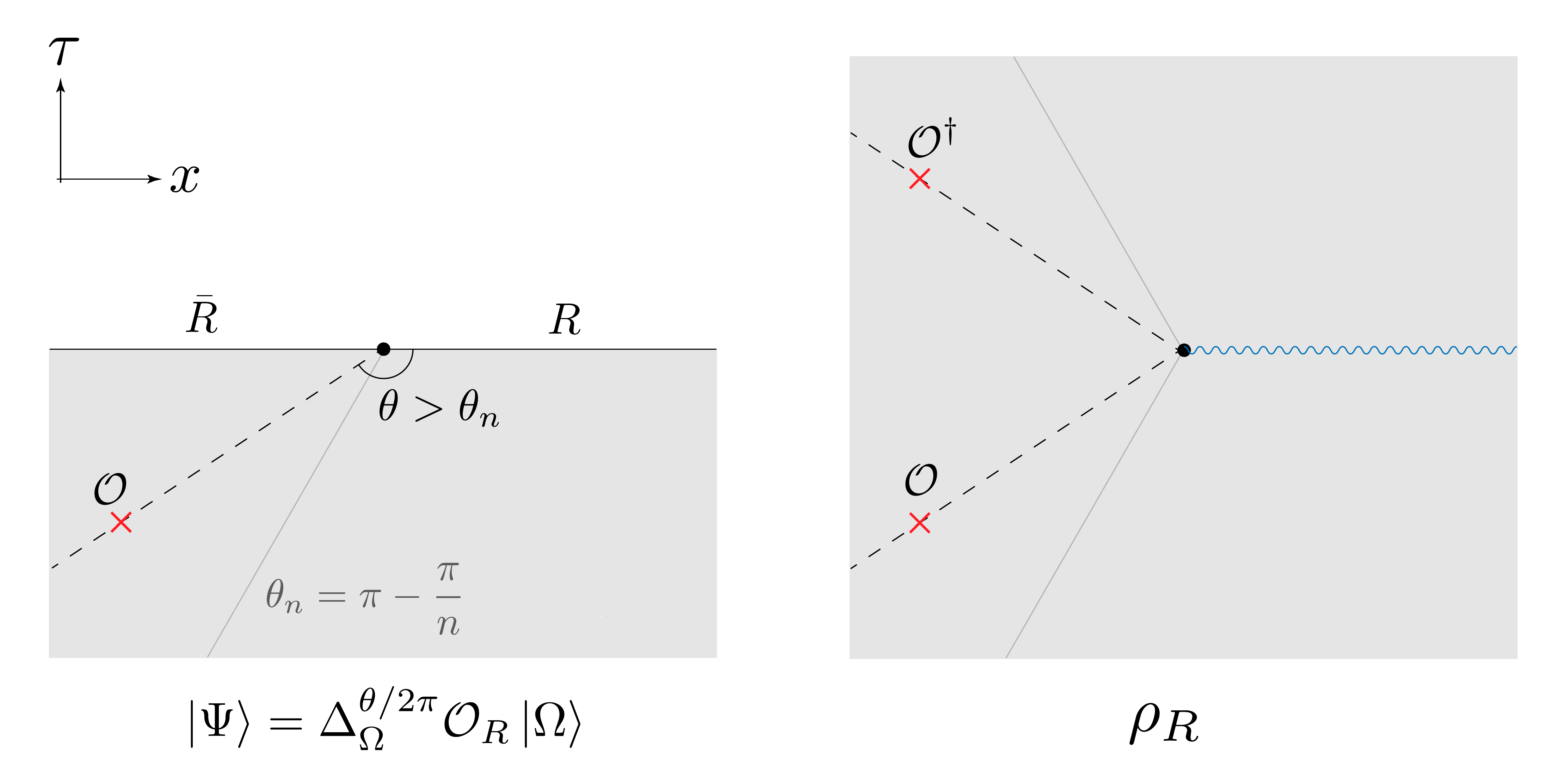}
    \caption{State Preparation. We consider a dense set of states that are prepared by a Euclidean path integral in the lower half plane. In order to have finite SRD with respect to the vacuum, the operators must be inserted at an angle greater than $\theta_n = \pi - \frac{\pi}{n}$. }
    \label{fig:state_prep}
\end{figure}

For $n>1$, the SRD is defined to be infinite if the intersection of the domains of $\Delta_{\chi|\Phi}^{-1}$ for all $\ket{\chi}$ does not include $\ket{\Psi}$.
For $n<1$ on the other hand, the SRD is
finite whenever $\ket{\Phi}$ is cyclic and separating as we discuss in Appendix~(\ref{app-finite}).
This is the generalization of the support condition on density matrices discussed below Eq.~\eqref{srd_density}.
$\Delta_{\chi|\Phi}^{-1}$ is an unbounded operator and thus, its domain for all $\ket{\chi}$ is not the whole Hilbert space \cite{Witten:2018zxz}.
Thus, there exist states $\ket{\Psi}$ such that $S_n^{\mathcal{M}}(\Psi||\Phi)$ is infinite and in QFT, the best we can hope for such an unbounded object is to consider a dense set of states for which SRD is finite.
We will now demonstrate the existence of such a dense set of states motivated by the Euclidean path integral.

In $d$-dimensional Minkowski space, consider the subregion $R$ to be the right Rindler wedge $x>0$ and the reference state is chosen to be the vacuum state $\ket{\Omega}$.
A dense set of states can be prepared by inserting local operators with arbitrary source functions in the Euclidean path integral over the lower half plane $\tau<0$ as seen in Fig.~(\ref{fig:state_prep}).
However, requiring the SRD to be finite will put constraints on the source functions as we now demonstrate.

To see this, consider a simpler state which is constructed by the insertion of a single operator $\OO(r,\theta)$ in the path integral as shown in Fig.~(\ref{fig:state_prep}).
The state prepared this way is given by
\begin{align}
\ket{\Psi} \, &= \OO(r,\theta) \ket{\Omega}\, ,\\
&=\Delta_{\Omega}^{\theta/2\pi} \, \OO_{R} \, \Delta_{\Omega}^{-\theta/2\pi}\ket{\Omega} \, , \\
&=\Delta_{\Omega}^{\theta/2\pi} \, \OO_{R}\ket{\Omega} \, ,
\end{align}
where $0 \le \theta \le \pi$ and $\OO_{R} \in \mathcal{M}$. Further, we have used the fact that $\Delta_{\Omega} \equiv \Delta_{\Omega|\Omega}$, the modular operator of the vacuum state, generates Euclidean rotations \cite{Bisognano:1976za} and leaves the vacuum invariant, e.g.
\begin{align}
    \Delta_{\Omega} \, \ket{\Omega} \, = \, \ket{\Omega} \, .
\end{align}
Now using Eq.~\eqref{eq-SRD-TT-1}, we find that the SRD for $n>1$ between the state $\ket{\Psi}$ and the vacuum $\ket{\Omega}$ is given by
\begin{align}
    S_{n}^{\mathcal{M}}(\Psi||\Omega) \, =&\, \, \frac{n}{n-1} \, \sup_{\ket{\chi}\in\mathcal{H}} \, \log \, \bra{\Omega} \OO_{R}^{\dagger} \,  \Delta_{\Omega}^{\theta/2\pi} \, \left(\Delta_{\chi|\Omega}\right)^{\frac{1-n}{n}} \, \Delta_{\Omega}^{\theta/2\pi} \, \OO_{R} \ket{\Omega} \, . \label{eq-SRD-exam}
\end{align}
By evaluating the right hand side for $\ket{\chi} = \ket{\Omega}$, we can replace the supremum by a lower bound.
We thus obtain
\begin{align}
    S_{n}^{\mathcal{M}}(\Psi||\Omega) \, \ge&\,  \, \frac{n}{n-1} \,  \log \, \bra{\Omega} \OO_{R}^{\dagger} \left(\Delta_{\Omega}\right)^{\frac{\theta}{\pi} + \frac{1-n}{n}} \, \OO_{R} \ket{\Omega} \, ,\\
    =&\, \, \frac{n}{n-1} \,  \log \, \left\Vert \left(\Delta_{\Omega}\right)^{\frac{\theta}{2\pi} + \frac{1-n}{2n}} \, \OO_{R} \ket{\Omega} \right\Vert^{2} \, \label{eq-SRD-norm}.
\end{align}

\begin{figure}
    \centering
    \includegraphics[width=\textwidth]{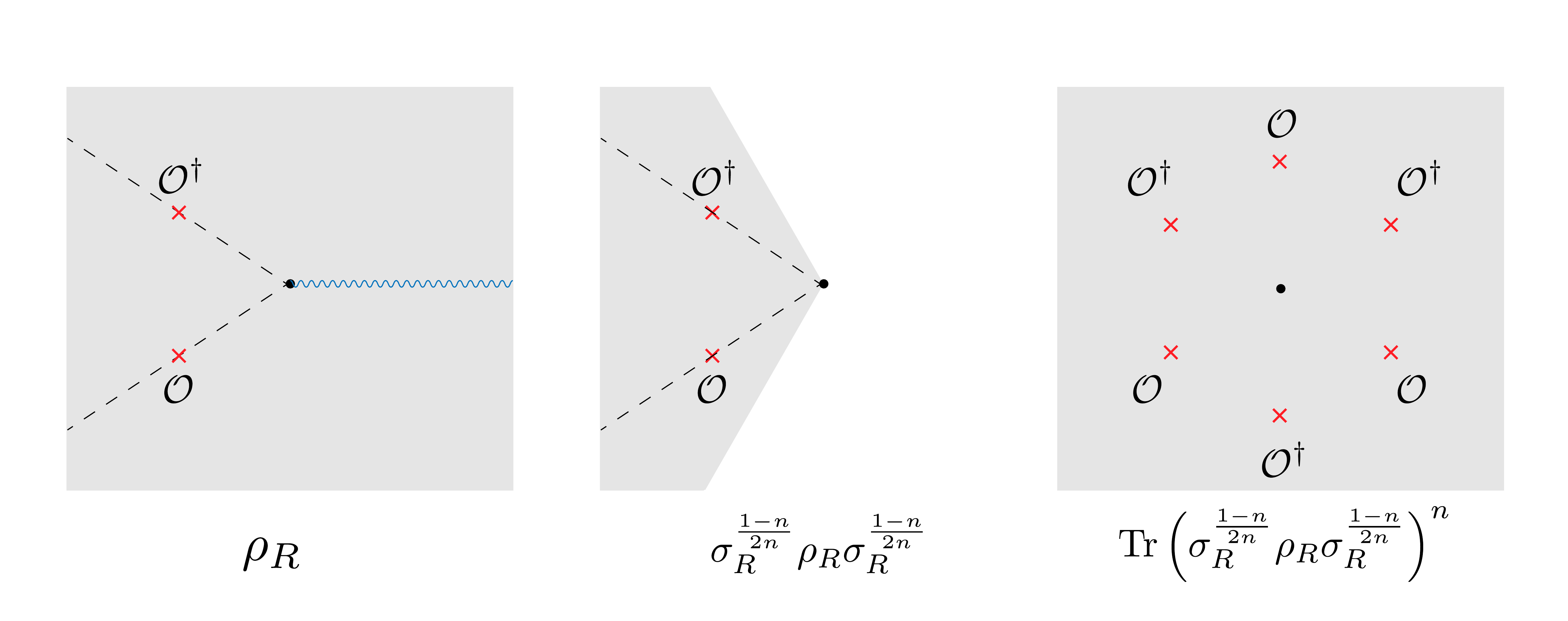}
    \caption{SRD as a correlation function for integer $n=3$ (adapted from Ref.~\cite{Lashkari:2018nsl}). We start with the reduced density matrix on the subregion $R$. Multiplying by inverse powers of the vacuum $\sigma_R$ corresponds to removing a portion of the path integral on either side of the cut, leaving a path integral whose angular size is $2\pi/n$. Finally, taking a cyclical trace of $n$ copies results in a $2n$ point function.}
    \label{fig:sandwich}
\end{figure}
Now, it is known that the state $\ket{\phi_{\alpha}}=\Delta^{\alpha}_{\Omega} \, \OO_{R}\ket{\Omega}$ has finite norm for $0 \le \alpha \le 1/2$ and generically infinite norm outside this range \cite{Witten:2018zxz, Borchers:2000pv, Bisognano:1976za}.
This can be seen from the fact that the state $\ket{\phi_{\alpha}}$ for $0\leq\alpha\leq\frac{1}{2}$ can be prepared using a Euclidean path integral, which can then be used to compute its norm.
This implies that the norm in Eq.~\eqref{eq-SRD-norm} and hence, the SRD in Eq.~\eqref{eq-SRD-exam} is infinite if $\theta < \pi - \frac{\pi}{n}$.
Note that this condition does not put any constraints when $n<1$ as should be expected from the fact that SRD is finite for a cyclic and separating vector like the vacuum $\ket{\Omega}$.
Conversely, for integer $n$, one can compute the SRD using the path integral when the source functions are supported only on $\theta \geq \pi - \frac{\pi}{n}$.
This amounts to computing a $2n$ point correlation function as seen in Fig.~(\ref{fig:sandwich}) that is manifestly finite \cite{Lashkari:2018nsl}.
We expect this conclusion to remain unchanged when we consider non-integer $n$.
Thus, we learn that for SRD for $n>1$ to be finite, we can consider states prepared by the Euclidean path integral with source functions that have support only in the wedge $\theta\geq \pi - \frac{\pi}{n}$.
States prepared in such a way are dense in the Hilbert space since in particular they include states prepared by acting with operators $\OO_{\bar{R}}$ in the algebra $\mathcal{M}'$.
This follows from the Reeh-Schlieder theorem which tells us that operators in $\mathcal{M}'$ already create a dense set of states due to the cyclic property of $\ket{\Omega}$.

Thus, we see that for a given $n$, finite SRD requires that the source functions of local operators vanish outside a wedge of angle $\frac{\pi}{n}$ around the complementary subregion $\bar{R}$.
Intuitively, the reason is that operators inserted closer to $\bar{R}$, i.e., for $\theta$ close to $\pi$, affect the subregion $\bar{R}$ more than they affect the subregion $R$.
The SRD for higher $n$ are more sensitive to the presence of excitations above the vacuum $\ket{\Omega}$ and thus, SRD grows as the insertions are brought closer to the subregion $R$ and eventually blows up for insertions outside the wedge.

The wedge shrinks to a vanishing angle as $n\to\infty$.
Thus, if we are interested in finite SRD for arbitrary $n$, we can consider a dense set of states of the form $\OO'\ket{\Omega}$ where $\OO'\in\mathcal{A'}$ \cite{Lashkari:2018nsl}.\footnote{This was in fact shown for arbitrary subregions $R$ \cite{Lashkari:2018nsl}.
We also expect analogous wedge constraints for a given $n$ to hold for arbitrary $R$, although the exact form may be complicated owing to the lack of symmetry.}

This concludes our discussion of the relative entropy and SRD in QFT.
The main result of this section is that a dense set of states with finite SRD for a given $n$ can be prepared by the Euclidean path integral with the constraint that source functions of local operators vanish outside a wedge of angle $\frac{\pi}{n}$ around the complementary subregion $\bar{R}$.
Having found these states from a rigorous algebraic QFT perspective, we will use them in later sections to analyze the Renyi QNEC in terms of density matrices as argued earlier.

\section{Renyi QNEC in Free Field Theories} \label{sec-free-QNEC}

Consider a null plane $N$ in $d$-dimensional Minkowski space with an entangling surface $\partial R$, that splits the subregions $R$ and $\bar{R}$, defined by an arbitrary cut of the null plane.
The Renyi QNEC is a condition on the second shape derivative of the SRD, of a state reduced to the subregion $R$ with respect to the vacuum, in the direction along $N$.
In particular, we are interested in the diagonal part of the second SRD variation which involves deforming the subregion locally near a given point $\tilde{p}$.
In this section, we will explain how this shape derivative can be computed in free field theories in $d>2$ using the technique of null quantization.
This technique is also applicable to superrenormalizable deformations of a free field theory and thus, our results also extend to such theories.
Hereafter, we include the possibility of such superrenormalizable interactions in our discussion of free field theories.

In Sec.~(\ref{sec-nq}) we review how the QFT state in free field theories can be described by quantizing it directly on the null plane $N$, which is discretized in the transverse direction into pencils of area $\mathcal{A}$.
The QFT state on $N$ has the special property that the vacuum $\ket{\Omega}$ factorizes into a product state on each of the individual pencils.
Since any state looks approximately like $\ket{\Omega}$ at short scales, this allows us to find a perturbative expansion in $\mathcal{A}$ for the reduced density matrix on the pencil $p$ containing the point $\tilde{p}$ that we are interested in.
In Sec.~(\ref{sec-null-QNEC-stat}) we use this perturbative expansion to extract the leading contribution that determines the Renyi QNEC.
Despite the fact that there is a perturbative expansion for the state on the pencil, the rest of the state is completely arbitrary.
Nevertheless, we show in Sec.~(\ref{sec-identity}) that the computation can be simply related to a calculation of SRD between two nearby states.

\subsection{Null Quantization} \label{sec-nq}

\begin{figure}
    \begin{subfigure}{.49\textwidth}
    \centering
    \includegraphics[width=\textwidth]{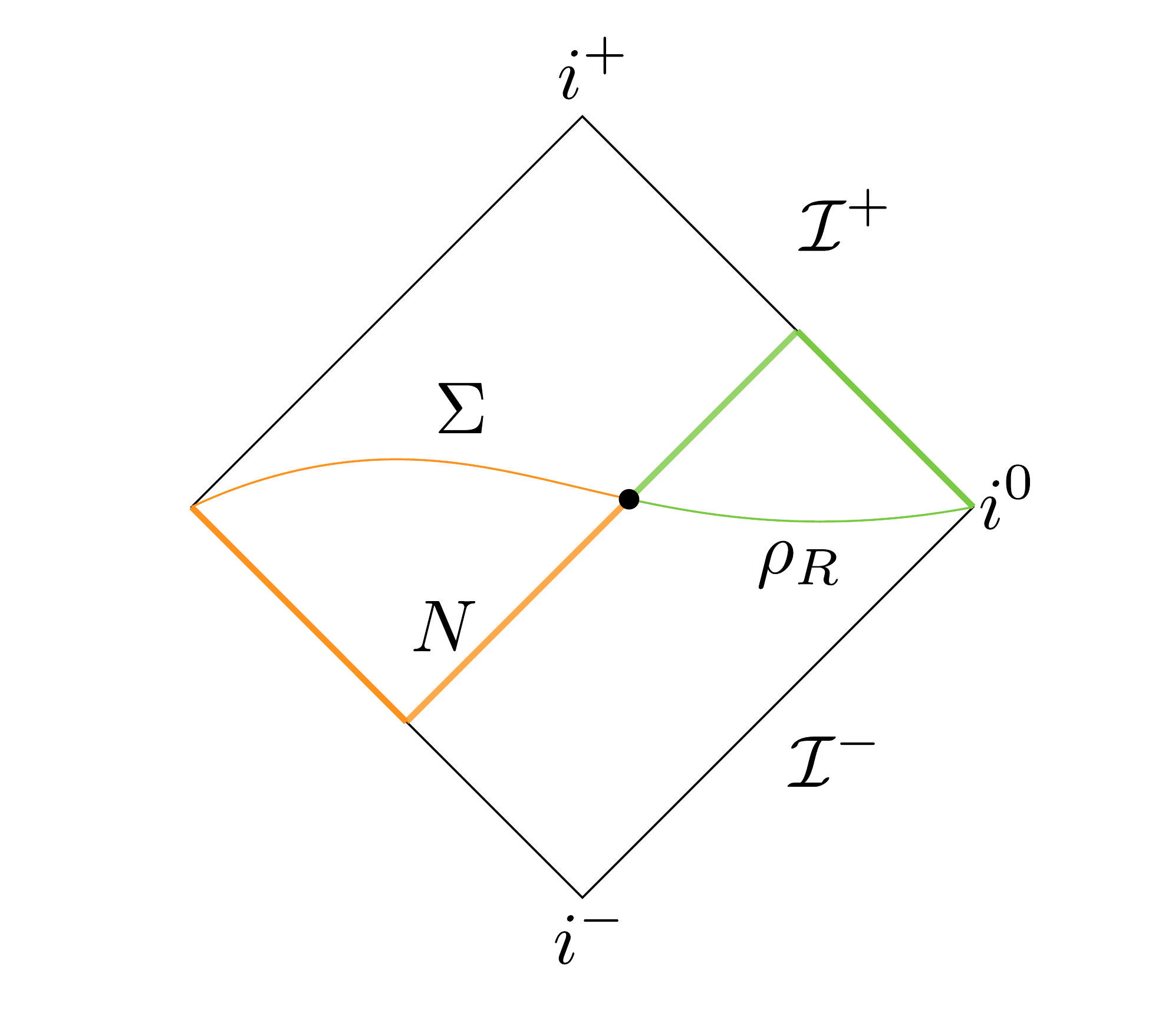}
    \end{subfigure}
    \begin{subfigure}{.49\textwidth}
    \centering
    \includegraphics[width=\textwidth]{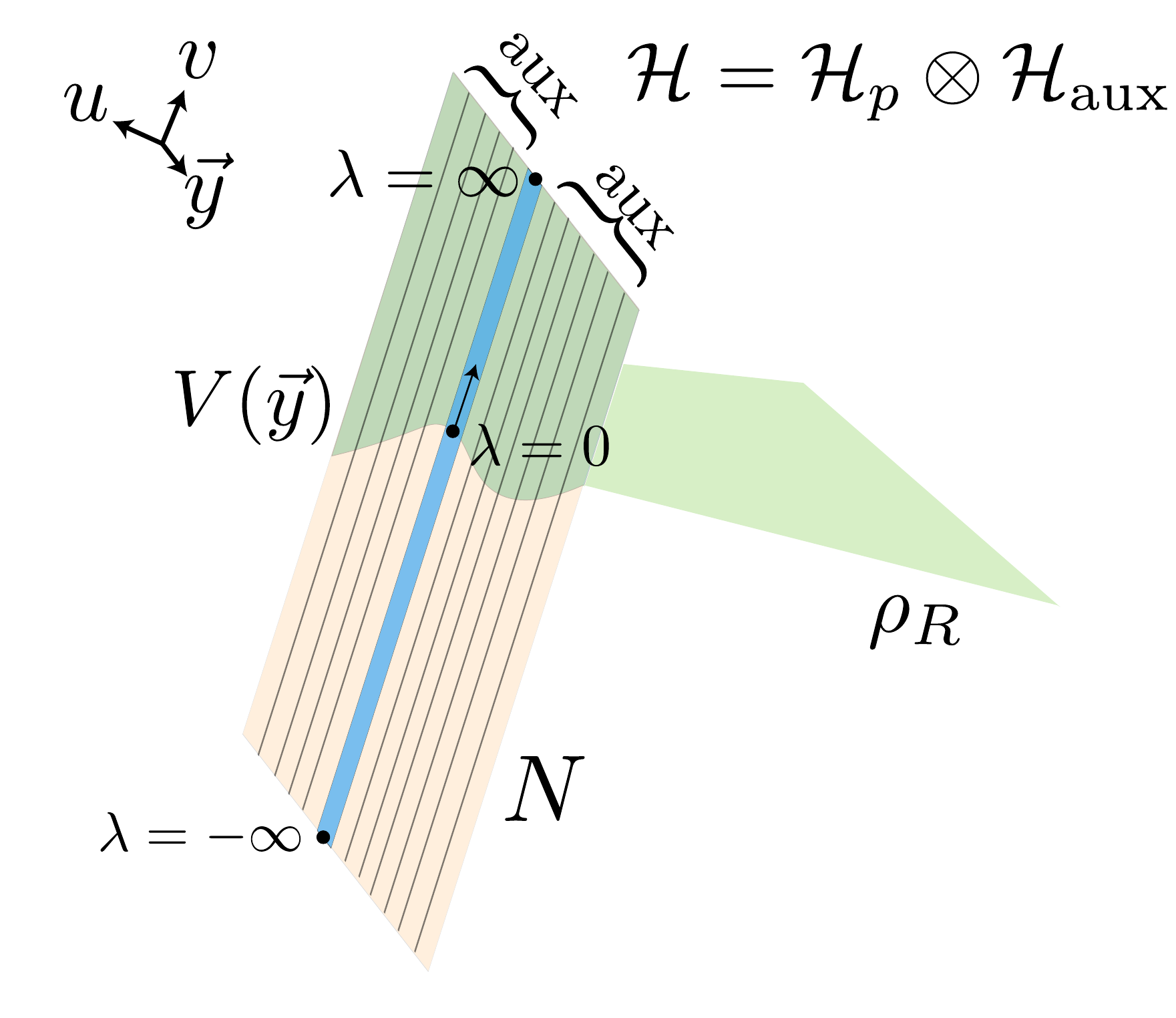}
    \end{subfigure}
    \caption{Null Quantization. The reduced state $\rho_R$ on a region $R$ (light green) is unitarily equivalent to part of $N$ along with part of null infinity (bold green). On the right, we show the null quantization of $N$, featuring transverse pencils of transverse area $\mathcal{A}$. Deformations of $V(\vec{y})$ around the point $\tilde{p}$ are equivalent to $\lambda$ derivatives along the pencil $p$. The auxiliary system includes both, the other pencils and relevant portions of null infinity.}
    \label{fig:null-quant}
\end{figure}
Consider $d$-dimensional Minkowski space $\mathbb{R}^{d-1,1}$ in lightcone coordinates with a metric
\begin{align}
    ds^2 &= -du\,dv+d\vec{y}^2,
\end{align}
where $u$ and $v$ are null coordinates while $\vec{y}$ represents all the transverse spatial coordinates.
In order to discuss the Renyi QNEC, we can pick a null plane $N$ to be the hypersurface $u=0$.
The QFT state on a Cauchy slice $\Sigma$ is unitarily equivalent to the state on $N$ and certain portions of $\mathcal{I}^{+}$ and $\mathcal{I}^{-}$ as shown in Fig.~(\ref{fig:null-quant}).
In general interacting QFTs, this is only formally true but in free field theories this can be made rigorous with the formalism of null quantization \cite{Wall:2011hj}.

In this section, we briefly review the setup of null quantization, which was also used in proving the QNEC for free field theories in \cite{Bousso:2015wca}.
We refer the reader to Sec.~($3$) of \cite{Bousso:2015wca} for further details.

In order to describe null quantization of a free scalar field theory\footnote{Our results can easily be extended to theories with spin \cite{Bousso:2015wca}, or to free fermionic theories \cite{Malik:2019dpg}.} in $d > 2$, we first discretize $N$ into null generators, called \textit{pencils}, each occupying a transverse region of area $\mathcal{A}$.
Given this, the Hilbert space on $N$ factorizes as a product of Hilbert spaces for each pencil.
On each null generator, we have a $(1+1)$-dimensional free CFT of a chiral boson $\Phi(v)$ \cite{Wall:2011hj,Bousso:2015wca}.
Moreover, the vacuum state when restricted on $N$ factorizes as
\begin{align}
    \ket{\Omega} \, = \, \bigotimes_{\vec{y}} \, \ket{\Omega_{\vec{y}}} \, ,
\end{align}
where $\ket{\Omega_{\vec{y}}}$ is the vacuum of the free chiral CFT on the null generator at transverse position $\vec{y}$.
Now, we can define the subregions $R$ and $\bar{R}$ to be divided by an entangling surface $\partial R$ which is given by an arbitrary cut, $v=V(\vec{y})$, of $N$ as seen in Fig.~(\ref{fig:null-quant}).
The vacuum state reduced to the subregion $R$ is then given by
\begin{align}
    \sigma \, = \, \bigotimes_{\vec{y}} \, \sigma_{\vec{y}} \, ,
\end{align}
where $\sigma_{\vec{y}}$ is the density matrix of the free chiral CFT on the pencil labelled by $\vec{y}$, reduced to the region $v>V(\vec{y})$.
Note that we henceforth suppress the dependence of $\sigma$ on the subregion $R$ to simplify the notation.
We remind the reader that we will be working with density matrices in this proof, and expect this to not affect the conclusion.

Since we are interested in a specific pencil $p$, we will instead consider the Hilbert space decomposition
\begin{align}
    \mathcal{H}&=\mathcal{H}_p \otimes \mathcal{H}_{\text{aux}},
\end{align}
where $\mathcal{H}_{\text{aux}}$ corresponds to all the degrees of freedom in the remaining pencils of $N$ and the relevant portions of null infinity as shown in Fig.~(\ref{fig:null-quant}).
The vacuum reduced density matrix can then be written as
\begin{align}
    \sigma(\lambda) \, = \, \sigma_{p}(\lambda) \, \otimes \, \sigma_{\aux} \, , \label{eq-null-vac}
\end{align}
where $\lambda$ denotes the affine parameter of the entangling surface on the pencil $p$.
The important point is that $\sigma_{p}$ is unentangled with the rest of the system.

The diagonal part of the Renyi QNEC is expressed in terms of null shape derivatives at a point $\tilde{p}$, so the privileged pencil $p$ in our analysis will be the one that contains $\tilde{p}$.
Eventually we will take the continuum limit $\mathcal{A} \to 0$ in computing the shape derivatives required for the Renyi QNEC.
However, before doing so, we use $\mathcal{A}$ as a small expansion parameter for our analysis which simplifies the description of the state on the pencil and allows us to compute the derivatives using perturbation theory.

To describe the state on the pencil $p$ we make use of the fact that any given state reduced to the pencil resembles the vacuum in the limit $\mathcal{A}\to 0$.
Further, the theory on $p$ is a free theory with a well defined Fock basis of states labelled by the number of particle excitations, $\{\ket{m}\}$.
Following \cite{Bousso:2015wca}, the probability of $m$ particle excitations on $p$ should scale as $\mathcal{A}^m$ since it should behave extensively in the $\mathcal{A}\to0$ limit.
The amplitude for $m$ particle excitations should then scale as $\mathcal{A}^{m/2}$, and thus, the coefficient of $\ket{m_1}\bra{m_2}$ in the Fock basis should scale as $\mathcal{A}^{(m_1+m_2)/2}$.
Thus in the $\mathcal{A}\to0$ limit, the leading correction to the vacuum comes from off diagonal terms of the form $(\ket{0}\bra{1} + \ket{1}\bra{0})$ with prefactor $\mathcal{A}^{1/2}$.

\begin{figure}
    \centering
    \includegraphics[width=.8\textwidth]{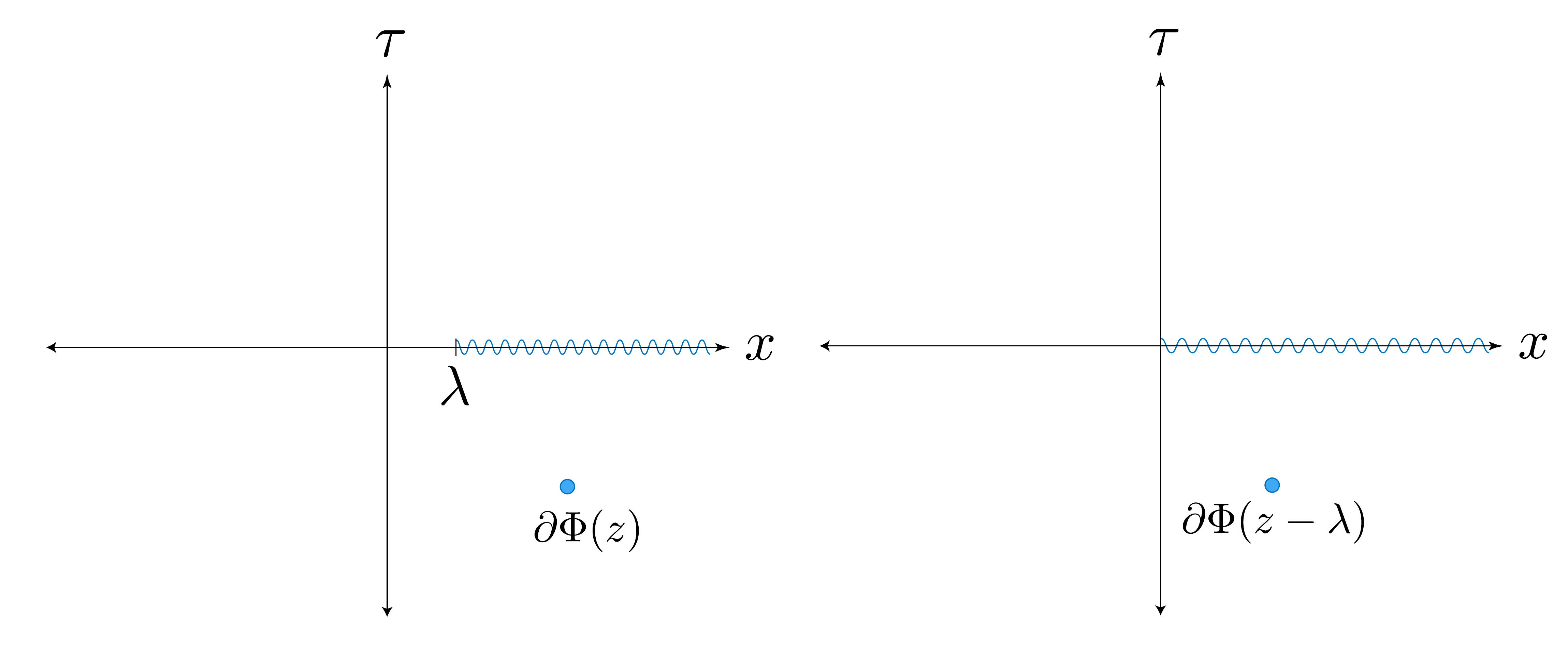}
    \caption{Here we illustrate two equivalent ways of constructing a one-particle state. In the first picture, we imagine that we are tracing out part of the state up to affine parameter $\lambda$. In the right figure, we view the same state as moving the location of the operator insertion. The relation between them is the difference between an active and a passive transformation. In this work, we prefer to keep the entangling surface fixed and shift the operator insertion.}
    \label{fig:2d_cft}
\end{figure}
Further, since the theory on the pencil is chiral, null translations along $v$ are equivalent to spatial translations along $x$.
This follows from the fact that all the fields are purely functions of $v$ and independent of $u$.
Thus, although we are interested in null derivatives for the Renyi QNEC, we can instead compute spatial derivatives.
In a chiral theory, the quantization surface $u=0$ is equivalent to the quantization surface $t=0$ by a similar argument.
Thus, we can describe the most general one particle states by a Euclidean path integral over the lower half plane, $\tau<0$, with a single smeared insertion of the lowest primary operator $\partial \Phi$ as seen in Fig.~(\ref{fig:2d_cft}).\footnote{One could instead consider an insertion of the operator $\Phi(v)$, but it leads to zero-mode subtleties discussed in \cite{Bousso:2015wca}.}

The affine parameter $\lambda=V(\tilde{p})$ describes the location of the entangling surface $\partial R$ on the pencil $p$.
From the previous discussion, we can equivalently describe it as the location $x=\lambda$ on the $\tau=0$ slice in the Euclidean path integral.
By translation symmetry, we can always move the entangling surface to $x=0$ by adjusting the operator insertions appropriately.
Using the path integral, we can then compute the reduced density matrix by tracing out the region $x<0$ corresponding to $\bar{R}$.
Shape derivatives can then be computed by deforming the location of the entangling surface.
Equivalently, we can take a passive perspective where we deform the locations of the operator insertions while holding the subregion fixed.
We will use the latter perspective in our analysis.
With this convention, the vacuum density matrix on $p$, i.e., $\sigma_p$ and hence, $\sigma$ in Eq.~\eqref{eq-null-vac} are independent of $\lambda$.

Now consider an arbitrary state on $N$.
The reduced state on the region $v>V(\vec{y})$ can be expanded as \cite{Bousso:2015wca}
\begin{align}
    \rho(\lambda) \, = \, \rho^{(0)} \, + \, \mathcal{A}^{1/2} \, \rho^{(1)}(\lambda) \, + \, O(\mathcal{A}) \, . \label{eq-null-state}
\end{align}
Note that an arbitrary state does not factorize because of entanglement between the pencil and auxiliary system in general states.
In this equation, $\rho^{(0)}$ is given by
\begin{align}
    \rho^{(0)} \, = \, \sigma_{p} \, \otimes \, \rho^{(0)}_{\aux} \, , \label{eq-null-state-0}
\end{align}
where $\rho^{(0)}_{\aux}$ is an arbitrary state of the auxiliary system, not necessarily the vacuum.
Moreover, $\rho^{(1)}(\lambda)$ in Eq.~\eqref{eq-null-state}, as discussed before, is given by a sum over 1-particle states on the pencil entangled with arbitrary states of the auxiliary system.
We can write this succinctly as
\begin{align}
    \rho^{(1)}(\lambda) \, = \, \sum_{\alpha\beta} \, \left( \sigma_{p} \, \int dr d\theta \, f_{\alpha\beta}(r,\theta) \, \partial\Phi(r e^{i\theta} \, - \, \lambda) \, \right) \, \otimes \, \ket{\alpha}\bra{\beta} \, ,\label{eq-null-state-1-int}
\end{align}
where $\{\ket{\alpha}\}$ is an arbitrary basis of $\mathcal{H}_{\text{aux}}$.
In \cite{Bousso:2015wca}, this basis was chosen to be the one in which $\rho^{(0)}_{\aux}$ is diagonal.
However, we will make a different choice in Sec.~(\ref{sec-identity}) which will be more convenient for the Renyi QNEC.

Before we proceed, we note that the source function $f_{\alpha\beta}(r,\theta)$ satisfies
\begin{align}
    f_{\alpha\beta}(r,2\pi-\theta) \, = \, f_{\beta\alpha}^{*}(r,\theta) \, . \label{eq-null-reality-1}
\end{align}
This condition ensures that the reduced density matrix is Hermitian.
We also require that $f_{\alpha\beta}(r,\theta)$ vanishes at the quantization surface, $\theta=0$ and $\theta=\pi$, so that the state is normalizable.

Further, in order to ensure finite SRD with respect to the vacuum, we restrict the source functions to have support in the wedge $\theta > \pi -\frac{\pi}{n}$ as discussed in Sec.~(\ref{sec-SRDQFT}).
For the auxiliary system on the other hand, we require that the support of $\rho^{(0)}_{\aux}$ be contained in the support of $\sigma_{\text{aux}}$ for $n>1$ as discussed in Sec.~(\ref{sec-SRDdensity}).
Apart from this, we will not require any additional structure on $\mathcal{H}_{\text{aux}}$, and it can be thought of as an arbitrary quantum mechanical system.

\subsection{Statement of Renyi QNEC} \label{sec-null-QNEC-stat}

Having reviewed the setup for null quantization of a free scalar theory, we now use it to find an equivalent formulation of the Renyi QNEC in this context.
As reviewed in Sec.~(\ref{sec-SRD}), the sandwiched Renyi divergence (SRD) is given by
\begin{align}
    S_{n}(\rho||\sigma) \, = \, \frac{1}{n-1} \, \log \, \widehat{Z}_{n}(\rho||\sigma) \, ,
\end{align}
where
\begin{align}
    \widehat{Z}_{n}(\rho||\sigma) \, \equiv \, \tr \, \left(\sigma^{\frac{1-n}{2n}} \, \rho \, \sigma^{\frac{1-n}{2n}}\right)^{n} \, .\label{eq-null-hatz-def}
\end{align}

The Renyi QNEC is a condition on the SRD of an arbitrary reduced density matrix $\rho(\lambda)$ with respect to the vacuum density matrix $\sigma$.
We are considering reduced density matrices for the subregion $v>V(\vec{y})$ and the label $\lambda$ represents the affine parameter along a specific pencil $p$ as seen in Fig.~(\ref{fig:null-quant}).
The Renyi QNEC involves computing the diagonal part of the second SRD variation, Eq.~\eqref{eq-Renyi}, where one deforms $V(y)$ in an infinitesimal transverse patch of area $\mathcal{A}$, around the point $\tilde{p}$, along pencil $p$.
In the above setup, where we have discretized the null plane into pencils, this is equivalent to a deformation of the affine parameter $\lambda$.
Thus, the Renyi QNEC becomes the the statement that
\begin{align}
    \lim_{\mathcal{A}\to 0} \, \frac{1}{\mathcal{A}} \, \frac{d^{2}}{d\lambda^{2}} \, S_{n}(\rho(\lambda)||\sigma) \Big|_{\lambda=0} \, \ge \, 0 \, ,\label{eq-null-qnec-conj}
\end{align}
where $\mathcal{A}$ is the transverse area of $p$ and the limit $\mathcal{A}\to0$ is taken while holding the overall state fixed.

Using Eq.~\eqref{eq-null-state}, $\rho(\lambda)$ can be expanded as
\begin{align}
    \rho(\lambda) \, = \, \sigma_{p}\otimes\rho^{(0)}_{\aux} \, +& \, \mathcal{A}^{1/2} \, \rho^{(1)}(\lambda) \, + \, \mathcal{A} \, \rho^{(2)}(\lambda) \, + \, O(\mathcal{A}^{3/2}) \, ,
\end{align}
where the explicit form of $\rho^{(1)}(\lambda)$ is given in Eq.~\eqref{eq-null-state-1-int}. 
Now using this perturbative expansion, we can expand $\widehat{Z}_{n}$ as
\begin{align}
    \widehat{Z}_{n}(\rho(\lambda)||\sigma) \, = \, \widehat{Z}^{(0)}_{n}(\lambda) \, + \, \mathcal{A}^{1/2} \, \widehat{Z}^{(1)}_{n}(\lambda) \, + \, \mathcal{A} \, \widehat{Z}^{(2)}_{n}(\lambda) \, + \, \mathcal{A} \, \widehat{Z}^{(1,1)}_{n}(\lambda) \, + \, O(\mathcal{A}^{3/2}) \, .\label{eq-null-sn-exp}
\end{align}
Note that we get two contributions at order $\mathcal{A}$, a contribution linear in $\rho^{(2)}$ that is denoted by $\widehat{Z}^{(2)}_{n}(\lambda)$ and a contribution quadratic in  $\rho^{(1)}$ that is denoted by $\widehat{Z}^{(1,1)}_{n}(\lambda)$.
We will now show that the first three terms on the right hand side of Eq.~\eqref{eq-null-sn-exp} are independent of $\lambda$ and hence, do not contribute to the Renyi QNEC.

Let us start by considering the first term.
At order $\mathcal{A}^{0}$, we have
\begin{align}
    \widehat{Z}^{(0)}_{n}(\lambda) \, =& \,\, \tr \, \left( \big(\sigma_{p}\otimes\sigma_{\aux}\big)^{\frac{1-n}{2n}} \, \big(\sigma_{p}\otimes\rho^{(0)}_{\aux}\big) \, \big(\sigma_{p}\otimes\sigma_{\aux}\big)^{\frac{1-n}{2n}}\right)^{n} \, ,\\
    =& \,\, \tr_{\aux} \, \left(\sigma_{\aux}^{\frac{1-n}{2n}} \, \rho^{(0)}_{\aux} \, \sigma_{\aux}^{\frac{1-n}{2n}}\right)^{n} \, ,
\end{align}
which is manifestly independent of $\lambda$.
Now we consider the second term in Eq.~\eqref{eq-null-sn-exp}.
For integer $n$, the form of the expression can be deduced by using the cyclicity of trace and collecting terms with a single power of $\rho^{(1)}$, i.e.,
\begin{align}
    \widehat{Z}_{n}^{(1)}(\lambda) \, =& \,\,  n \, \tr \left( \Big(\sigma^{\frac{1-n}{2n}} \, \rho^{(0)}(\lambda) \, \sigma^{\frac{1-n}{2n}}\Big)^{n-1} \, \cdot \, \Big(\sigma^{\frac{1-n}{2n}} \, \rho^{(1)}(\lambda) \, \sigma^{\frac{1-n}{2n}}\Big) \right) \, . \label{eq-null-z1-int}
\end{align}
In fact, the above result is also true for non-integer $n$, as we show in Appendix~(\ref{sec-app-iden}).
Using Eq.~\eqref{eq-null-vac} and Eq.~\eqref{eq-null-state-0}, we have
\begin{align}
    \sigma^{\frac{1-n}{2n}} \,  \Big(\sigma^{\frac{1-n}{2n}} \, \rho^{(0)} \, \sigma^{\frac{1-n}{2n}}\Big)^{n-1} \,  \sigma^{\frac{1-n}{2n}} \, = \, \sigma_{\aux}^{\frac{1-n}{2n}} \,  \Big(\sigma_{\aux}^{\frac{1-n}{2n}} \, \rho_{\aux}^{(0)} \, \sigma_{\aux}^{\frac{1-n}{2n}}\Big)^{n-1} \,  \sigma_{\aux}^{\frac{1-n}{2n}} \, .
\end{align}
Now using this result and the cyclic property of the trace, we write Eq.~\eqref{eq-null-z1-int} as
\begin{align}
    \widehat{Z}_{n}^{(1)}(\lambda) \, =& \,\,  n \, \tr \left( \rho^{(1)}(\lambda) \, \cdot \,  \sigma_{\aux}^{\frac{1-n}{2n}} \,  \Big(\sigma_{\aux}^{\frac{1-n}{2n}} \, \rho_{\aux}^{(0)} \, \sigma_{\aux}^{\frac{1-n}{2n}}\Big)^{n-1} \,  \sigma_{\aux}^{\frac{1-n}{2n}}  \right) \, ,\\
    =& \,\,  n \, \tr_{\aux} \left( \rho^{(1)}(\infty)\ \, \cdot \,  \sigma_{\aux}^{\frac{1-n}{2n}} \,  \Big(\sigma_{\aux}^{\frac{1-n}{2n}} \, \rho_{\aux}^{(0)} \, \sigma_{\aux}^{\frac{1-n}{2n}}\Big)^{n-1} \,  \sigma_{\aux}^{\frac{1-n}{2n}}  \right) \, ,
\end{align}
where $\rho^{(1)}(\infty) \, = \, \tr_{p}\rho^{(1)}(\lambda)$.
This implies that $\widehat{Z}_{n}^{(1)}$ is independent of $\lambda$.
By an identical argument, we can also show that $\widehat{Z}_{n}^{(2)}$ is also independent of $\lambda$.
This verifies our claim that the first three terms in Eq.~\eqref{eq-null-sn-exp} are independent of $\lambda$.

The above analysis implies that only $\widehat{Z}_{n}^{(1,1)}(\lambda)$ contributes to the left hand side of Eq.~\eqref{eq-null-qnec-conj}.
Expanding the logarithm and using the $\lambda$-independence of the remaining
terms, we find that the statement of the Renyi QNEC is equivalent to
\begin{align}
\frac{d^{2}}{d\lambda^{2}} \, {Z}_{n}^{(1,1)}(\lambda) \Big|_{\lambda=0} \, &\ge \, 0 \, , \label{eq-null-renyi-qnec}
\end{align}
where
\begin{align}
    {Z}_{n}^{(1,1)}(\lambda) \, = \, \frac{1}{n-1} \, \widehat{Z}_{n}^{(1,1)}(\lambda) \, , \label{eq-null-z-def}
\end{align}
and we have left the dependence on the states implicit since they are fixed for our analysis.
\subsection{Free Field Calculation as a Perturbative Calculation}\label{sec-identity}

In Sec.~(\ref{sec-proof}), our goal will be to prove the inequality in Eq.~\eqref{eq-null-renyi-qnec} for arbitrary $n>1$ and demonstrate counterexamples for $n<1$.
In order to do so, we will need to compute Eq.~\eqref{eq-null-renyi-qnec} for an arbitrary state which is in general a difficult task.
However, the main result of this section will be Eq.~\eqref{eq-nearvac}, which relates the calculation for an arbitrary state, far from the vacuum, to a perturbative calculation of SRD around a nearby state.
This relation will then allow us to use perturbative calculations of relative entropy, and more generally SRD, in Sec.~(\ref{sec-proof}) which have been performed extensively in the past \cite{Faulkner:2014jva,Lashkari:2015hha,Faulkner:2017tkh,May:2018tir,Ugajin:2018rwd,Bao:2019aol}.\footnote{Similar perturbative calculations for entanglement/Renyi entropies were done in Refs.~\cite{Rosenhaus:2014woa,Rosenhaus:2014ula,Allais:2014ata,Lewkowycz:2014jia,Rosenhaus:2014zza,Mezei:2014zla,Carmi:2015dla,Faulkner:2015csl,Leichenauer:2016rxw,Belin:2018juv,Agon:2020fqs}.}

As discussed in Sec.~(\ref{sec-null-QNEC-stat}), the $O(\mathcal{A})$ terms in the state $\rho(\lambda)$ in Eq.~\eqref{eq-null-state} do not contribute to the Renyi QNEC.
Therefore, we will ignore those terms from now on and will simply write the state $\rho(\lambda)$ as
\begin{align}
    \rho(\lambda) \, = \, \rho^{(0)} \, + \, \mathcal{A}^{1/2} \, \rho^{(1)}(\lambda) \, , \label{eq-null-state-sim}
\end{align}
where $\rho^{(0)}$ and $\rho^{(1)}(\lambda)$ are given in Eq.~\eqref{eq-null-state-0} and Eq.~\eqref{eq-null-state-1-int} respectively.
Despite the fact that we have a perturbative expansion in $\mathcal{A}$, we
still need to compute the SRD between the states $\rho(\lambda)$ and $\sigma$
which are not necessarily ``close''.
Namely, the state $\rho^{(0)}$ in Eq.~\eqref{eq-null-state-sim} is not the vacuum state unless $\rho_{\aux}^{(0)} \, = \, \sigma_{\aux}$.
Therefore, known perturbative results for SRD are not directly applicable.\footnote{Instead, if we were interested in the $\lambda$-derivative of the relative entropy, we could simply use $\rho^{(0)}$ instead of $\sigma$ as the reference state.
This is because $S_{\text{rel}}(\rho(\lambda)||\sigma)$ and $S_{\text{rel}}(\rho(\lambda)||\rho^{(0)})$ only differ by a $\lambda$-independent `constant':
\begin{equation*}
    S_{\text{rel}}(\rho(\lambda)||\sigma) \, - \, S_{\text{rel}}(\rho(\lambda)||\rho^{(0)}) \, = \, \tr \Big( \rho(\lambda) \, \big(\log\rho_{\aux}^{0} - \log\sigma_{\aux} \big) \Big) \, = \, \tr_{\aux} \Big( \rho(\infty) \, \big(\log\rho_{\aux}^{0} - \log\sigma_{\aux} \big) \Big) \, .
\end{equation*}
Hence, we could have used the known relative entropy formula for nearby states to prove the QNEC as was done in \cite{Balakrishnan:2019gxl}.}

However, we will now derive an identity for $Z_{n}(\lambda)$ which relates it to a calculation in a perturbatively close state.
In particular, we show that ${Z}_{n}(\lambda)$, defined in Eq.~\eqref{eq-null-hatz-def} and Eq.~\eqref{eq-null-z-def}
 can be written as
\begin{align}
    {Z}_{n}(\lambda) \, = \, \frac{1}{n-1} \, \tr \, \left( \big(\trho^{(0)} \big)^{1/n} \, + \, \mathcal{A}^{1/2} \, \big(\trho^{(0)}\big)^{\frac{1-n}{2n}} \cdot \, \trho^{(1)}(\lambda) \, \cdot \, \big(\trho^{(0)}\big)^{\frac{1-n}{2n}} \right)^{n} \, , \label{eq-null-zn-eqv}
\end{align}
where we have analogous to Eq.~\eqref{eq-null-state-0}
\begin{align}
    \trho^{(0)} \, \equiv \, \sigma_{p} \, \otimes \, \trho^{(0)}_{\aux} \, \quad\quad\quad\quad\quad \, \trho_{\aux}^{(0)} \, \equiv \, \left( \sigma_{\aux}^{\frac{1-n}{2n}} \, \rho_{\aux}^{(0)} \, \sigma_{\aux}^{\frac{1-n}{2n}} \right)^{n} \, , \label{eq-null-sand-0}
\end{align}
and analogous to Eq.~\eqref{eq-null-state-1-int},
\begin{align}
    \trho^{(1)}(\lambda)  \, = \, \sum_{\alpha\beta} \, \trho^{(0)} \, \int dr d\theta \, \tf_{\alpha\beta}(r,\theta) \, \left( \partial\Phi(r e^{i\theta} \, - \, \lambda) \,  \otimes \, E_{\alpha\beta}(\theta) \right) \, . \label{eq-null-sand-1}
\end{align}
The precise definition of $\tf_{\alpha\beta}$ and $E_{\alpha\beta}(\theta)$ are given later in Eq.~\eqref{eq-null-tf-def} and in Eq.~\eqref{eq-null-def-Eab} respectively.
Note that the $n$ dependence of quantities like $\trho^{(0)}$ and $\tf_{\alpha\beta}(r,\theta)$ will be kept implicit henceforth to simplify the notation.

Before we derive Eq.~\eqref{eq-null-zn-eqv}, it is worth pointing out that the usefulness of Eq.~\eqref{eq-null-zn-eqv} stems from the
observation that we can use Eq.~\eqref{eq-null-z-def} to write Eq.~\eqref{eq-null-zn-eqv} as
\begin{align}\label{eq-nearvac}
    {Z}_{n}(\lambda) \, = \, {Z}_{n}\left(\trho^{(0)}  \, + \, \mathcal{A}^{1/2} \, \trho^{(1)}(\lambda) \Big|\Big|\trho^{(0)}\right) \, .
\end{align}
In this sense, we have reduced our calculation of SRD of an arbitrary state $\rho(\lambda)$ with respect to the vacuum $\sigma$ to the calculation of two perturbatively close density matrices.
The price that we have to pay is that the density matrix $\trho^{(0)}$ is not normalized and hence, is not a physical state.
Nevertheless, this will allow us to use the known perturbative results for SRD in our analysis \cite{Faulkner:2014jva,Lashkari:2015hha,Faulkner:2017tkh,May:2018tir,Ugajin:2018rwd,Bao:2019aol}.


To derive Eq.~\eqref{eq-null-zn-eqv}, we start with the definition of ${Z}_{n}$ in Eq.~\eqref{eq-null-hatz-def}
 and use Eq.~\eqref{eq-null-state-sim} to get
\begin{align}
    {Z}_{n}(\lambda) \, = \, \frac{1}{n-1} \, \tr \, \left( \sigma^{\frac{1-n}{2n}} \, \rho^{(0)} \, \sigma^{\frac{1-n}{2n}} \, + \, \mathcal{A}^{1/2} \, \sigma^{\frac{1-n}{2n}} \, \rho^{(1)}(\lambda) \, \sigma^{\frac{1-n}{2n}} \right)^{n} \, . \label{eq-null-zn-sim-exp}
\end{align}
Now, using Eq.~\eqref{eq-null-vac} and Eq.~\eqref{eq-null-state-0}, we can write the first term in the parenthesis as
\begin{align}
    \sigma^{\frac{1-n}{2n}} \, \rho^{(0)} \, \sigma^{\frac{1-n}{2n}} \, = \, \sigma_{p}^{1/n} \otimes \left( \sigma_{\aux}^{\frac{1-n}{2n}} \, \rho_{\aux}^{(0)} \, \sigma_{\aux}^{\frac{1-n}{2n}} \right) \, = \, \left( \sigma_{p} \otimes \trho_{\aux}^{(0)} \right)^{1/n} \, = \, \big(\trho^{(0)} \big)^{1/n}  \, ,
\end{align}
where $\trho_{\aux}^{(0)}$ and $\trho^{(0)}$ are defined in Eq.~\eqref{eq-null-sand-0}.

We now consider the second term inside the parenthesis in Eq.~\eqref{eq-null-zn-sim-exp}.
Using Eq.~\eqref{eq-null-vac} and Eq.~\eqref{eq-null-state-1-int}, we get
\begin{align}
    \sigma^{\frac{1-n}{2n}} \, \rho^{(1)}(\lambda) \, \sigma^{\frac{1-n}{2n}} \, = \, \sum_{\alpha\beta} \, &\left( \sigma_{p}^{\frac{1-n}{2n}} \cdot \sigma_{p} \, \int dr d\theta \, f_{\alpha\beta}(r,\theta) \, \partial\Phi(r e^{i\theta} \, - \, \lambda) \, \cdot \sigma_{p}^{\frac{1-n}{2n}} \right) \, \nonumber\\ \otimes& \, \left(\sigma_{\aux}^{\frac{1-n}{2n}}\ket{\alpha}\bra{\beta}\sigma_{\aux}^{\frac{1-n}{2n}}\right) \, .
\end{align}
By inserting complete sets of states, we can write this as
\begin{align}
    \sigma^{\frac{1-n}{2n}} \, \rho^{(1)}(\lambda) \, \sigma^{\frac{1-n}{2n}} \, = \, \sum_{\alpha'\beta'}\sum_{\alpha\beta} \, &\left( \sigma_{p}^{\frac{1-n}{2n}} \cdot \sigma_{p} \, \int dr d\theta \, f_{\alpha\beta}(r,\theta) \, \partial\Phi(r e^{i\theta} \, - \, \lambda) \, \cdot \sigma_{p}^{\frac{1-n}{2n}} \right) \, \nonumber\\ \otimes& \, \left(\bra{\alpha'}\sigma_{\aux}^{\frac{1-n}{2n}}\ket{\alpha} \bra{\beta}\sigma_{\aux}^{\frac{1-n}{2n}} \ket{\beta'} \, \cdot \, \ket{\alpha'}\bra{\beta'} \right) \, . \label{eq-null-sand-1-int}
\end{align}
We now define
\begin{align}
    \tf_{\alpha\beta}(r,\theta) \, \equiv \, \sum_{\alpha'\beta'} \, f_{\alpha'\beta'}(r,\theta) \, \bra{\alpha}\sigma_{\aux}^{\frac{1-n}{2n}}\ket{\alpha'} \, \bra{\beta'}\sigma_{\aux}^{\frac{1-n}{2n}} \ket{\beta} \, , \label{eq-null-tf-def}
\end{align}
where, the support condition required in our choice of states for SRD to be finite, ensures that $\tf_{\alpha\beta}$ is well defined.
It is straightforward to check that $\tf_{\alpha\beta}$ satisfies the reality condition in Eq.~\eqref{eq-null-reality-1}, i.e.,
\begin{align}
    \tf_{\alpha\beta}(r,2\pi-\theta) \, = \, \tf_{\beta\alpha}^{*}(r,\theta) \, . \label{eq-null-reality}
\end{align}
Additionally, $\tf_{\alpha\beta}(r,\theta)$ also satisfies the same wedge condition discussed below Eq.~\eqref{eq-null-reality-1}, which was required for SRD to be finite.
With this definition, we simplify Eq.~\eqref{eq-null-sand-1-int} to get
\begin{align}
    \sigma^{\frac{1-n}{2n}} \, \rho^{(1)}(\lambda) \, \sigma^{\frac{1-n}{2n}} \, = \, \sum_{\alpha\beta}  \left( \sigma_{p}^{\frac{1-n}{2n}} \cdot \sigma_{p} \, \int dr d\theta \, \tf_{\alpha\beta}(r,\theta) \, \partial\Phi(r e^{i\theta}  -  \lambda) \, \cdot \sigma_{p}^{\frac{1-n}{2n}} \right) \, \otimes \, \ket{\alpha}\bra{\beta}  \, . \label{eq-null-sand-1-int-2}
\end{align}
So far we have not fixed the choice of the basis $\{\ket{\alpha}\}$.
Now we choose $\{\ket{\alpha}\}$ to be the basis in which $\trho^{(0)}_{\aux}$ is diagonal, i.e.,
\begin{align}
    \trho^{(0)}_{\aux} \, \ket{\alpha} \, = \, e^{-2\pi K_{\alpha}} \, \ket{\alpha} \, .  \label{eq-null-trho-eigen}
\end{align}
Moreover we follow \cite{Bousso:2015wca} and define:
\begin{align}
    E_{\alpha\beta}(\theta) \, \equiv \, e^{\theta(K_{\alpha}-K_{\beta})} \, \ket{\alpha}\bra{\beta} \, . \label{eq-null-def-Eab}
\end{align}
It is then easy to check that
\begin{align}
   \left(\trho_{\aux}^{(0)}\right)^{\frac{1-n}{2n}} \, \trho_{\aux}^{(0)} \, E_{\alpha\beta}(\theta) \, \left(\trho_{\aux}^{(0)}\right)^{\frac{1-n}{2n}} \, = \, e^{-(2\pi-\theta)K_{\alpha}} \, e^{-\theta K_{\beta}} \, \, e^{ \frac{\pi(n-1)}{n} (K_{\alpha}+K_{\beta})} \,  \ket{\alpha}\bra{\beta} \, . \label{eq-null-eab-ex}
\end{align}
Also note that $\tf_{\alpha\beta}(r,\theta)$ is arbitrary as long as it satisfies the reality condition in Eq.~\eqref{eq-null-reality}.
This means we can rescale $\tf_{\alpha\beta}(r,\theta)$ by
\begin{align}
    \tf_{\alpha\beta}(r,\theta) \, \longrightarrow \, e^{(2\pi-\theta)K_{\alpha}} \, e^{\theta K_{\beta}} \, \, e^{ \frac{\pi(1-n)}{n} (K_{\alpha}+K_{\beta})} \, \tf_{\alpha\beta}(r,\theta) \, .\label{eq-null-tf-rescale}
\end{align}
With this definition, the rescaled $\tf_{\alpha\beta}(r,\theta)$ also satisfies the reality condition, Eq.~\eqref{eq-null-reality}, and the wedge condition for finiteness of SRD.
Finally, we can combine Eq.~\eqref{eq-null-eab-ex} with Eq.~\eqref{eq-null-tf-rescale} to write Eq.~\eqref{eq-null-sand-1-int-2} as
\begin{align}
    \sigma^{\frac{1-n}{2n}} \, \rho^{(1)}(\lambda) \, \sigma^{\frac{1-n}{2n}} \, = \, \sum_{\alpha\beta} \, &\left( \sigma_{p}^{\frac{1-n}{2n}} \cdot \sigma_{p} \, \int dr d\theta \, \tf_{\alpha\beta}(r,\theta) \, \partial\Phi(r e^{i\theta} \, - \, \lambda) \, \cdot \sigma_{p}^{\frac{1-n}{2n}} \right) \, \nonumber\\ \otimes& \, \left( \left(\trho_{\aux}^{(0)}\right)^{\frac{1-n}{2n}} \, \cdot \, \trho_{\aux}^{(0)} \, E_{\alpha\beta}(\theta) \, \cdot \,  \left(\trho_{\aux}^{(0)}\right)^{\frac{1-n}{2n}} \right) \, . \label{eq-null-sand-1-int-3}
\end{align}
Equivalently, this can be written in the form
\begin{align}
    \sigma^{\frac{1-n}{2n}} \, \rho^{(1)}(\lambda) \, \sigma^{\frac{1-n}{2n}} \, = \, \left(\trho^{(0)}\right)^{\frac{1-n}{2n}} \cdot \, \trho^{(1)}(\lambda) \, \cdot \, \left(\trho^{(0)}\right)^{\frac{1-n}{2n}} \, ,
\end{align}
where $\trho^{(1)}(\lambda)$ is defined in Eq.~\eqref{eq-null-sand-1}.

This completes our derivation of Eq.~\eqref{eq-null-zn-eqv}.
We will use this result in the following section to derive an expression for $Z_{n}^{(1,1)}$ which will then allow us to prove the Renyi QNEC for $n>1$.

\section{Proof of Renyi QNEC for Free Field Theories}
\label{sec-proof}
Having setup the problem, we now arrive at our main section where we will prove the Renyi QNEC for $n>1$ and disprove it for $n<1$ in free field theories.
As a warm-up, we prove the Renyi QNEC for integer $n>1$ in Sec.~(\ref{sec-int-n-proof}).
This proof is simpler to understand than the proof for arbitrary $n$ since it follows directly from reflection positivity of Euclidean correlation functions without having to do an explicit computation.
In Sec.~(\ref{sec-null-zn-corr}), we shift our focus to arbitrary values of $n$ and perform an explicit computation of the second shape derivative of SRD.
Using the result of this computation, we prove that the Renyi QNEC holds for arbitrary $n>1$ in Sec.~(\ref{sec-proof-Renyi-QNEC}).
In Sec.~(\ref{sec-violation-Renyi-QNEC}), we show using a simple counterexample that the Renyi QNEC is violated for $n<1$.

\subsection{Proving the Renyi QNEC for integer \texorpdfstring{$n>1$}{n > 1}} \label{sec-int-n-proof}

In this subsection, we will assume that $n > 1$ is an integer and prove the Renyi QNEC, i.e., Eq.~\eqref{eq-null-renyi-qnec} for this simpler case.
Despite the fact that this calculation is a special case of a more general analysis for arbitrary $n$ that we will present later, it is still useful and interesting to consider it separately.
In particular, as we will show in this subsection, the Renyi QNEC for integer $n>1$ follows from reflection positivity of Euclidean correlation functions.
This also gives us a better understanding of when the Renyi QNEC can be saturated for integer $n>1$, a statement that we will not be able to make rigorously for arbitrary $n$.
Moreover, we will establish some notation in this subsection that will also be useful in later subsections.

To prove the Renyi QNEC, we start with the identity Eq.~\eqref{eq-null-zn-eqv} that we derived in Sec.~(\ref{sec-identity}), i.e.,
\begin{align}\label{eq-identity-2}
    Z_{n}(\lambda) \, = \, \frac{1}{n-1} \, \tr \, \left( \big(\trho^{(0)} \big)^{1/n} \, + \, \mathcal{A}^{1/2} \, \left(\trho^{(0)}\right)^{\frac{1-n}{2n}} \cdot \, \trho^{(1)}(\lambda) \, \cdot \, \left(\trho^{(0)}\right)^{\frac{1-n}{2n}} \right)^{n} \, .
\end{align}
For integer $n$, we have the identity
\begin{align}
    \tr \left(A_{0} + A_{1} \right)^{n} \, = \, \tr A_{0}^{n} \, + \, n \, \tr \left(A_{0}^{n-1} \, A_{1} \right) \, + \, \frac{n}{2} \, \sum_{k=0}^{n-2} \, \tr \left(A_{0}^{k} \, A_{1} \, A_{0}^{n-2-k} \, A_{1} \right) \, + \, ... \, .
\end{align}
Using this identity in Eq.~\eqref{eq-identity-2} and comparing with Eq.~\eqref{eq-null-sn-exp}, we deduce that for integer $n>1$ we have
\begin{align}
    Z_{n}^{(1,1)}(\lambda) \, = \, \frac{n}{2(n-1)} \, \sum_{k=1}^{n-1} \, \tr \left( \left(\trho^{(0)}\right)^{-1+\frac{k}{n}}  \, \trho^{(1)}(\lambda) \, \left(\trho^{(0)}\right)^{-\frac{k}{n}}  \, \trho^{(1)}(\lambda) \, \right) \, . \label{eq-null-z11-int}
\end{align}

Now for the Renyi QNEC, Eq.~\eqref{eq-null-renyi-qnec}, we need to compute the second derivative of $Z_{n}^{(1,1)}(\lambda)$ with respect to $\lambda$.
This yields
\begin{align}
    \ddot{Z}_{n}^{(1,1)}(\lambda) \, = \, \frac{n}{2(n-1)} \, \sum_{k=1}^{n-1} \, \Big[ 2  &\tr \left( \left(\trho^{(0)}\right)^{-1+\frac{k}{n}}  \, \dot{\trho}^{(1)}(\lambda) \, \left(\trho^{(0)}\right)^{-\frac{k}{n}}  \, \dot{\trho}^{(1)}(\lambda) \, \right) \nonumber\\ + \, &\tr \left( \left(\trho^{(0)}\right)^{-1+\frac{k}{n}}  \, \ddot{\trho}^{(1)}(\lambda) \, \left(\trho^{(0)}\right)^{-\frac{k}{n}}  \, \trho^{(1)}(\lambda) \, \right) \nonumber\\ + \, &\tr \left( \left(\trho^{(0)}\right)^{-1+\frac{k}{n}}  \, \trho^{(1)}(\lambda) \, \left(\trho^{(0)}\right)^{-\frac{k}{n}}  \, \ddot{\trho}^{(1)}(\lambda) \, \right) \, \Big]\, , \label{eq-null-int-n-ddz11}
\end{align}
where dot represents a derivative with respect to $\lambda$.
Using the expression for $\trho^{(1)}$ given in Eq.~\eqref{eq-null-sand-1} and evaluating the above expression at $\lambda \, = \, 0$, we obtain
\begin{align}
    \ddot{Z}_{n}^{(1,1)} \, = \, \frac{n}{2(n-1)}  \sum_{k=1}^{n-1} \, \int d\mu \, \Big[ 2  &\tr \left( \left(\trho^{(0)}\right)^{1-\frac{k}{n}}  \, \dot{\OO}_{\alpha'\beta'}(r_{2},\theta_{2}) \, \left(\trho^{(0)}\right)^{\frac{k}{n}}  \, \dot{\OO}_{\alpha\beta}(r_{1},\theta_{1}) \, \right) \label{eq-null-int-n-ddz-2}\\ + \, &\tr \left( \left(\trho^{(0)}\right)^{1-\frac{k}{n}}  \, \ddot{\OO}_{\alpha'\beta'}(r_{2},\theta_{2}) \, \left(\trho^{(0)}\right)^{\frac{k}{n}}  \, {\OO}_{\alpha\beta}(r_{1},\theta_{1}) \, \right) \nonumber\\ + \, &\tr \left( \left(\trho^{(0)}\right)^{1-\frac{k}{n}}  \, {\OO}_{\alpha'\beta'}(r_{2},\theta_{2}) \, \left(\trho^{(0)}\right)^{\frac{k}{n}}  \, \ddot{\OO}_{\alpha\beta}(r_{1},\theta_{1}) \, \right) \, \Big]\, , \nonumber
\end{align}
where we are using the following notation
\begin{align}
    \OO_{\alpha\beta}(r,\theta) \, &\equiv \, \partial\Phi(re^{i\theta}) \, \otimes \, E_{\alpha\beta}(\theta) \, , \label{eq-OO-0} \\
    \dot{\OO}_{\alpha\beta}(r,\theta) \, &\equiv \, \partial^{2}\Phi(re^{i\theta}) \, \otimes \, E_{\alpha\beta}(\theta) \, ,\label{eq-dOO-0}\\
    \ddot{\OO}_{\alpha\beta}(r,\theta) \, &\equiv \, \partial^{3}\Phi(re^{i\theta}) \, \otimes \, E_{\alpha\beta}(\theta) \, , \label{eq-ddOO-0}
\end{align}
and
\begin{align}\label{eq-dmu}
   \int d\mu \, = \, \sum_{\alpha'\beta'}\sum_{\alpha\beta} \, &\int dr_{1}dr_{2} \, \int_{\pi-\pi/n}^{\pi+\pi/n} d\theta_{1}d\theta_{2} \,\, \tf_{\alpha\beta}(r_{1},\theta_{1}) \, \tf_{\alpha'\beta'}(r_{2},\theta_{2}) \, .
\end{align}

To simplify Eq.~\eqref{eq-null-int-n-ddz-2}, we need to determine how the operators $\OO_{\alpha\beta}(r,\theta)$ transform under conjugation by $\trho^{(0)} \, = \, \sigma_{p}\otimes\trho_{\aux}^{(0)}$.
Conjugation by real powers of $\sigma_p$ generates Euclidean rotations and since $\partial^{m}\Phi$ has conformal dimension $(m,0)$, it transforms as
\begin{align}
    \sigma_{p}^{-\frac{k}{2n}} \, \partial^{m}\Phi(r e^{i\theta}) \, \sigma_{p}^{\frac{k}{2n}} \, = \, e^{i m\pi \frac{k}{n}} \, \partial^{m}\Phi(r e^{i\theta+i\pi k/n}) \, . \label{eq-phi-trans}
\end{align}
Moreover using the definition of $E_{\alpha\beta}$ in Eq.~\eqref{eq-null-def-Eab}, we deduce that
\begin{align}
    \left(\trho_{\aux}^{(0)}\right)^{-\frac{k}{2n}} \, E_{\alpha\beta}(\theta) \, \left(\trho_{\aux}^{(0)}\right)^{\frac{k}{2n}} \, =& \, E_{\alpha\beta}(\theta+\pi k/n) \, . \label{eq-Eab-trans}
\end{align}
Using these transformation properties, Eq.~\eqref{eq-null-int-n-ddz-2} can be simplified to obtain
\begin{align}
    \ddot{Z}_{n}^{(1,1)} \, = \, \frac{n}{2(n-1)} \sum_{k=1}^{n-1} \, \int d\mu \, \Big[ &2  \tr \left( \trho^{(0)}  \, \dot{\OO}_{\alpha'\beta'}(r_{2},\theta_{2}+\pi k/n) \, \dot{\OO}_{\alpha\beta}(r_{1},\theta_{1}-\pi k/n) \, \right) \label{eq-null-int-n-ddz-3}\\ + \, &e^{i \frac{2\pi k}{n}} \, \tr \left( \trho^{(0)} \, \ddot{\OO}_{\alpha'\beta'}(r_{2},\theta_{2}+\pi k/n)  \, {\OO}_{\alpha\beta}(r_{1},\theta_{1}-\pi k/n) \, \right) \nonumber\\ + \, &e^{-i \frac{2\pi k}{n}} \, \tr \left( \trho^{(0)} \, {\OO}_{\alpha'\beta'}(r_{2},\theta_{2}+\pi k/n) \, \ddot{\OO}_{\alpha\beta}(r_{1},\theta_{1}-\pi k/n) \, \right) \, \Big]\, . \nonumber
\end{align}
The wedge condition ensures that the operator insertions are angle-ordered. Hence,
each of the three terms above can be written as a two-point correlation function
of smeared operators.
Thus, Eq.~\eqref{eq-null-int-n-ddz-3} can be written as\footnote{Note that the density matrix on the pencil is chosen to be trace normalized and thus, the trace on the pencil can be written in terms of a correlation function. On the other hand, the correlation function on the auxiliary portion is defined in terms of a trace like in Eq.~\eqref{eq-null-int-n-ddz-3}.}
\begin{align}
    \ddot{Z}_{n}^{(1,1)} \, = \, \frac{n}{2(n-1)} \sum_{k=1}^{n-1} \, \int d\mu \, \Big[ &2 \left\langle  \dot{\OO}_{\alpha'\beta'}(r_{2},\theta_{2}+\pi k/n) \, \dot{\OO}_{\alpha\beta}(r_{1},\theta_{1}-\pi k/n) \, \right\rangle \label{eq-null-int-n-ddz-4}\\ + \, &e^{i \frac{2\pi k}{n}} \, \left\langle  \ddot{\OO}_{\alpha'\beta'}(r_{2},\theta_{2}+\pi k/n)  \, {\OO}_{\alpha\beta}(r_{1},\theta_{1}-\pi k/n) \, \right\rangle \nonumber\\ + \, &e^{-i \frac{2\pi k}{n}} \,  \left\langle {\OO}_{\alpha'\beta'}(r_{2},\theta_{2}+\pi k/n) \, \ddot{\OO}_{\alpha\beta}(r_{1},\theta_{1}-\pi k/n) \, \right\rangle \, \Big]\, . \nonumber
\end{align}

To simplify this expression further, we note that these correlation functions are in fact proportional to each other in a free scalar theory.
For example, we have
\begin{align}
    \left\langle  \ddot{\OO}_{\alpha'\beta'}(r_{2},\theta_{2})  \, {\OO}_{\alpha\beta}(r_{1},\theta_{1}) \, \right\rangle \, =& \,\, \big\langle  \partial^{3}\Phi(r_{2} e^{i\theta_{2}})  \, \partial\Phi(r_{1}e^{i\theta_{1}}) \, \big\rangle_{p} \, \, \, \left\langle  E_{\alpha'\beta'}(\theta_{2})  \, E_{\alpha\beta}(\theta_{1}) \, \right\rangle_{\aux} \, ,\nonumber\\
    =& \, - \big\langle  \partial^{2}\Phi(r_{2} e^{i\theta_{2}})  \, \partial^{2}\Phi(r_{1}e^{i\theta_{1}}) \, \big\rangle_{p} \, \, \, \left\langle  E_{\alpha'\beta'}(\theta_{2})  \, E_{\alpha\beta}(\theta_{1}) \, \right\rangle_{\aux} \, ,\nonumber\\
    =& \, - \left\langle  \dot{\OO}_{\alpha'\beta'}(r_{2},\theta_{2})  \, \dot{\OO}_{\alpha\beta}(r_{1},\theta_{1}) \, \right\rangle \, ,
\end{align}
where we have used
\begin{align}
    \big\langle  \partial^{3}\Phi(z)  \, \partial\Phi(w) \, \big\rangle_{p} \, = \, - \big\langle  \partial^{2}\Phi(z)  \, \partial^{2}\Phi(w) \, \big\rangle_{p} \, .
\end{align}
By a similar argument, we also have
\begin{align}
    \left\langle  {\OO}_{\alpha'\beta'}(r_{2},\theta_{2})  \, \ddot{\OO}_{\alpha\beta}(r_{1},\theta_{1}) \, \right\rangle \, = \, - \left\langle  \dot{\OO}_{\alpha'\beta'}(r_{2},\theta_{2})  \, \dot{\OO}_{\alpha\beta}(r_{1},\theta_{1}) \, \right\rangle \, .
\end{align}
With these observations, we can simplify Eq.~\eqref{eq-null-int-n-ddz-4} to write $\ddot{Z}_{n}^{(1,1)}$ as
\begin{align}
    \ddot{Z}_{n}^{(1,1)} \, = \, \frac{2n}{n-1}  \sum_{k=1}^{n-1} \, \sin^{2}\left(\frac{\pi k}{n}\right) \, \int d\mu \,  \left\langle  \dot{\OO}_{\alpha'\beta'}(r_{2},\theta_{2}+\pi k/n) \, \dot{\OO}_{\alpha\beta}(r_{1},\theta_{1}-\pi k/n) \, \right\rangle \, . \label{eq-null-int-n-ddz-5} 
\end{align}

\begin{figure}
    \centering
    \includegraphics[width=.6\textwidth]{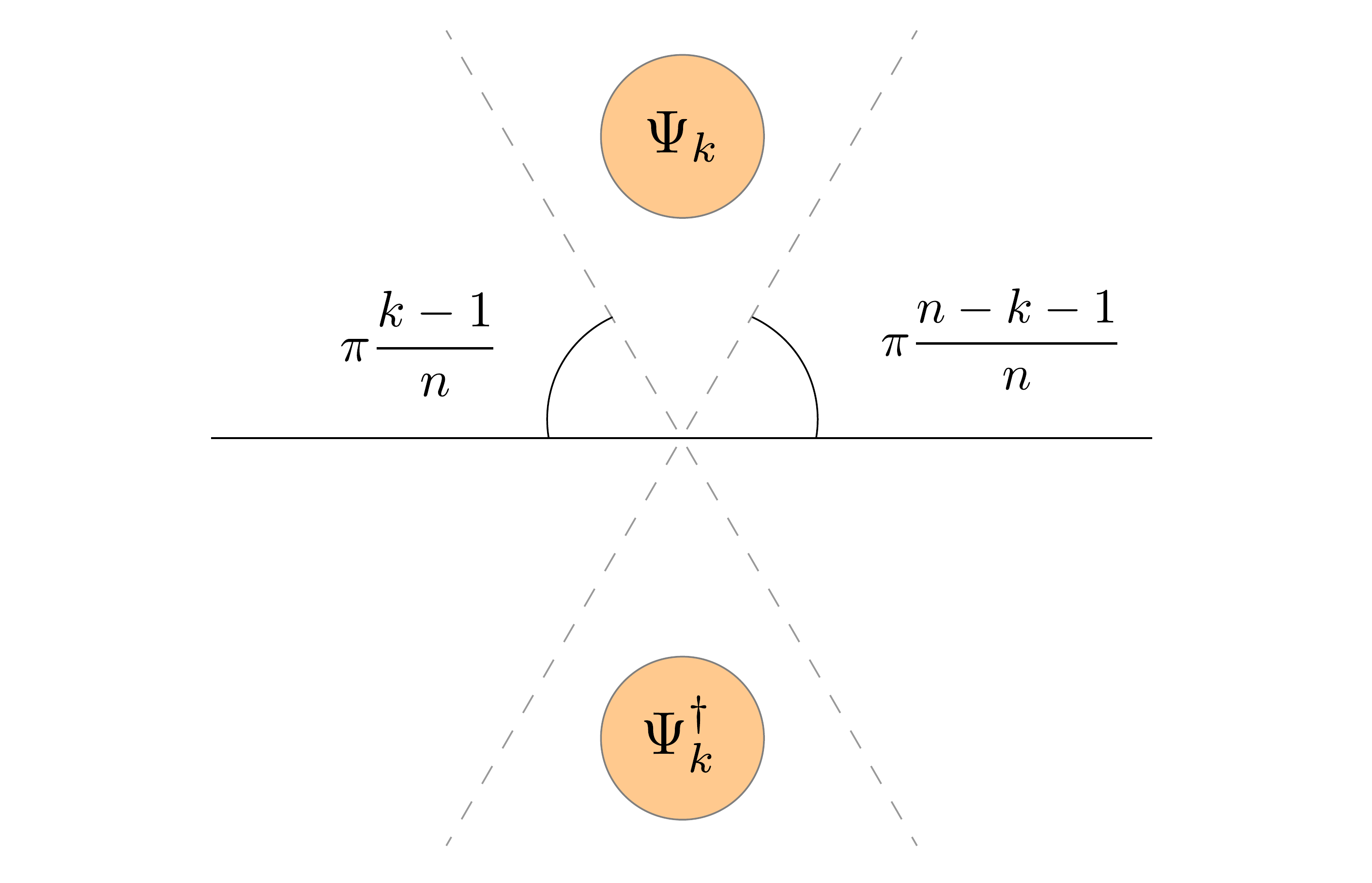}
    \caption{Reflection Positivity. For integer $n$, the second SRD variation is a sum of correlation functions of the operator insertions $\Psi_{k}$ and $\overline{\Psi}_{k}$, where $k$ ranges from $1$ to $n-1$. These correspond to insertions in the original wedge, now rotated by an amount $\frac{k\pi}{n}$ clockwise and anti-clockwise respectively. This results in a configuration where $\Psi_{k}$ is inserted in the upper-half plane, whereas $\overline{\Psi}_{k}$ is inserted symmetrically in the lower-half plane corresponding to the hermitian conjugate $\Psi_{k}^{\dagger}$. The correlation function $\big\langle \,   {\Psi}_{k} \, \Psi_{k}^{\dagger} \, \big\rangle$ is positive due to reflection positivity. The pictorial representation of $\big\langle \,   {\Psi}_{k} \, \Psi_{k}^{\dagger} \, \big\rangle$ is shown here for $n=6$ and $k=3$.}
    \label{fig-pi-rp}
\end{figure}

In order to prove that $ \ddot{Z}_{n}^{(1,1)}$ in Eq.~\eqref{eq-null-int-n-ddz-5} is positive, we write it as
\begin{align}
    \ddot{Z}_{n}^{(1,1)} \, = \, \frac{2n}{n-1}  \sum_{k=1}^{n-1} \, \sin^{2}\left(\frac{\pi k}{n}\right) \,\,  \big\langle \,   \overline{\Psi}_{k} \, \Psi_{k} \, \big\rangle \, , \label{eq-null-int-n-ddz-6}
\end{align}
where we have defined
\begin{align}
    \Psi_{k} \, \equiv& \,\, \sum_{\alpha\beta} \, \int dr \, \int_{\pi-\pi/n}^{\pi+\pi/n} d\theta \, \tf_{\alpha\beta}(r,\theta) \,\,  \dot{\OO}_{\alpha\beta}(r,\theta-\pi k/n) \, , \\
    \overline{\Psi}_{k} \, \equiv& \,\, \sum_{\alpha\beta} \, \int dr \, \int_{\pi-\pi/n}^{\pi+\pi/n} d\theta \, \tf_{\alpha\beta}(r,\theta) \,\,  \dot{\OO}_{\alpha\beta}(r,\theta+\pi k/n) \, .
\end{align}
Intuitively, the operators $\Psi_k$ and $\overline{\Psi}_k$ correspond to insertions in a wedge of size $\frac{2\pi}{n}$ as in Fig.~(\ref{fig:state_prep}), but now rotated by an angle $\frac{k\pi}{n}$ clockwise and anti-clockwise respectively.
This results in a configuration where $\Psi_k$ is inserted in the upper-half plane, whereas $\overline{\Psi}_k$ is inserted symmetrically in the lower-half plane as seen in Fig.~(\ref{fig-pi-rp}).

Now by changing the integration variable $\theta \to 2\pi-\theta$, we find that $\overline{\Psi}_{k}$ can be written as
\begin{align}
    \overline{\Psi}_{k} \, = \, (\trho^{(0)})^{-1} \, \Psi_{k}^{\dagger} \,\, (\trho^{(0)}) \, ,
\end{align}
where we have used the reality condition for $\tf_{\alpha\beta}(r,\theta)$ in Eq.~\eqref{eq-null-reality} and
\begin{align}
    \dot{\OO}_{\alpha\beta}(r,2\pi-\theta) \, = \, (\trho^{(0)})^{-1} \, \dot{\OO}_{\beta\alpha}^{\dagger}(r,\theta) \,\, (\trho^{(0)})
\end{align}
which follows from Eq.~\eqref{eq-dOO-0}, Eq.~\eqref{eq-phi-trans}, and Eq.~\eqref{eq-Eab-trans}.
Reflection positivity then implies that the two-point function in Eq.~\eqref{eq-null-int-n-ddz-6} is positive, i.e., 
\begin{align}
    \big\langle \,   \overline{\Psi}_{k} \, \Psi_{k} \, \big\rangle \, = \, \big\langle \,   (\trho^{(0)})^{-1} \, \Psi_{k}^{\dagger} \,\, (\trho^{(0)}) \, \Psi_{k} \, \big\rangle \, = \, \big\langle \,   {\Psi}_{k} \, \Psi_{k}^{\dagger} \, \big\rangle \, > \, 0 \, ,
\end{align}
where we have used the KMS condition in the second equality. Combining this with Eq.~\eqref{eq-null-int-n-ddz-6}, we deduce that
\begin{align}
    \ddot{Z}_{n}^{(1,1)} \, \geq \, 0 \, .
\end{align}
Notably, equality holds if and only if $\tf_{\alpha\beta}(r,\theta)$ vanishes identically.
This implies that the Renyi QNEC for integer $n>1$ is saturated if and only if the state on the pencil is approximated by the vacuum up to $O(\mathcal{A})$ terms instead of the generic $O(\mathcal{A}^{1/2})$ corrections that show up in Eq.~\eqref{eq-null-state-sim}.
We highlight that this is quite unlike the case of the QNEC as we review later in Eq.~\eqref{eq-nto1limitQNEC}, where there are many non-trivial choices of $\tf_{\alpha\beta}(r,\theta)$ that lead to saturation of the QNEC.

This finishes our proof of the Renyi QNEC for the simpler case that $n>1$ is an integer.
We will extend this proof to arbitrary $n>1$ in the following subsections.

\subsection{Calculating Second SRD Variation for arbitrary \texorpdfstring{$n$}{n}} \label{sec-null-zn-corr}

In the previous subsection, we found that $\ddot{Z}_{n}^{(1,1)}$ for integer $n$ can be written as a two-point function.
In this subsection, our goal is to generalize this result to non-integer $n$.
In particular, we will derive an expression for $\ddot{Z}_{n}^{(1,1)}$ in terms of a two-point function and will use this expression to calculate $\ddot{Z}_{n}^{(1,1)}$ explicitly in a free field theory.
This result will be used later to prove the Renyi QNEC for $n > 1$ in Sec.~(\ref{sec-proof-Renyi-QNEC}) and to disprove it for $n<1$ in Sec.~(\ref{sec-violation-Renyi-QNEC}).

Before considering arbitrary $n$, let us assume for a moment that $n$ is an integer.
In this case, we found in Eq.~\eqref{eq-null-z1-int} that
\begin{align}
    Z_{n}^{(1,1)}(\lambda) \, = \, \frac{n}{2(n-1)} \, \sum_{k=1}^{n-1} \, \tr \left( \left(\trho^{(0)}\right)^{-1+\frac{k}{n}}  \, \trho^{(1)}(\lambda) \, \left(\trho^{(0)}\right)^{-\frac{k}{n}}  \, \trho^{(1)}(\lambda) \, \right) \, , \label{eq-null-z11-int-new}
\end{align}
We now pick a basis in which $\trho^{(0)}$ is diagonal,
\begin{align}
    \trho^{(0)} \, \ket{\kappa} \, = \, e^{-2\pi \kappa} \, \ket{\kappa} \, ,
\end{align}
where we remind the reader that $\trho^{(0)}$ is unnormalized and thus, $\kappa$ need not be positive.
Evaluating the trace in Eq.~\eqref{eq-null-z11-int-new} in this basis, we obtain
\begin{align}
    Z_{n}^{(1,1)}(\lambda) \, =& \, \frac{n}{2(n-1)} \, \int d\kappa \int d\kappa' \,\, \sum_{k=1}^{n-1} \, e^{2\pi\kappa} \, e^{-2\pi(\kappa-\kappa')\frac{k}{n}} \, \left|\bra{\kappa}\trho^{(1)}(\lambda)\ket{\kappa'}\right|^{2} \, ,\\
    =& \, \frac{1}{2} \, \int d\kappa \int d\kappa' \,\,   e^{2\pi\kappa'} \, F_{n}(\kappa-\kappa') \, \left|\bra{\kappa}\trho^{(1)}(\lambda)\ket{\kappa'}\right|^{2} \, ,  \label{eq-null-z11-int-2}
\end{align}
where
\begin{align}
    F_{n}(x) \, = \, \frac{n}{1-n} \, \frac{e^{2\pi \left(\frac{n-1}{n}\right) x}  - 1}{e^{-2\pi x/n}  - 1} \, . \label{eq-null-Fn-def}
\end{align}
Despite the fact that Eq.~\eqref{eq-null-z11-int-2} was derived by assuming $n$ is an integer, the result is in fact valid for all $n$ \cite{May:2018tir}.
For completeness, we present a derivation of Eq.~\eqref{eq-null-z11-int-2} for arbitrary $n$ in Appendix~(\ref{sec-app-iden}).

In order to study the Renyi QNEC, Eq.~\eqref{eq-null-renyi-qnec}, we take the second derivative of Eq.~\eqref{eq-null-z11-int-2} with respect to $\lambda$.
This yields
\begin{align}
    \ddot{Z}_{n}^{(1,1)}(\lambda) \, = \, \int d\kappa \int d\kappa' \,\, e^{2\pi\kappa'} \, &F_{n}(\kappa-\kappa')
    \,\,  \Big(\bra{\kappa}{\trho}^{(1)}(\lambda)\ket{\kappa'}\bra{\kappa'}\ddot{\trho}^{(1)}(\lambda)\ket{\kappa} \nonumber\\ &+ \bra{\kappa}\dot{\trho}^{(1)}(\lambda)\ket{\kappa'}\bra{\kappa'}\dot{\trho}^{(1)}(\lambda)\ket{\kappa}\Big) \, ,\label{eq-null-z11-dd-int}
\end{align}
where we remind the reader that dots represent derivatives with respect to $\lambda$.
Now using the expression for $\trho^{(1)}$ in Eq.~\eqref{eq-null-sand-1}, we obtain 
\begin{align}
    \ddot{Z}_{n}^{(1,1)}(\lambda=0) \, = \, \int d\mu \int d\kappa \int &d\kappa' \, F_{n}(\kappa-\kappa') \, e^{-2\pi\kappa} \, \Big(\bra{\kappa}{\OO}^{\alpha\beta}(r_{1},\theta_{1})\ket{\kappa'}\bra{\kappa'}\ddot{\OO}^{\alpha'\beta'}(r_{2},\theta_{2})\ket{\kappa} \nonumber\\ &  + \, \bra{\kappa}\dot{\OO}^{\alpha\beta}(r_{1},\theta_{1})\ket{\kappa'}\bra{\kappa'}\dot{\OO}^{\alpha'\beta'}(r_{2},\theta_{2})\ket{\kappa}\Big) \, , \label{eq-null-z11-dd-int-3}
\end{align}
where we have used the notation introduced in Eqs.~\eqref{eq-OO-0}-\eqref{eq-ddOO-0}, and $d\mu$ includes the smearing function for both operator insertions as in Eq.~\eqref{eq-dmu}.

We remind the reader that for $n>1$, we restrict our Euclidean source functions $\tf_{\alpha\beta}(r,\theta)$ to only have support in a wedge of angular size $2\pi/n$ centered around $\theta = \pi$.
As discussed in Sec.~(\ref{sec-SRDQFT}), this is a necessary condition to guarantee that the SRD is finite.
It is instructive to understand why this condition is indispensable at the level of Eq.~\eqref{eq-null-z11-dd-int-3} .
We can see this from the fact that the matrix element $\bra{\kappa}\OO_{\alpha\beta}(r,\theta)\ket{\kappa'}$ scales as $e^{\theta(\kappa-\kappa')}$\footnote{This follows from writing $\OO_{\alpha\beta}(r,\theta) \, \sim \, ( \trho^{(0)} )^{-\theta/2\pi} \, \OO_{\alpha\beta}(r,0) \,  ( \trho^{(0)} )^{\theta/2\pi}$.}.
This implies that the integrand in Eq.~\eqref{eq-null-z11-dd-int-3} scales like $e^{(\theta_{1}-\theta_{2}-2\pi/n)\kappa}$ for large $\kappa$.
Therefore, in order for the integral in Eq.~\eqref{eq-null-z11-dd-int-3} to converge, we must demand that the source functions vanish outside a wedge of angular size $2\pi/n$.
Moreover, the reality condition for $\tf_{\alpha\beta}(r,\theta)$ in  Eq.~\eqref{eq-null-reality} ensures that this wedge must be centered around $\theta = \pi$.

In order to compute Eq.~\eqref{eq-null-z11-dd-int-3}, we will write it as a correlation function.
Since this requires being careful about angular-ordering, we first split the integral into two parts, i.e., $\theta_{1}>\theta_{2}$ and $\theta_{2}>\theta_{1}$ respectively, to get
\begin{align}
    \ddot{Z}_{n}^{(1,1)} \, = \, \int_{\theta_{1}>\theta_{2}} d\mu \int d\kappa \int &d\kappa' \, F_{n}(\kappa-\kappa') \, e^{-2\pi\kappa} \, \Big(\bra{\kappa}{\OO}^{\alpha\beta}(r_{1},\theta_{1})\ket{\kappa'}\bra{\kappa'}\ddot{\OO}^{\alpha'\beta'}(r_{2},\theta_{2})\ket{\kappa} \nonumber\\ &  + \, \bra{\kappa}\dot{\OO}^{\alpha\beta}(r_{1},\theta_{1})\ket{\kappa'}\bra{\kappa'}\dot{\OO}^{\alpha'\beta'}(r_{2},\theta_{2})\ket{\kappa}\Big) \, \nonumber\\
    + \int_{\theta_{2}>\theta_{1}} d\mu  \int d\kappa \int &d\kappa' \, F_{n}(\kappa'-\kappa) \, e^{-2\pi\kappa'} \, \Big(\bra{\kappa}{\OO}^{\alpha\beta}(r_{1},\theta_{1})\ket{\kappa'}\bra{\kappa'}\ddot{\OO}^{\alpha'\beta'}(r_{2},\theta_{2})\ket{\kappa} \nonumber\\ &  + \, \bra{\kappa}\dot{\OO}^{\alpha\beta}(r_{1},\theta_{1})\ket{\kappa'}\bra{\kappa'}\dot{\OO}^{\alpha'\beta'}(r_{2},\theta_{2})\ket{\kappa}\Big) \, , \label{eq-null-z11-dd-int-7}
\end{align}
where, in the integral for $\theta_{2}>\theta_{1}$, we have used the property
\begin{align}
    F_{n}(x) \, = \, e^{2\pi x} \, F_{n}(-x) \, . \label{eq-null-F-x}
\end{align}
Now, to simplify this result, we follow \cite{Faulkner:2017tkh,May:2018tir} and define the Fourier transform of $F_{n}$,
\begin{align}
    F_{n}(x) \, \equiv \,  \int_{-\infty}^{\infty} ds \, e^{is x} \, \mathcal{F}_{n}(s) \, . \label{eq-Fn-fourier}
\end{align}
Since $F_{n}$ grows exponentially for $n>1$, $\mathcal{F}_{n}$ in fact has to be defined as a distribution over test functions which decay faster than any exponential.
The space of such distributions, which is larger than the more familiar space of tempered distributions defined over Schwartz functions, is studied in \cite{gordon10}.
The explicit expression of $\mathcal{F}_{n}$ will not be important for our analysis and we will only use it as an intermediate tool.
Expressing Eq.~\eqref{eq-null-z11-dd-int-7} in terms of $\mathcal{F}_{n}(s)$, we obtain
\begin{align}
    \ddot{Z}_{n}^{(1,1)} \, = \, \int_{\theta_{1}>\theta_{2}} d\mu \int_{-\infty}^{\infty} ds \,  \mathcal{F}_{n}(s) \,  \Big[ &\tr \left( \left(\trho^{(0)}\right)^{1-{is}/{2\pi}} {\OO}^{\alpha\beta}(r_{1},\theta_{1})  \left(\trho^{(0)}\right)^{{is}/{2\pi}}\ddot{\OO}^{\alpha'\beta'}(r_{2},\theta_{2}) \right) \nonumber\\ + & \tr \left( \left(\trho^{(0)}\right)^{1-{is}/{2\pi}} \dot{\OO}^{\alpha\beta}(r_{1},\theta_{1}) \left(\trho^{(0)}\right)^{{is}/{2\pi}} \dot{\OO}^{\alpha'\beta'}(r_{2},\theta_{2}) \right) \Big] \, \nonumber\\
    + \int_{\theta_{2}>\theta_{1}} d\mu \int_{-\infty}^{\infty} ds \,  \mathcal{F}_{n}(-s) \,  \Big[ &\tr \left( \left(\trho^{(0)}\right)^{1+{is}/{2\pi}} \ddot{\OO}^{\alpha'\beta'}(r_{2},\theta_{2})  \left(\trho^{(0)}\right)^{-{is}/{2\pi}} {\OO}^{\alpha\beta}(r_{1},\theta_{1}) \right) \nonumber\\ + & \tr \left( \left(\trho^{(0)}\right)^{1+{is}/{2\pi}} \dot{\OO}^{\alpha'\beta'}(r_{2},\theta_{2}) \left(\trho^{(0)}\right)^{-{is}/{2\pi}} \dot{\OO}^{\alpha\beta}(r_{1},\theta_{1}) \right) \Big] \, .
\end{align}
Given that correlation functions are defined to be angular ordered in Euclidean time, we can now write the above expression as
\begin{align}
     \ddot{Z}_{n}^{(1,1)} \, = \, \int d\mu \,   \int_{-\infty}^{\infty} ds \, \mathcal{F}_{n}\Big(\text{sgn}(\theta_{1}-\theta_{2}) \, s \Big) \, \mathcal{G}(s) \, , \label{eq-null-z11-corr-1}
\end{align}
where we have defined
\begin{align}
    \mathcal{G}(s) \, \equiv \, &\left\langle \big( \trho^{(0)} \big)^{-is/2\pi} \, {\OO}^{\alpha\beta}(r_{1},\theta_{1}) \, \big( \trho^{(0)} \big)^{is/2\pi} \,   \ddot{\OO}^{\alpha'\beta'}(r_{2},\theta_{2}) \right\rangle \,\nonumber\\ \, + \, &\left\langle \big( \trho^{(0)} \big)^{-is/2\pi}  \, \dot{\OO}^{\alpha\beta}(r_{1},\theta_{1}) \, \big( \trho^{(0)} \big)^{is/2\pi} \, \dot{\OO}^{\alpha'\beta'}(r_{2},\theta_{2}) \right\rangle \, . \label{eq-null-def-Gs}
\end{align}

Thus, we have obtained an expression for $\ddot{Z}_{n}^{(1,1)}$ in terms of correlation functions in the free field theory.
However, the formula in Eq.~\eqref{eq-null-z11-corr-1} is not very useful for computing $\ddot{Z}_{n}^{(1,1)}$ since it depends on the distribution $\mathcal{F}_{n}(s)$.
To remedy this, we write the Fourier transformed version of Eq.~\eqref{eq-null-z11-corr-1}, i.e.,
\begin{align}
     \ddot{Z}_{n}^{(1,1)} \, = \, \int d\mu \,   \int_{-\infty}^{\infty} dw \, {F}_{n}\Big(\text{sgn}(\theta_{1}-\theta_{2}) \, \omega \Big) \, G(\omega) \, , \label{eq-null-z11-corr-2}
\end{align}
where
\begin{align}
    G(\omega) \, \equiv \, \frac{1}{2\pi} \, \int_{-\infty}^{\infty} ds \, e^{-is\omega} \, \mathcal{G}(s) \, . \label{eq-null-def-Gw}
\end{align}
Equivalently, we can write Eq.~\eqref{eq-null-z11-corr-2} as
\begin{align}
     \ddot{Z}_{n}^{(1,1)} \, = \, \int d\mu \,   \int_{-\infty}^{\infty} d\omega \, \tF_{n}(\omega) \,  e^{\pi \, \text{sgn}(\theta_{1}-\theta_{2}) \, \omega} \, G(\omega) \, , \label{eq-null-z11-corr}
\end{align}
where we have defined
\begin{align}
     \tF_{n}(\omega) \, \equiv& \,\, e^{-\pi\omega} \, F_{n}(\omega) \, = \, \frac{n}{n-1} \, \frac{ \sinh\Big( \pi \big(\frac{n-1}{n}\big) \omega \Big)}{ \sinh\big(\pi \omega/n\big)} \, . \label{eq-null-tFn-def}
\end{align}

The correlation function $\mathcal{G}(s)$ and its Fourier transform $G(\omega)$ can be calculated using the known correlation functions of the $1+1$ free CFT of a chiral boson.
We leave the details of this calculation to Appendix~(\ref{app-int-Gw}) and simply present the final answer here:
\begin{align}
    {G}(\omega) \, = \, &\, \, \frac{1}{2} \, \delta_{\alpha\beta'} \, \delta_{\beta\alpha'} \, e^{-\pi\left(K_{\alpha}+K_{\beta}\right)} \, \frac{1}{\big(r_{1} e^{i\theta_{1}}\big)^{2} \, \big(r_{2} e^{i\theta_{2}}\big)^{2}} \, \left(\frac{r_{1}}{r_{2}}\right)^{iv_{\alpha\beta}} \, e^{-\pi \, \text{sgn}(\theta_{1}-\theta_{2}) \, \omega} \, \label{eq-null-Gw-final}\\ &\quad\times \Bigg[ Q\big(v_{\alpha\beta}-\omega\big) \, \left(\frac{r_{1}e^{i\theta_{1}}}{r_{2}e^{i\theta_{2}}}\right)^{-i\omega} \, + \, Q\big(v_{\alpha\beta}-\omega-i\big) \, \left(\frac{r_{1}e^{i\theta_{1}}}{r_{2}e^{i\theta_{2}}}\right)^{1-i\omega} \Bigg] \, ,\nonumber
\end{align}
where $v_{\alpha\beta} \, \equiv \, K_{\alpha} \, - \, K_{\beta}$, and  
\begin{align}
    Q(x) \, \equiv \, \frac{x(x^{2}+1)}{\sinh (\pi x)} \, . \label{eq-Q-def}
\end{align}
Using the result for $\mathcal{G}(\omega)$ in Eq.~\eqref{eq-null-z11-corr}, we find that $\ddot{Z}_{n}^{(1,1)}$ becomes
\begin{align}
    \ddot{Z}_{n}^{(1,1)} \, = \,&\,  \frac{1}{2} \, \int d\tilde{\mu} \,\, e^{-\pi\left(K_{\alpha}+K_{\beta}\right)} \, \frac{1}{\big(r_{1} e^{i\theta_{1}}\big)^{2} \, \big(r_{2} e^{i\theta_{2}}\big)^{2}} \, \left(\frac{r_{1}}{r_{2}}\right)^{iv_{\alpha\beta}} \, \int_{-\infty}^{\infty} d\omega \, \tF_{n}(\omega)
    \, \label{eq-null-dd-z11-final}\\ &\quad\times\Bigg[ Q\big(v_{\alpha\beta}-\omega\big) \, \left(\frac{r_{1}e^{i\theta_{1}}}{r_{2}e^{i\theta_{2}}}\right)^{-i\omega} \, + \, Q\big(v_{\alpha\beta}-\omega-i\big) \, \left(\frac{r_{1}e^{i\theta_{1}}}{r_{2}e^{i\theta_{2}}}\right)^{1-i\omega} \Bigg] \, , \nonumber
\end{align}
where
\begin{align}
    \int d\tilde{\mu} \, \equiv \, \sum_{\alpha\beta} \, \int dr_{1}dr_{2} \, \int d\theta_{1}d\theta_{2} \,\, \tf_{\alpha\beta}(r_{1},\theta_{1}) \, \tf_{\beta\alpha}(r_{2},\theta_{2}) \, .\label{eq-null-dtmu}
\end{align}
Interestingly, we find that all the dependence on $\text{sgn}(\theta_{1}-\theta_{2})$ finally drops out.

The explicit expression for the second shape derivative of SRD, Eq.~\eqref{eq-null-dd-z11-final}, is the main result of this section.
In the next subsection, we will use this expression to show that $\ddot{Z}_{n}^{(1,1)}$ is non-negative for $n>1$, thus proving that the Renyi QNEC holds.
Having done that, we will focus on the case of $n<1$ in Sec.~(\ref{sec-violation-Renyi-QNEC}) and show that there exists states for which $\ddot{Z}_{n}^{(1,1)}$ is negative.
This implies that the Renyi QNEC is not generally true for $n<1$.

\subsection{Proving the Renyi QNEC for \texorpdfstring{$n>1$}{n > 1}} \label{sec-proof-Renyi-QNEC}

In this section, our goal is to show that $\ddot{Z}_{n}^{(1,1)}$ in Eq.~\eqref{eq-null-dd-z11-final} is non-negative for $n>1$ and thus to prove the Renyi QNEC for $n>1$ for free theories.

Recall that for $n>1$,  in order to have finite SRD, the source functions $\tf_{\alpha\beta}(r,\theta)$ are chosen to be non-vanishing only for $|\theta-\pi| < \pi/n$.
This means that we can write Eq.~\eqref{eq-null-dd-z11-final} as
\begin{align}
    \ddot{Z}_{n}^{(1,1)} \, = \,&\,  \frac{1}{2} \, \int d\tilde{\mu}_{n} \,\, e^{-\pi\left(K_{\alpha}+K_{\beta}\right)} \, \frac{1}{\big(r_{1} e^{i\theta_{1}}\big)^{2} \, \big(r_{2} e^{i\theta_{2}}\big)^{2}} \, \left(\frac{r_{1}}{r_{2}}\right)^{iv_{\alpha\beta}} \, \int_{-\infty}^{\infty} d\omega \, \tF_{n}(\omega)
    \, \label{eq-null-qnec-int}\\ &\quad\times\Bigg[ Q\big(v_{\alpha\beta}-\omega\big) \, \left(\frac{r_{1}e^{i\theta_{1}}}{r_{2}e^{i\theta_{2}}}\right)^{-i\omega} \, + \, Q\big(v_{\alpha\beta}-\omega-i\big) \, \left(\frac{r_{1}e^{i\theta_{1}}}{r_{2}e^{i\theta_{2}}}\right)^{1-i\omega} \Bigg] \, , \nonumber
\end{align}
where
\begin{align}
    \int d\tilde{\mu}_{n} \, \equiv \, \sum_{\alpha\beta} \, \int dr_{1}dr_{2} \, \int_{\pi-\pi/n}^{\pi+\pi/n} d\theta_{1}d\theta_{2} \,\, \tf_{\alpha\beta}(r_{1},\theta_{1}) \, \tf_{\beta\alpha}(r_{2},\theta_{2}) \, \label{eq-null-dtmun}
\end{align}
involves an integral with restricted angular bounds.

We will now simplify the integral in Eq.~\eqref{eq-null-qnec-int} by performing a contour deformation.
In order to do so, we note from Eq.~\eqref{eq-Q-def} that $Q\big(v_{\alpha\beta}-\omega\big)$ does not have any poles in the strip $-1 \, \le \, \text{Im}(\omega) \, \le \, 1 \,$.
Moreover, from Eq.~\eqref{eq-null-tFn-def}, we see that $\tF_{n}(\omega)$ has poles at $\omega \, = \, i\,n\,p$ where $p$ is a non-zero integer.
Therefore, for $n > 1$, a simple contour deformation from $\omega \to (\omega+i)$ implies that
\begin{align}
    \int_{-\infty}^{\infty} \, d\omega \,  \tF_{n}(\omega \, - \, i) \, Q\big(v_{\alpha\beta}-\omega\big) \, \left(\frac{r_{1}e^{i\theta_{1}}}{r_{2}e^{i\theta_{2}}}\right)^{-i\omega}  \, = \, \int_{-\infty}^{\infty} \, d\omega \, \tF_{n}(\omega) \, Q\big(v_{\alpha\beta}-\omega-i\big) \, \left(\frac{r_{1}e^{i\theta_{1}}}{r_{2}e^{i\theta_{2}}}\right)^{1-i\omega} \, ,\label{eq-null-int-rel}
\end{align}
where we have used the fact that the integrand vanishes for $\text{Re}(\omega) \, = \, \pm\infty$, since the bounds on integration restrict us to the domain $(\theta_{1}-\theta_{2})\le 2\pi/n$.
Using this observation, we write Eq.~\eqref{eq-null-qnec-int} as
\begin{align}
    \ddot{Z}_{n}^{(1,1)} \, = \,&\,  \frac{1}{2} \, \int d\tilde{\mu}_{n} \,\, e^{-\pi\left(K_{\alpha}+K_{\beta}\right)} \, \frac{1}{\big(r_{1} e^{i\theta_{1}}\big)^{2} \, \big(r_{2} e^{i\theta_{2}}\big)^{2}} \, \left(\frac{r_{1}}{r_{2}}\right)^{iv_{\alpha\beta}} \, \int_{-\infty}^{\infty} d\omega \, \left(\frac{r_{1}e^{i\theta_{1}}}{r_{2}e^{i\theta_{2}}}\right)^{-i\omega} \, \label{eq-null-qnec-int-2}\\ &\quad\times Q\big(v_{\alpha\beta}-\omega\big) \, \Big( \tF_{n}(\omega)  + \tF_{n}(\omega-i)   \Big) \, . \nonumber
\end{align}

Now, we are free to make a change of variables, $\alpha \leftrightarrow \beta$, $(r_{1},\theta_{1}) \leftrightarrow (r_{2},\theta_{2})$, and $\omega \to -\omega$ in Eq.~\eqref{eq-null-qnec-int-2}.
Under this change of dummy variables, we obtain
\begin{align}
    \ddot{Z}_{n}^{(1,1)} \, = \,&\,  \frac{1}{2} \, \int d\tilde{\mu}_{n} \,\, e^{-\pi\left(K_{\alpha}+K_{\beta}\right)} \, \frac{1}{\big(r_{1} e^{i\theta_{1}}\big)^{2} \, \big(r_{2} e^{i\theta_{2}}\big)^{2}} \, \left(\frac{r_{1}}{r_{2}}\right)^{iv_{\alpha\beta}} \, \int_{-\infty}^{\infty} d\omega \, \left(\frac{r_{1}e^{i\theta_{1}}}{r_{2}e^{i\theta_{2}}}\right)^{-i\omega} \, \label{eq-null-qnec-int-3}\\ &\quad\times Q\big(v_{\alpha\beta}-\omega\big) \, \Big( \tF_{n}(\omega)  + \tF_{n}(\omega+i)   \Big) \, , \nonumber
\end{align}
where we have used the fact that $Q(x)$ in Eq.~\eqref{eq-Q-def} and $\tF_{n}(\omega)$ in Eq.~\eqref{eq-null-tFn-def} are even functions: $Q(-x)=Q(x)$ and $\tF_{n}(-\omega) = \tF_{n}(\omega)$.
By combining Eqs.~\eqref{eq-null-qnec-int-2} and \eqref{eq-null-qnec-int-3}, and using Eq.~\eqref{eq-null-dtmun}, we obtain the expression
\begin{align}
    \ddot{Z}_{n}^{(1,1)} \, = \,\frac{1}{4} &\,  \sum_{\alpha\beta} \, e^{-\pi \left(K_{\alpha}+K_{\beta}\right)}  \,  \int_{-\infty}^{\infty} d\omega \, \widehat{F}_{n}(\omega) \, Q(v_{\alpha\beta} - \omega) \nonumber\\
    &\times \, \int dr_{1} \, \int_{\pi-\pi/n}^{\pi+\pi/n} d\theta_{1}\, \tf_{\alpha\beta}(r_{1},\theta_{1}) \, \left(r_{1}\right)^{iv_{\alpha\beta}-i\omega-2} \, e^{-(2i-\omega)\theta_{1}} \label{eq-null-qnec-int-4}\\
    &\times \, \int dr_{2} \, \int_{\pi-\pi/n}^{\pi+\pi/n} d\theta_{2}\, \tf_{\beta\alpha}(r_{2},\theta_{2}) \, \left(r_{2}\right)^{-iv_{\alpha\beta}+i\omega-2} \, e^{-(2i+\omega)\theta_{2}}\,  , \nonumber
\end{align}
where
\begin{align}
    \widehat{F}_{n}(\omega) \, \equiv& \,\, 2\tF_{n}\big(\omega\big) \, + \, \tF_{n}\big(\omega+i\big) \, + \, \tF_{n}\big(\omega-i\big) \, \\
    =&\,\, \frac{2n}{n-1} \, \frac{\sinh(\pi\omega)}{\tanh\big(\pi\omega/n\big)} \, \frac{\sin^{2}\big(\pi/n\big)}{\sinh^{2}\big(\pi\omega/n\big)+\sin^{2}\big(\pi/n\big)} \, ,\label{eq-null-fnw-final}
\end{align}
which is non-negative for $n>1$.
Finally, we take $\theta_{2} \to (2\pi-\theta_{2})$ and use the reality condition in Eq.~\eqref{eq-null-reality} to get
\begin{align}
    \ddot{Z}_{n}^{(1,1)} \, = \,\frac{1}{4} &\,  \sum_{\alpha\beta} \, e^{-\pi \left(K_{\alpha}+K_{\beta}\right)}  \,  \int_{-\infty}^{\infty} d\omega \, e^{-2\pi\omega} \, \widehat{F}_{n}(\omega) \, Q(v_{\alpha\beta} - \omega) \, \Big|M_{\alpha\beta}(\omega)\Big|^{2} \, , \label{eq-null-qnec-fin}
\end{align}
where we have defined
\begin{align}
    M_{\alpha\beta}(\omega) \, \equiv \, \int dr \, \int_{\pi-\pi/n}^{\pi+\pi/n} d\theta\, \tf_{\alpha\beta}(r,\theta) \, r^{iv_{\alpha\beta}-i\omega-2} \, e^{-(2i-\omega)\theta} \, . \label{eq-smearing-integral}
\end{align}
We deduce from Eq.~\eqref{eq-null-qnec-fin} that
\begin{align}\label{eq-finalQNEC}
    \ddot{Z}_{n}^{(1,1)} \, \ge \, 0 \, .
\end{align}
Hence, we have proved the Renyi QNEC for all $n>1$ in free field theories.
Further, we speculate that it is in fact impossible to have an equality in Eq.~\eqref{eq-finalQNEC} without choosing $\tf_{\alpha\beta}(r,\theta)$ to identically vanish.
This appears to be the case since vanishing of $M_{\alpha \beta}(\omega)$ in Eq.~\eqref{eq-null-qnec-fin} is a very constraining requirement.

It is interesting to note that we can recover the original result of \cite{Bousso:2015wca} by taking the $n \to 1^+$ limit of our result in Eq.~\eqref{eq-null-qnec-fin}.
This follows from the fact that 
\begin{align}
     \lim_{n\to 1^{+}} \, \widehat{F}_{n}\big(\omega\big) \, = \, 2\pi \, \delta\big(\omega\big) \, , \label{eq-null-fnw-lim-3}
\end{align}
which we derive in Appendix~(\ref{app-n-1}). Using this, we find that the $n\to 1^{+}$ limit of Eq.~\eqref{eq-null-qnec-fin} is
\begin{align}\label{eq-nto1limitQNEC}
    \lim_{n\to 1^{+}} \, \ddot{Z}_{n}^{(1,1)} \, = \, \frac{\pi}{2} \,  \sum_{\alpha\beta} \, e^{-\pi \left(K_{\alpha}+K_{\beta}\right)}  \,  Q(v_{\alpha\beta}) \, \Bigg|\int dr \, \int_{0}^{2\pi} d\theta\, \tf_{\alpha\beta}(r,\theta) \, r^{iv_{\alpha\beta}-2} \, e^{-2i\theta}\Bigg|^{2} \, .
\end{align}
This precisely matches the known result from \cite{Bousso:2015wca} which was used to prove the QNEC for free field theories (see also \cite{Balakrishnan:2019gxl}).
Note that since the integral here localizes to $\omega=0$, saturation of the QNEC follows from $M_{\alpha\beta}(0)$ vanishing.
This is a much weaker requirement than what was needed for the Renyi QNEC to be saturated.
As seen from Eq.~\eqref{eq-nto1limitQNEC}, any non-trivial smearing functions with vanishing second Fourier mode will in fact saturate the QNEC.

\subsection{Violation of Renyi QNEC for \texorpdfstring{$n<1$}{n < 1}} \label{sec-violation-Renyi-QNEC}
For $n < 1$, we now show, by providing an explicit example, that the Renyi QNEC can be violated in a suitably chosen state.
We again focus on the diagonal part of the second SRD variation, $\ddot{Z}_{n}^{(1,1)}$, but now we will demonstrate that this quantity need not be positive.
In fact, by choosing states where different pencils are unentangled with each other, the off diagonal SRD variation vanishes and thus, the same counterexample disproves the more general Renyi QNEC proposed in Ref.~\cite{Lashkari:2018nsl}.

Our starting point is the explicit expression for $\ddot{Z}_{n}^{(1,1)}$ in Eq~\eqref{eq-null-dd-z11-final}.
As noted previously, $\widetilde{F}_n(\omega)$ has poles at $\omega = i\,n\,p$ for non-zero integer values of $p$.
Thus, deformation of the integration contour $(\omega - i)\to \omega $ now results in an additional contribution from poles of $\widetilde{F}_n(\omega)$.
For the SRD, we are interested in the domain $1/2 < n < 1$, for which we only receive a contribution from a single pole at $\omega = -i \,n$.
The relation between the two terms in Eq. \eqref{eq-null-int-rel} obtained from contour deformation is now modified to
\begin{align}
    \int_{-\infty}^{\infty}  d\omega \,  \tF_{n}(\omega  -  i)  Q\big(v_{\alpha\beta}-\omega\big) \, \left(\frac{r_{1}e^{i\theta_{1}}}{r_{2}e^{i\theta_{2}}}\right)^{-i\omega}  \, =& \, \int_{-\infty}^{\infty}  d\omega \, \tF_{n}(\omega)  Q\big(v_{\alpha\beta}-\omega-i\big) \, \left(\frac{r_{1}e^{i\theta_{1}}}{r_{2}e^{i\theta_{2}}}\right)^{1-i\omega} \, \nonumber\\
    +& \,\, \frac{2 n^{2} \sin(\pi n)}{n-1} \,  Q\big(v_{\alpha\beta} - i(1-n) \big) \, \left(\frac{r_{1}e^{i\theta_{1}}}{r_{2}e^{i\theta_{2}}}\right)^{1-n}  \, ,
\end{align}
where we have used
\begin{align}
    \text{Res}\left[ \tF_n(\omega) \, ; \, \omega =  -i n \right] \, = \, \frac{i}{\pi} \, \frac{n^2}{1-n}\sin(\pi n) \, ,
\end{align}
and the fact that $Q\big(v_{\alpha\beta}-\omega\big)$ is analytic in the strip $-1 \, \le \, \text{Im}(\omega) \, \le \, 1 \,$.

Following similar manipulations as in the $n>1$ case, we arrive at the following expression,
\begin{align}
    \ddot{Z}_{n}^{(1,1)} \, = \,\frac{1}{4} \, \sum_{\alpha\beta} \, &e^{-\pi \left(K_{\alpha}+K_{\beta}\right)}  \, \Bigg[\int_{-\infty}^{\infty} d\omega \, e^{-2\pi\omega} \, \widehat{F}_{n}(\omega) \, Q(v_{\alpha\beta} - \omega) \, \Big|M_{\alpha\beta}(\omega)\Big|^{2} \, \label{eq-qnec-pole} \\
    & + \, \frac{4 n^2\sin(\pi n)}{1-n} \, \Big[Q\big(v_{\alpha\beta} - \omega\big) M_{\alpha\beta}(\omega)M_{\beta\alpha}(-\omega) \Big]\bigg|_{\omega = i(1-n)} \, \Bigg], \nonumber
\end{align}
which is importantly modified by the additional pole contribution.
Contrasting this with the $n>1$ case, we see that the kernel $\widehat{F}_n(\omega)$, given in Eq.~\eqref{eq-null-fnw-final}, is non-positive for $n<1$.
Thus, the first term now contributes negatively to $\ddot{Z}_{n}^{(1,1)}$ independent of the choice of state.
Thus, in order to show a violation of the Renyi QNEC for $n<1$, it suffices to show that the second term, which came from the additional pole contribution, can in fact be negative for a suitable choice of state.

The choice of state is encapsulated by the source function $\tilde{f}_{\alpha\beta}(r,\theta)$ and violations can be found by explicitly computing the second term in Eq.~\eqref{eq-qnec-pole} for various choices of $\tilde{f}_{\alpha\beta}(r,\theta)$.
A simple way, for example, to construct a state that leads to a violation is to consider a state in which there is no entanglement between the pencil and the auxillary system.
Namely, consider a state in Eq.~\eqref{eq-null-state-1-int} that is of the form
\begin{align}
    \rho^{(1)}(\lambda) \, = \, \left( \sigma_{p} \, \int dr d\theta \, f(r,\theta) \, \partial\Phi(r e^{i\theta} \, - \, \lambda) \, \right) \, \otimes \, \rho_{\aux}^{(0)} \, .
\end{align}
In this case, we can ignore the auxillary system in our analysis.
The result in Eq.~\eqref{eq-qnec-pole} is now restricted to a single allowed value of $\alpha$ and $\beta$.
With this product state in mind, consider the source function
\begin{align}
    {f}(r, \theta) \, = \, \delta(r-r_0) \,  \big(1-\cos(4\theta)\big) \, . \label{eq-violation-source}
\end{align}
For this choice of state, we find that the pole term in Eq.~\eqref{eq-qnec-pole} is
\begin{align}\label{eq-nless1-explicit}
    \ddot{Z}_{n}^{(1,1)}\Big|_{\text{pole}} \, = \, - \, \frac{ (2-n) \, n^{3} \, \sin^{2}(n\pi) \,}{(7-n) \, (3-n)^{2} \, (1+n)^{2} \, (5+n) \,} \, \left(\frac{8}{r_{0}}\right)^{4} \, .
\end{align}
This is manifestly negative for all $n \in [1/2 , 1)$ and hence, provides a simple violation to the Renyi QNEC for this range of $n$.
We see that the violation turns off as one takes the limit $n\to1^{-}$, where we recover the QNEC.
In fact, from Eq.~\eqref{eq-nto1limitQNEC}, we see that this choice of source function leads to QNEC saturation as we take the limit $n\to1^{-}$.
As discussed earlier, this in fact follows from the fact that our choice of source function has a vanishing second Fourier mode.

Given that $\ddot{Z}_{n}^{(1,1)}$ is necessarily non-negative for $n\ge 1$ whereas it can be negative for $n<1$, one can ask if $\ddot{Z}_{n}^{(1,1)}$ can jump discontinuously around $n = 1$.
We will now show that this does not in fact happen and $\ddot{Z}_{n}^{(1,1)}$ is continuous at $n=1$.
To see this, we take the limit of $\ddot{Z}_{n}^{(1,1)}$ in Eq.~\eqref{eq-qnec-pole} as $n \to 1^{-}$.
In Appendix~(\ref{app-n-1}), we show that
\begin{align}
      \lim_{n\to 1^{-}} \, \widehat{F}_{n}\big(\omega\big) \, = \, - \, 2\pi \, \delta\big(\omega\big) \, . \label{eq-null-fnw-lim-below}
\end{align}
Using this result, we find that Eq.~\eqref{eq-qnec-pole} in the limit $n \to 1^{-}$ reduces to
\begin{align}
    \lim_{n\to 1^{-}} \, \ddot{Z}_{n}^{(1,1)} \, = \, \frac{\pi}{2} \,  \sum_{\alpha\beta} \, e^{-\pi \left(K_{\alpha}+K_{\beta}\right)}  \,  Q(v_{\alpha\beta}) \, \Bigg|\int dr \, \int_{0}^{2\pi} d\theta\, \tf_{\alpha\beta}(r,\theta) \, r^{iv_{\alpha\beta}-2} \, e^{-2i\theta}\Bigg|^{2} \, .
\end{align}
This is the same as the known result from Ref.~\cite{Bousso:2015wca}, which we reviewed in Eq.~\eqref{eq-nto1limitQNEC}.
This shows that $\ddot{Z}_{n}^{(1,1)}$ is continuous at $n=1$, and violations of the Renyi QNEC must vanish as we take the limit $n\to1^{-}$ as we saw in our example, Eq.~\eqref{eq-nless1-explicit}.


\section{Generalizations and Future Directions}\label{sec-disc}

In summary, we have proved the diagonal part of the $n>1$ Renyi QNEC for free field theories in spacetime dimensions $d>2$.
We have also demonstrated counterexamples to show it does not generally hold in the case $n<1$.
Given this, one could consider various generalizations of our result which we now discuss in this section.

In Sec.~(\ref{sec-interacting}), we provide some evidence to show that the Renyi QNEC could extend to interacting QFTs as well.
In particular, we demonstrate using a perturbative calculation that to first non-trivial order the Renyi QNEC is in fact saturated in theories with a twist gap, similar to the QNEC \cite{Balakrishnan:2017bjg}.
In Sec.~(\ref{sec-offdiagonal}), we discuss the off-diagonal part of the Renyi QNEC.
In Sec.~(\ref{sec=2dQNEC}), we provide some numerical evidence for the Renyi QNEC in $d=2$.
In Sec.~(\ref{sec-alphaz}), we suggest generalizations of the Renyi QNEC to other measures of distinguishability beyond the SRD.

\subsection{Renyi QNEC in Interacting Theories}\label{sec-interacting}

Having proved the Renyi QNEC for $n>1$ in free theories, we now make a preliminary investigation into the Renyi QNEC in interacting theories.
Although, we will not be able to provide a general proof, we will show some evidence in favour of the Renyi QNEC being saturated  in interacting theories.
In this section, we consider states of interacting QFTs in $d>2$ that are perturbatively close to the vacuum.
Using a perturbative analysis, we will show that the Renyi QNEC is saturated to leading non-trivial order in interacting theories.
This calculation is a generalization of a similar result for the QNEC which we closely follow \cite{Balakrishnan:2019gxl}.

Consider a state perturbatively close to the vacuum, so that its reduced density matrix to a Rindler region $R$, $u<0$ {and} $v > 0$, is of the form
\begin{align}
    \rho_R \, = \, \sigma_R \, + \, \epsilon \, \delta\rho_R \, + \, O(\epsilon^{2}) \, ,\label{eq-rho-nv}
\end{align}
where $\sigma$ is the vacuum state reduced to region $A$ and $\delta\rho$ is given by
\begin{align}
    \delta\rho_R \, = \, \sigma_R \, \int dr d\theta d^{d-2}y \, f(r,\theta,y) \, \OO(r,\theta,y) \, , \label{eq-rho1-nv}
\end{align}
prepared by a Euclidean path integral with source function $f(r,\theta,y)$ restricted to the wegde, like in Fig.~(\ref{fig:state_prep}).
Note that $\{r,\theta\}$ are the Euclidean polar coordinates around the entangling surface $\partial R$, and $y$ denotes the $(d-2)$ transverse coordinates on $\partial R$.
In addition to the wedge condition, the source function $f(r,\theta,y)$ also satisfies a reality condition
\begin{align}
    f(r,\theta,y) \, = \, f^{*}(r,2\pi-\theta,y) \, .
\end{align}

Now consider the SRD between the state $\rho_R$ in Eq.~\eqref{eq-rho-nv} and the vacuum $\sigma_R$.
Since $\rho_R$ is perturbatively close to $\sigma_R$, we can expand the SRD in a perturbation series in the small parameter $\epsilon$.
Since both states are indistinguishable at $O(\epsilon^{0})$, the SRD vanishes at this order.
Moreover, since SRD is non-negative, it must vanish at $O(\epsilon)$ as well, a result termed the first law of Renyi divergence in Ref.~\cite{Faulkner:2020iou}.
Therefore, the first non-zero contribution to SRD appears at $O(\epsilon^{2})$.
This leading non-trivial contribution has been studied previously in \cite{May:2018tir} (see also Appendix~(\ref{sec-app-iden})), where it was found that
\begin{align}
    S_{n}(\rho_R||\sigma_R) \, = \, \frac{\epsilon^{2}}{2} \, \frac{n}{1-n} \, \int d\kappa \int d\kappa' \,\,   \frac{e^{2\pi \left(\frac{n-1}{n}\right) \kappa}  - e^{2\pi \left(\frac{n-1}{n}\right) \kappa'}}{e^{-2\pi \kappa/n}  - e^{-2\pi \kappa'/n}} \, \left|\bra{\kappa}\delta\rho_R\ket{\kappa'}\right|^{2} \, + \, O(\epsilon^{3}) \, ,\label{eq-sn-pert-nv}
\end{align}
where $\ket{\kappa}$ represents the eigenvectors of the reduced state $\sigma_R$, i.e.,
\begin{align}
    \sigma_R \ket{\kappa} \, = \, e^{-2\pi\kappa} \, \ket{\kappa} \, .
\end{align}
Eq.~\eqref{eq-sn-pert-nv} holds for an arbitrary state $\sigma_R$ and perturbation $\delta\rho_R$ in a given Hilbert space $\mathcal{H}_R$.
We now apply it to our case where $\sigma_R$ is the vacuum state of the QFT reduced to a Rindler region $R$, and the perturbation $\delta\rho_R$ is given by Eq.~\eqref{eq-rho1-nv}.
Thus, we obtain
\begin{align}
    S_{n}(\rho_R||\sigma_R) \, = \, \frac{\epsilon^{2}}{2} \, \int d\mu  \int d\kappa \int d\kappa' \, F_{n}(\kappa-\kappa') \, e^{-2\pi\kappa} \,  \bra{\kappa}\OO(r_{1},\theta_{1},y_{1})\ket{\kappa'} \bra{\kappa'}\OO(r_{2},\theta_{2},y_{2})\ket{\kappa} \, ,\label{eq-sn2-nv}
\end{align}
where
\begin{align}
    F_{n}(x) \, = \, \frac{n}{1-n} \, \frac{e^{2\pi \left(\frac{n-1}{n}\right) x}  - 1}{e^{-2\pi x/n}  - 1} \, \label{eq-Fn-vn}
\end{align}
is the same kernel that appeared in our free field calculation, Eq.~\eqref{eq-null-Fn-def}.
We have ignored $O(\epsilon^{3})$ term in Eq.~\eqref{eq-sn2-nv} since we are working to leading non-trivial order, and following Ref.~\cite{Balakrishnan:2019gxl}, we have used the shorthand notation
\begin{align}
    \int d\mu \, = \, \prod_{i=1}^{2} \, \left[ \int dr_{i} \,  d\theta_{i} \, d^{d-2}y_{i} \, f(r_{i},\theta_{i},y_{i}) \right] \, .
\end{align}

Now our goal is to write $S_{n}(\rho_R||\sigma_R)$ in Eq.~\eqref{eq-sn2-nv} in terms of a correlation function, analogous to Eq.~\eqref{eq-null-z11-corr-1} in the free field case.
To do this, we repeat the analysis of our free field theory calculation in Sec.~(\ref{sec-null-zn-corr}) and find that Eq.~\eqref{eq-sn2-nv} can be written as
\begin{align}
    S_{n}(\rho_R||\sigma_R) \, = \, \frac{\epsilon^{2}}{2} \, \int d\mu \int_{-\infty}^{\infty}ds \,
    \mathcal{F}_{n}\Big(\text{sgn}(\theta_{1}-\theta_{2}) \, s\Big) \, \left\langle \sigma_R^{-is/2\pi} \, \OO(r_{1},\theta_{1},y_{1}) \, \sigma_R^{is/2\pi} \, {\OO}(r_{2},\theta_{2},y_{2}) \right\rangle , \label{eq-sn-corr-s}
\end{align}
where $\mathcal{F}_{n}(s)$ is formally defined as the Fourier transform
\begin{align}
    \mathcal{F}_{n}(s) \, \equiv \, \frac{1}{2\pi} \, \int_{-\infty}^{\infty} dx \, e^{-isx} \, F_{n}(x) \, .
\end{align}
The exact expression for $\mathcal{F}_{n}(s)$ is not important for the following discussion.
In fact, as discussed in Sec.~(\ref{sec-null-zn-corr}), $F_{n}$ grows exponentially for $n>1$  and hence, $\mathcal{F}_{n}$ has to be defined as a distribution for test functions which decay faster than any exponential. 

For an operator $\OO$ supported purely in the region $R$, we have
\begin{align}
    \sigma_R^{-is/2\pi} \, \OO \, \sigma_R^{is/2\pi} \, = \, \Delta_{\Omega}^{-is/2\pi} \, \OO \, \Delta_{\Omega}^{is/2\pi} \, ,
\end{align}
which follows from Eq.~\eqref{eq-Delta-mat} and we have kept the $R$ dependence of $\Delta_{\Omega}$ implicit.
Moreover, $\Delta_{\Omega}$ leaves the vacuum invariant.
These two observations allow us to write Eq.~\eqref{eq-sn-corr-s} as
\begin{align}
    S_{n}(\rho_R||\sigma_R) \, = \, \frac{\epsilon^{2}}{2} \, \int d\mu \int_{-\infty}^{\infty}ds \,
    \mathcal{F}_{n}\Big(\text{sgn}(\theta_{1}-\theta_{2}) \, s\Big) \, \left\langle  \OO_{1} \, \Delta_{\Omega}^{is/2\pi} \, {\OO}_{2} \right\rangle , \label{eq-sn-corr-Delta}
\end{align}
where we have introduced the notation $\OO_{i} \, = \, \OO(r_{i},\theta_{i},y_{i})$.
The above correlation function with $\Delta_{\Omega}^{is/2\pi}$ has appeared previously in the literature.
For example, it appears in the formula for the relative entropy of near vacuum states in \cite{Balakrishnan:2019gxl} (see also \cite{Lashkari:2018oke}).
In fact, our result reduces to the known perturbative formula for relative entropy in the limit $n\to 1$.
Importantly, despite the fact that we derived Eq.~\eqref{eq-sn-corr-Delta} for $\partial R$ being a flat cut of a null plane, there is in fact evidence that it holds for an arbitrary cut of a null plane \cite{Balakrishnan:2019gxl}.
This involves subtleties with angular ordering in going from
Eq.~\eqref{eq-sn-corr-s} to Eq.~\eqref{eq-sn-corr-Delta} since the Euclidean modular flow no longer acts locally.
However, these issues can be fixed by insertions of the modular conjugation operator which do not affect the final result as suggested in \cite{Balakrishnan:2019gxl}.
We expect the same to hold in the case of SRD with the only difference coming from the integration kernel.
Thus, from now onwards we assume that Eq.~\eqref{eq-sn-corr-Delta} in fact holds for $\partial R$ being an arbitrary cut of the null plane, i.e., $u=0$ and $v=V(y)$.


We will now apply the theory of half-sided modular inclusions to study the second null deformation of SRD using the perturbative expression in Eq.~\eqref{eq-sn-corr-Delta}.
We first note that the shape dependence of SRD in Eq.~\eqref{eq-sn-corr-Delta} is completely captured by the dependence of the modular operator, $\Delta_{\Omega}$, on the subregion $R$.
Now consider a null cut $v = V(y)$ on the null plane $u = 0$ and let $\tilde{V}(y) > V(y)$.
Then the region $Q \, : \, u<0$ {and} $v > \tilde{V}(y)$ is a subregion of $R$.
Moreover, the algebra of operators in $R$ and in $Q$ satisfy the properties of {half-sided modular inclusion} \cite{wiesbrock1993}.
In this case, we have\footnote{This result is true more generally for any state $\ket{\psi}$ which is cyclic and separating for algebras of operators in $Q$ and $R$.
More generally, the operator $P$ is defined as $\frac{1}{2\pi} \, \left( \log\Delta_{\psi|R}-\log\Delta_{\psi|Q}\right) $ \cite{wiesbrock1993}.}  \cite{wiesbrock1993}
\begin{align}
    \Delta_{\Omega|R}^{is/2\pi} \, \Delta_{\Omega|Q}^{-is/2\pi} \, = \,  \exp\big( i (-1 + e^{-s}) \, P \big) \, , \label{eq-nested-mod-flow}
\end{align}
where \cite{Casini:2017roe} 
\begin{align}
    P \, \equiv \, \int d^{d-2}y \, \left(\Tilde{V}(y) \, - V(y) \right) \, \mathcal{E}(y) \, ,
\end{align}
is a positive operator and $\mathcal{E}$ is the averaged null energy (ANE) operator defined by
\begin{align}
    \mathcal{E}(y) \, = \, \int_{-\infty}^{\infty} \, dv \, T_{vv}(v,u=0,y) \, .
\end{align}

Now by using Eq.~\eqref{eq-nested-mod-flow} to compute null shape deformations of SRD in Eq.~\eqref{eq-sn-corr-Delta}, we get
\begin{align}
    \frac{\delta^{2} \, S_{n}(\rho_R||\sigma_R)}{\delta V(y')\delta V(y)} \,  = \, -\frac{\epsilon^{2}}{2} \, \int d\mu \int_{-\infty}^{\infty}ds& \, (-1 + e^{-s})^{2} \,
    \mathcal{F}_{n}\Big(\text{sgn}(\theta_{1}-\theta_{2}) \, s\Big) \, \nonumber\\ \times& \left\langle  \OO_{1} \, \mathcal{E}(y) \, \mathcal{E}(y') \, \Delta_{\Omega}^{is/2\pi} \, {\OO}_{2} \right\rangle . \label{eq-d2sn2}
\end{align}
In order to study the ``diagonal" part of the Renyi QNEC, we need to take the limit $y' \to y$ in Eq.~\eqref{eq-d2sn2} and extract the contribution with a $\delta^{d-2}(y-y')$ dependence.
Equivalently, we need to extract the delta function piece from the operator product expansion (OPE) of two ANE operators.

The OPE of two ANE operators can be written as \cite{Hofman:2008ar,Kologlu:2019mfz}
\begin{align}
    \mathcal{E}(y) \, \mathcal{E}(y') \, \sim \, \sum_{i} \, \frac{c_{i} \,  \mathbb{O}_{i}(y)}{|y-y'|^{2(d-2)-\tau_{i}}} \, , \label{eq-ope-anec}
\end{align}
where $\mathbb{O}_{i}$ are spin-$3$ light-ray operators with twist $\tau_{i}$ \cite{Kologlu:2019mfz}.
We remind the reader that the twist $\tau$ of a primary operator in a CFT is defined by $\tau=h-l$, where $h$ is the conformal dimension and $l$ is the spin quantum number.
It was argued in Ref.~\cite{Balakrishnan:2019gxl} that there is no delta function in this OPE in theories with a twist gap.
Here, we review the argument of \cite{Balakrishnan:2019gxl} which mainly consists of two parts.
Firstly, we make use of the following representation of the delta function
\begin{align}
    \lim_{\xi\to 0} \, \frac{\xi}{|y-y'|^{(d-2)-\xi}} \, = \, S_{d-3} \, \delta^{d-2}(y-y') \, , \label{eq-delta-nv}
\end{align}
where $S_{d-3}$ is the volume of a $(d-3)$-dimensional sphere.
Comparing Eq.~\eqref{eq-ope-anec} with Eq.~\eqref{eq-delta-nv}, we deduce that a delta function can only appear in the OPE if there exists a spin-$3$ operator $\mathbb{O}_{i}$ with twist $\tau_{i} = d-2$ and if the OPE coefficient scale as $c_{i} \sim (\tau_{i} - d + 2)$.
Secondly, we consider the leading Regge trajectory, which is the set of operators of minimal conformal dimension for each spin.
The twist of the operators on the leading Regge trajectory increases monotonically with spin, i.e.,$\frac{d\tau(J)}{dJ} \ge 0$ \cite{Komargodski:2012ek,Costa:2017twz}.
Since the stress energy tensor has $\tau  = (d-2)$ and $J=2$, we deduce that there are no spin-$3$ operators with $\tau = (d-2)$ in theories with a twist gap.
Thus, it was argued in Ref.~\cite{Balakrishnan:2019gxl} that there is no delta function contribution to the OPE of two ANE operators in theories with the twist gap.

The absence of a delta function contribution in the OPE of two ANE operators then implies that
\begin{align}
    S''_{n}(\rho_R||\sigma_R) \, = \, \left( \frac{\delta^{2} \, S_{n}(\rho_R||\sigma_R)}{\delta V(y')\delta V(y)} \, \right)_{\text{diag}} \, = \, O(\epsilon^{3}) \, , \label{eq-Renyi-qnec-nv-fin}
\end{align}
where we have reintroduced the $O(\epsilon^{3})$ contribution to the perturbative SRD.
Therefore, we conclude that the diagonal part of the Renyi QNEC is in fact saturated at leading non-trivial order for near-vacuum states.
It is natural to study the subleading order contributions to the second null deformations of the SRD.
If the $O(\epsilon^{3})$ contribution in Eq.~\eqref{eq-Renyi-qnec-nv-fin} does not vanish, then this would imply that $S''_{n}(\rho_R||\sigma_R)$ does not have a definite sign.
Hence, this would give us a violation of the Renyi QNEC.
We expect the calculation of this section can be extended to all orders using the recent results of \cite{Lashkari:2018oke,Balakrishnan:2020lbp}.
However, we leave a detailed analysis for future investigation.

Beyond the perturbative analysis, one may try to use other techniques that have been used to prove the QNEC \cite{Koeller:2015qmn,Balakrishnan:2017bjg,Ceyhan:2018zfg}.
An important hurdle in doing so is the fact that for the relative entropy, we have the decomposition
\begin{align}\label{eq-srel-deco}
    S_{\text{rel}}''&=\langle T_{vv}\rangle -S'',
\end{align}
and thus, the QNEC can equivalently be phrased in terms of the stress energy tensor and entropy variation.
The holographic proof in Ref.~\cite{Koeller:2015qmn} and the causality based proof in Ref.~\cite{Balakrishnan:2017bjg} use this decomposition.
For example, the holographic proof involves physically distinct sources of contributions, the entropy variation from the shape of the Ryu-Takayanagi (RT) surface \cite{Ryu:2006bv,Hubeny:2007xt}, and the energy from the metric perturbation near the boundary.
Further, defect OPE techniques have been used to study the $S''$ contribution in Eq.~\eqref{eq-srel-deco} \cite{Balakrishnan:2017bjg}.
In the case of the Renyi QNEC, there is no straightforward decomposition into such objects that are easier to study and thus, new techniques would have to be used in order to prove it in general.
It is interesting to note that the proof of the QNEC in Ref.~\cite{Ceyhan:2018zfg} directly uses the formulation in terms of the relative entropy.
This proof relies on the Connes cocycle flow to generate states that saturate a certain optimization problem, often termed the ant conjecture \cite{Wall:2017blw}.
However, a crucial ingredient used to prove the QNEC is a sum rule that relates
the relative entropies to the ANE operator, which does not generalize straightforwardly to the Renyi QNEC.
The holographic dual of the Connes cocycle flow as a one-sided boost was recently proposed in Ref.~\cite{Bousso:2020yxi}.
It would be interesting to combine these ideas to have a general proof of the Renyi QNEC.

\subsection{Off-Diagonal Renyi QNEC}\label{sec-offdiagonal}

\begin{figure}
    \centering
    \includegraphics[width=.45\textwidth]{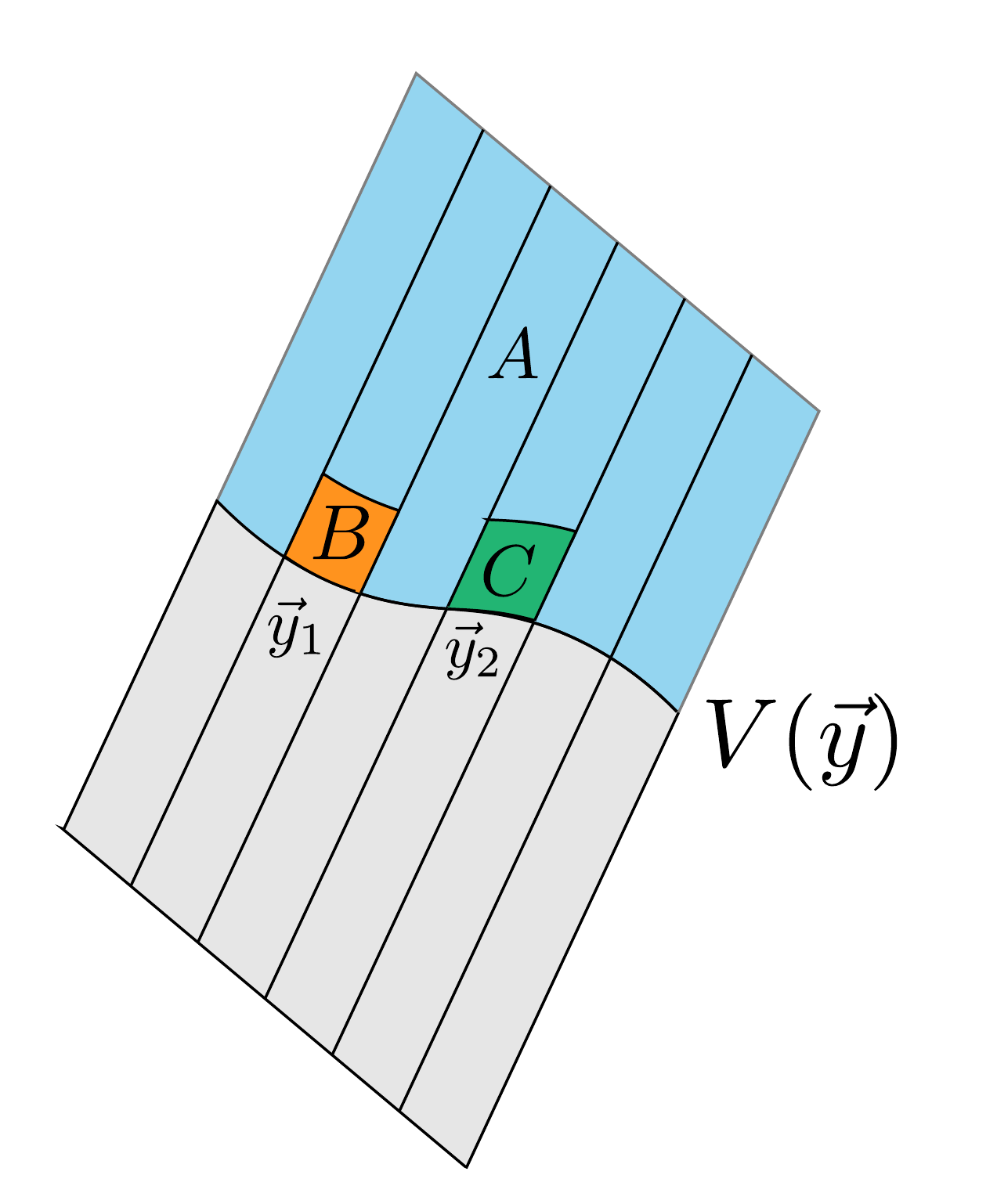}
    \caption{Off-Diagonal Variations. We show here the deformation of the profile of the entangling surface, $V(\vec{y})$, at two different locations $\vec{y}_1$ and $\vec{y}_2$. Under such deformations, the regions $B$ and $C$ are traced out of the state, with the remainder assigned to region $A$. For the QNEC, strong subadditivity of entropies guaranteed the inequality in Eq.~\eqref{eq-superadd} is true. For the Renyi QNEC, such an inequality is not true in general.}
    \label{fig:superadditivity}
\end{figure}

In this work, we have focused on the Renyi QNEC, which is a sign constraint on the diagonal part of the SRD variation.
The analogous condition on the off-diagonal part of the SRD variation can be written as
\begin{align}\label{eq-superadd}
    S_n(\rho_{A}||\sigma_{A})-S_n(\rho_{AB}||\sigma_{AB})-S_n(\rho_{AC}||\sigma_{AC})+S_n(\rho_{ABC}||\sigma_{ABC})\geq 0
\end{align}
where $A$, $B$ and $C$ are subregions of the null plane as seen in Fig.~(\ref{fig:superadditivity}).
This condition is called the strong superadditivity of SRD, which is not generally true.

On the other hand, the strong superadditivity of relative entropy, although not true in general, is in fact true for subregions of the null plane as shown in Ref.~\cite{Casini:2017roe}.
Importantly, it requires the fact that the vacuum modular Hamiltonian restricted to the null plane is local.
The locality of the modular Hamiltonian implies that the vacuum is in fact a Markov state where the contributions to relative entropy in Eq.~\eqref{eq-superadd} coming from the modular Hamiltonian cancel out.
This implies that the off-diagonal part of the QNEC is true, and in fact, by using the locality of the modular Hamiltonian simply follows from strong subadditivity of entropy \cite{Bousso:2015mna}.

Thus, one might hope that the Markov property also suffices to prove the off-diagonal part of the Renyi QNEC.
However, it is easy to find counterexamples to strong superadditivity of SRD for Markov states.
For example, in Eq.~\eqref{eq-superadd}, one can consider $\sigma_{ABC}=\frac{\mathbb{1}}{d_{ABC}}$, the maximally mixed density matrix on $\mathcal{H}_{ABC}$.
In this case, the SRD simplifies to
\begin{align}\label{eq-renyientropy}
    S_n(\rho_{ABC}||\sigma_{ABC})&=\log d_{ABC} - S_n(\rho_{ABC}),
\end{align}
where $S_n(\rho_{ABC})$ is the Renyi entropy of the density matrix $\rho_{ABC}$.
Using Eq.~\eqref{eq-renyientropy}, the condition in Eq.~\eqref{eq-superadd} becomes
\begin{align}
    S_n(\rho_{AB})+S_n(\rho_{AC})\geq S_n(\rho_A)+S_n(\rho_{ABC}),
\end{align}
which is the condition of strong subadditivity of Renyi entropy, known to not be true in general.
Thus, if true, proving the off-diagonal QNEC would require more specific properties of the vacuum in QFTs.

Another possible approach would be to use the techniques used in this paper to explicitly compute the off-diagonal part of the Renyi QNEC in free theories, or perturbatively in interacting theories.
Ref.~\cite{Lashkari:2018nsl} computed the second SRD variation for uniform deformations of a flat entangling surface in various examples and found that it was indeed positive.
Such examples involve both the diagonal and off-diagonal parts, and thus, it is consistent with a sign constraint on the off-diagonal part of the SRD variation.
We leave further analysis of this issue to future work.

\subsection{Renyi QNEC in \texorpdfstring{$d=2$}{d = 2}}\label{sec=2dQNEC}

In Sec.~(\ref{sec-proof-Renyi-QNEC}), we proved the $n>1$ Renyi QNEC in free theories for dimensions $d>2$.
This was importantly utilized in discretizing the null plane in the transverse direction.
Similarly, the perturbative analysis in Sec.~(\ref{sec-interacting}) was restricted to $d>2$ since we required a twist gap for the analysis to go through.
This raises the question of whether the Renyi QNEC holds in $d=2$.
Firstly, we note that the counterexample provided in Sec.~(\ref{sec-violation-Renyi-QNEC}) continues to work in $d=2$ in a straightforward manner.
Thus, we only need to consider the Renyi QNEC for $n>1$.
Further, note that in $d=2$, there is no distinction between the diagonal and off-diagonal part of the second SRD variation and thus, the issues we discussed in Sec.~(\ref{sec-offdiagonal}) come to the forefront again.

\begin{figure}
    \begin{subfigure}{.49\textwidth}
    \centering
    \includegraphics[width=\textwidth]{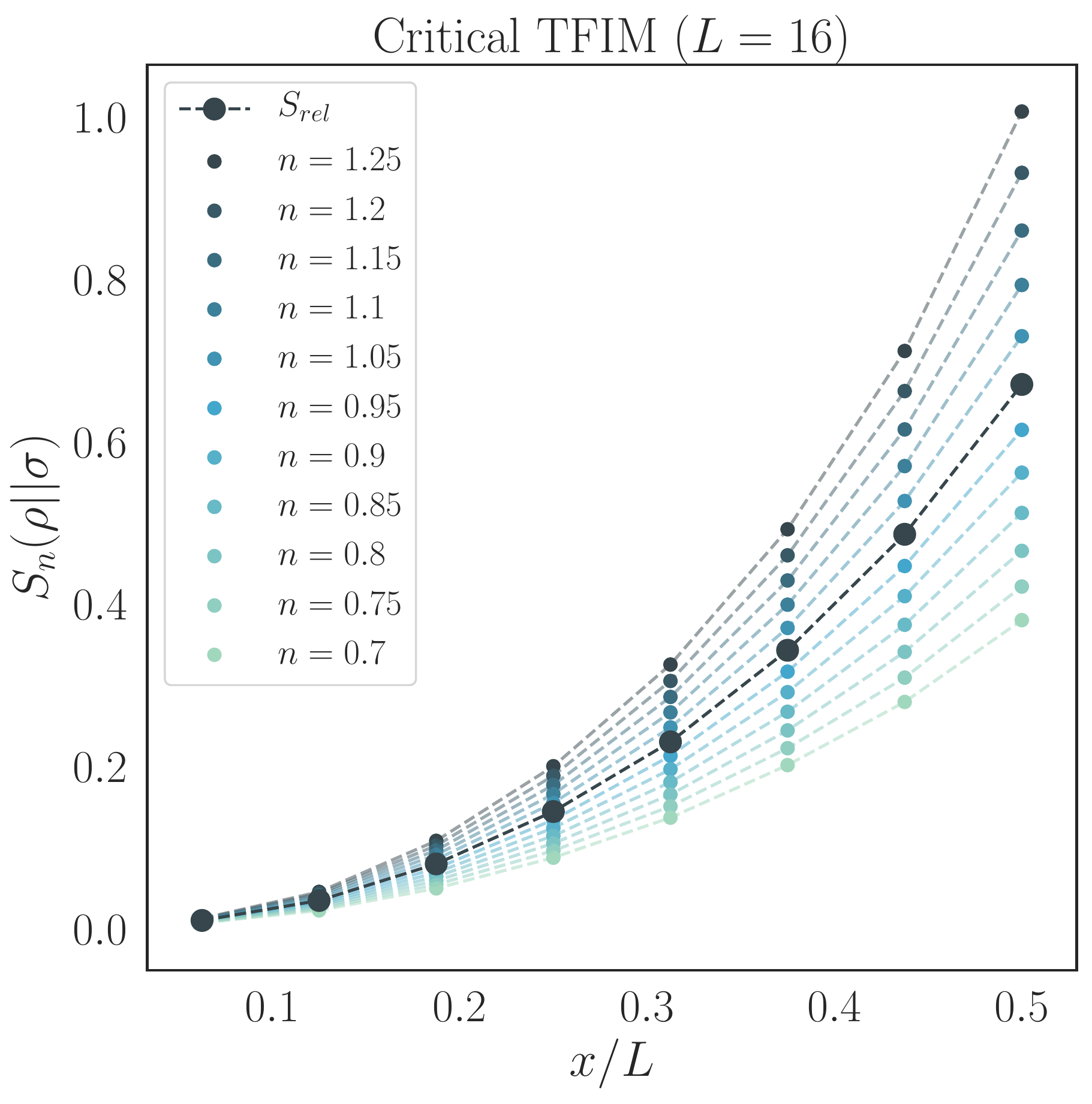}
    \end{subfigure}
    \begin{subfigure}{.50\textwidth}
    \centering
    \includegraphics[width=\textwidth]{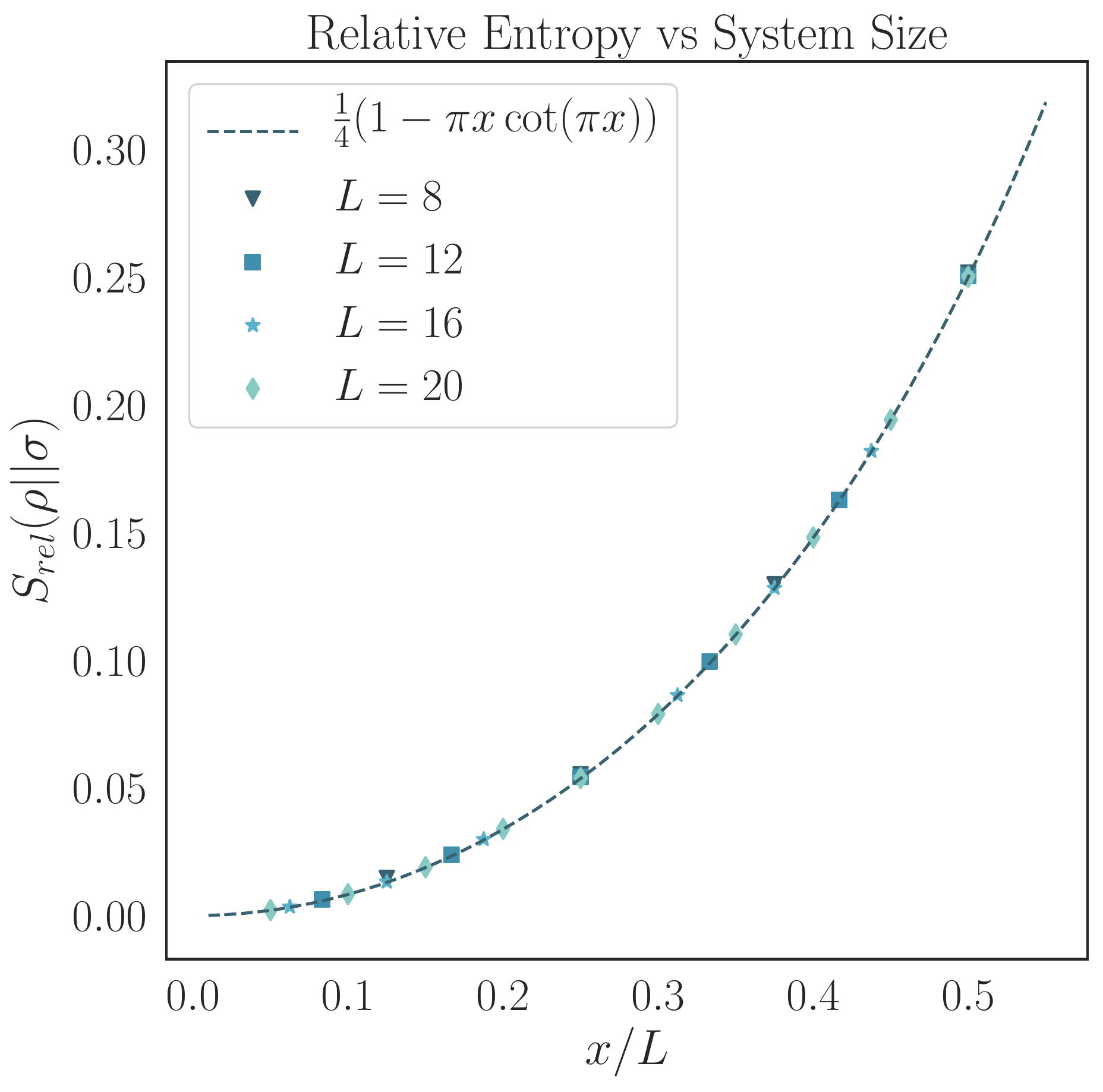}
    \end{subfigure}
    \caption{We use DMRG to produce the first excited state in the TFIM. The $x$ axis corresponds to the size of the reduced density matrix. As expected, monotonicity implies that SRD grows with system size. The positive second derivative (concavity) implies the Renyi QNEC is satisfied by this state. On the right, we show relative entropy of the first excited state for differing chain lengths, demonstrating that the finite size effects are negligible. }
    \label{fig:tfim_srd}
\end{figure}

In the case of the QNEC, the proof for free fields was dimensionally reduced to $d=2$ \cite{Bousso:2015wca}.
In order to do so, one can write the second shape variation of relative entropy in $d=2$ as a uniform integral over a $D$-dimensional second shape variation of relative entropy, i.e.,
\begin{align}\label{eq-dimred}
    S_{\text{rel,2d}}''&=\int d^{D-2}y\,d^{D-2}y'\,\frac{\delta^2 S_{\text{rel,D}}}{\delta V(y)\delta V(y')}\\
    &= (\text{off-diagonal})+(\text{diagonal})\geq 0,
\end{align}
where the off-diagonal part is positive by strong superadditivity of relative entropy and the diagonal part was proved to be positive by a calculation similar to that in Sec.~(\ref{sec-proof-Renyi-QNEC}).
Thus, we see that repeating the dimensional reduction argument for the Renyi QNEC requires the off-diagonal part of the Renyi QNEC in $d>2$.

Despite the fact that we have not been able to prove the Renyi QNEC in $d=2$, we will now provide some numerical evidence that it is in fact true.
We considered SRD in critical spin chains on a finite lattice, which although far from conformal, have a low energy limit described by a CFT \cite{Calabrese:2009qy}.
Studying low energy states below the lattice spacing will then serve as good approximations to states that exist in the continuum field theory.

The numerical study of relative entropy of states in various CFTs was done in
\cite{Nakagawa:2017fzo}.
The technique used there was exact diagonalization, but
since we are interested in the low lying states of the theory, the density matrix renormalization group (DMRG) \cite{White:1992zz} technique can also be employed.

\begin{figure}
    \begin{subfigure}{.49\textwidth}
    \centering
    \includegraphics[width=\textwidth]{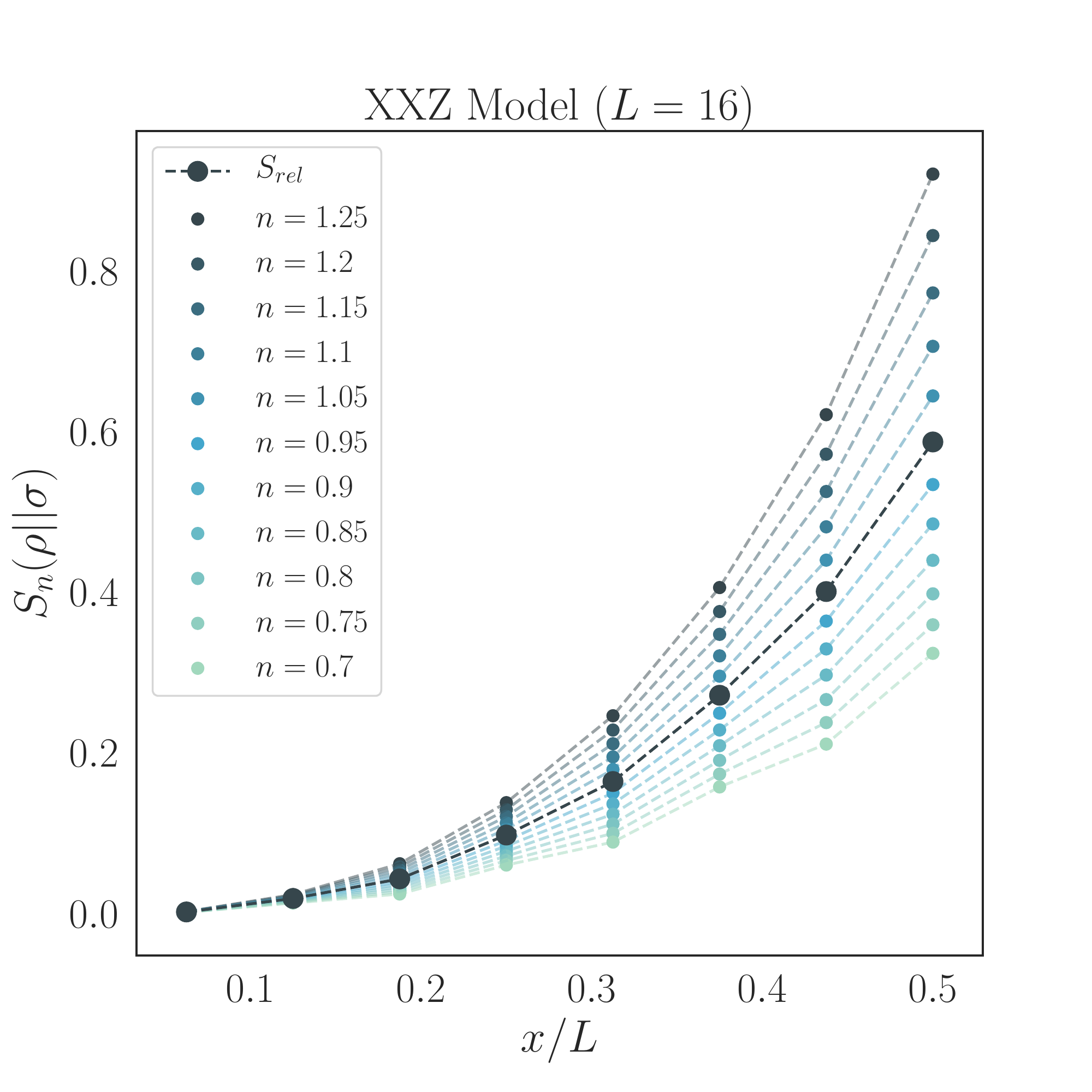}
    \end{subfigure}
    \begin{subfigure}{.49\textwidth}
    \centering
    \includegraphics[width=\textwidth]{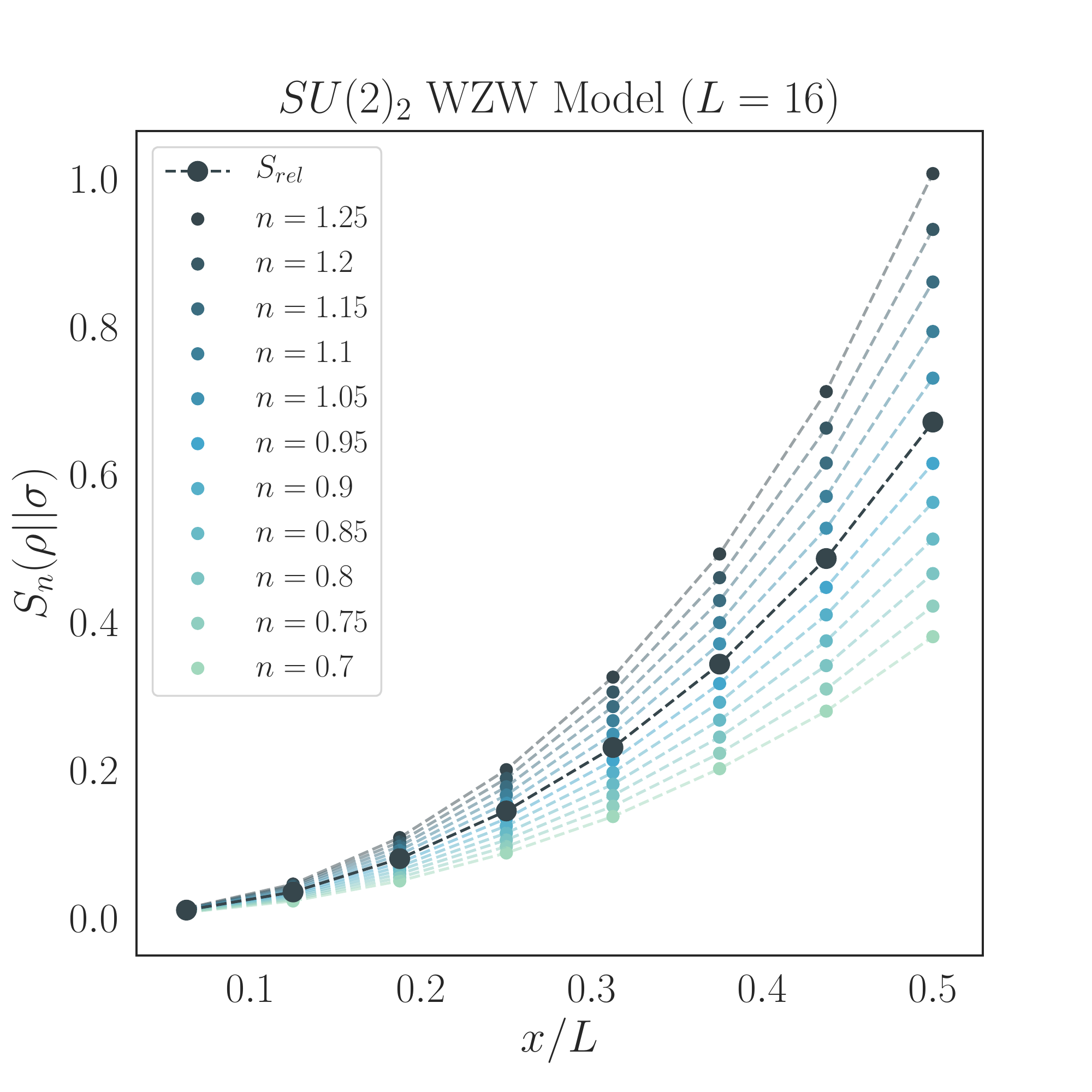}
    \end{subfigure}
    \caption{We investigated the SRD of the first excited state with respect to the vacuum for two other critical spin chains, the XXZ model with $J_z/J_x = \Delta=1/2$ and the $SU(2)_2$ WZW model. The Renyi QNEC is also satisfied in this case.}
    \label{fig:xxz_wzw}
\end{figure}

To verify that states in the CFT are really being probed in finite systems, we outline the following procedure.
Take a system on $L$ sites and compute the relative entropy of an excited state
with respect to the vacuum for a given choice of subsystem.
By normalizing by the number of sites, one can compute the relative entropy as a function of the interval length $x/L$, e.g. the fraction of the state.
By showing that this quantity collapses to the same function for arbitrary $L$ (and comparing to the analytical result derived in \cite{Lashkari:2014yva}), one confirms that the conformal physics is indeed being probed.

Following this procedure, we start by computing the SRD in the critical transverse field Ising model (TFIM), which is described by a $c=1/2$ CFT.\footnote{To carry out the numerics, we make use of the ITensor package in Julia (\url{https://github.com/ITensor/ITensors.jl}).
Our code is publicly available on Github (\url{https://github.com/vipasu/Renyi-QNEC}).}
In Fig.~(\ref{fig:tfim_srd}) we show plots of the SRD for varying $n$.
Notably, we do not see violations of the Renyi QNEC, i.e., the second derivative is always non-negative for all $n$ including $n<1$.
Note that we are using the fact that $x$ derivatives are interchangeable with $v$ derivatives for eigenstates of the Hamiltonian which are time independent.
 Numerically, because of the inverse powers of density matrices, there is a trade-off between the
numerical precision of the state and larger $n$.
For large $n$, the SRD is sensitive to the tail ends of the spectrum and thus cannot be trusted for eigenvalues which are below the error threshold of the DMRG approximation.


The critical TFIM is dual to a $c=1/2$ Ising model, which is dual to a free Majorana fermion \cite{Nakagawa:2017fzo}.
In \cite{Headrick:2012fk,Radicevic:2016tlt}, the mapping of entanglement entropies under such a duality was considered.
Up to subtleties with edge modes, the entropies were shown to be duality invariant and thus, we expect that the relative entropies computed here would also be characteristic of a free theory.


Additionally we were able to probe the low lying spectra of the XXZ and $SU(2)_2$ Wess Zumino Witten models in Fig.~(\ref{fig:xxz_wzw}).
The explicit Hamiltonians for the spin chains are described in Ref.~\cite{F_hringer_2008}.
The critical XXZ model is dual to a $c=1$ free boson \cite{alcaraz1994critical} while WZW models can in general be studied in terms of the Wakimoto free field representation \cite{DiFrancesco:1997nk}.
Thus, we expect these spin chains to also serve as examples of the Renyi QNEC in a free theory.
In both cases, we find numerical evidence that the Renyi QNEC is true for all $n$.

In Sec.~(\ref{sec-violation-Renyi-QNEC}), we showed that the Renyi QNEC is not generally true for $n<1$.
Despite this, we have not found such violations in our numerical examples and it would be interesting to probe this further.

\subsection{Other Renyi divergences}\label{sec-alphaz}

In this work, we have focused on the SRD which is a particular Renyi generalization of the relative entropy.
However, there is a different Renyi generalization called the \textit{Petz divergence} that is defined as \cite{petz1986quasi,Petz,Lashkari:2018nsl}
\begin{align}
    D_{n}(\rho_R||\sigma_R) \, \equiv \, \frac{1}{n-1} \, \log \, \text{tr} \left( \sigma_R^{1-n} \, \rho_R^{n} \right) \, .
\end{align}
Both the SRD and the Petz divergence are in fact special cases of a two-parameter family of Renyi divergences known as the $n$-$z$ divergence \cite{audenaert2013alpha,May:2018tir}:
\begin{align}
    D_{n,z}(\rho_R||\sigma_R) \, \equiv \, \frac{1}{n-1} \, \log \, \tr \, \left(\sigma_R^{\frac{1-n}{2z}} \, \rho_R^{\frac{n}{z}} \, \sigma_R^{\frac{1-n}{2z}}\right)^{z} \, ,
\end{align}
which satisfy all the properties of a measure of distinguishability for $z>|n-1|$ and $n\geq0$.
The Petz divergence corresponds to the case $z=1$ whereas SRD corresponds to the case $z=n$.

One could then consider analogues of the Renyi QNEC for the $n$-$z$ divergence as well, which we call the $n$-$z$ QNEC.
For near vacuum states such as those given by Eq.~\eqref{eq-rho-nv} and Eq.~\eqref{eq-rho1-nv}, the $n$-$z$ divergence was computed perturbatively in \cite{May:2018tir}.
It was found that at leading non-trivial order in the perturbation parameter, the $n$-$z$ divergence is given by
\begin{align}
    D_{n,z}(\rho_R||\sigma_R) \, = \, \frac{\epsilon^{2}}{2} \, \int d\mu  \int d\kappa \int d\kappa' \, F_{n,z}(\kappa-\kappa') \, e^{-2\pi\kappa} \,  \bra{\kappa}\OO(r_{1},\theta_{1},y_{1})\ket{\kappa'} \bra{\kappa'}\OO(r_{2},\theta_{2},y_{2})\ket{\kappa} \, ,\label{eq-nzD2-nv}
\end{align}
which is similar to Eq.~\eqref{eq-sn2-nv} apart from the kernel $F_{n,z}(\kappa-\kappa')$ which in this case is given by
\begin{align}
    F_{n,z}(x) \, = \, \frac{z}{1-n} \, \frac{e^{-2\pi n x/z}  - 1}{e^{-2\pi x}  - 1} \, \frac{e^{2\pi \left(\frac{n-1}{z}\right) x}  - 1}{e^{-2\pi x/z}  - 1} \, \label{eq-Fnz-vn}
\end{align}
Note that $F_{n,z}(x)$ has a similar transformation property under $x \to -x$ as $F_{n}(x)$, i.e., it satisfies the relation
\begin{align}
    F_{n,z}(-x) \, = \, e^{-2\pi x} \, F_{n,z}(x) \, .
\end{align}
This condition allows us to write the $n$-$z$ divergence in Eq.~\eqref{eq-nzD2-nv} as an angle-ordered correlation function by repeating the analysis of Sec.~(\ref{sec-interacting}).
This would give us an expression for perturbative $n$-$z$ divergence which is similar to Eq.~\eqref{eq-sn-corr-Delta} apart from the kernel.
By the same argument as Sec.~(\ref{sec-interacting}), we can deduce that there is no delta function in the second null deformation of the $n$-$z$ divergence at the leading order in interacting theories. That is,
\begin{align}
    D''_{n,z}(\rho_R||\sigma_R) \, = \, \left( \frac{\delta^{2} \, D_{n,z}(\rho_R||\sigma_R)}{\delta V(y')\delta V(y)} \, \right)_{\text{diag}} \, = \, O(\epsilon^{3}) \, . \label{eq-Dnz-nv-fin}
\end{align}

For free theories on the other hand, one can compute the correlation function and hence, the second SRD variation perturbatively using the techniques of Sec.~(\ref{sec-null-zn-corr}).
A similar calculation can then be used to show that the $n$-$z$ QNEC can be violated for $n<1$.
For other parameter ranges, we have neither been able to find counterexamples, nor prove positivity of the second SRD variation.
It would be interesting to prove/disprove the $n$-$z$ QNEC perturbatively using techniques similar to Sec.~(\ref{sec-proof-Renyi-QNEC}) and Sec.~(\ref{sec-violation-Renyi-QNEC}).

Another Renyi generalization of the relative entropy, called the \textit{refined} Renyi divergence was recently proposed in \cite{Bao:2019aol}.
The refined Renyi divergence is defined in terms of the sandwiched Renyi divergence as
\begin{align}
    \S_{n}(\rho||\sigma) \, \equiv \, n^{2} \, \partial_{n} \left( \frac{n-1}{n} \, S_{n} (\rho||\sigma) \right) \, . \label{eq-RRD-def}
\end{align}
For near vacuum states, we can deduce the expression for the perturbative refined Renyi divergence from our analysis of sandwiched Renyi divergence in Sec.~(\ref{sec-interacting}).
By comparing Eq.~\eqref{eq-RRD-def} with Eq.~\eqref{eq-sn-corr-Delta}, we find that the leading order contribution to the refined Renyi divergence is given by
\begin{align}
    \S_{n}(\rho_R||\sigma_R) \, = \, \frac{\epsilon^{2}}{2} \, \int d\mu \int_{-\infty}^{\infty}ds \,
    \widetilde{\mathcal{F}}_{n}\Big(\text{sgn}(\theta_{1}-\theta_{2}) \, s\Big) \, \left\langle  \OO_{1} \, \Delta_{\Omega}^{is/2\pi} \, {\OO}_{2} \right\rangle ,
\end{align}
where the kernel in this case is given by
\begin{align}
    \widetilde{\mathcal{F}}_{n}(s) \, = \, n^{2} \, \partial_{n} \left( \frac{n-1}{n} \, {\mathcal{F}}_{n}(s)  \right) \, .
\end{align}
Again by the argument of Sec.~(\ref{sec-interacting}), we find that
\begin{align}
    \S''_{n}(\rho_R||\sigma_R) \, = \, \left( \frac{\delta^{2} \, \S_{n}(\rho_R||\sigma_R)}{\delta V(y')\delta V(y)} \, \right)_{\text{diag}} \, = \, O(\epsilon^{3}) \, . \label{eq-Dnz-QNEC}
\end{align}

Thus, for near vacuum states, one in fact finds that the diagonal part of the second null deformation of \textit{any} of the above Renyi divergences vanishes at leading order in the perturbation parameter.
It would be interesting to see if this holds true at sub-leading orders.
More generally, these Renyi divergences provide interesting generalizations of the QNEC which can be analyzed by techniques that have been applied to studying the QNEC.

\section*{Acknowledgements}
We would like to thank Ven Chandrasekaran, Thomas Faulkner, Thomas Hartman, Nima Lashkari, Adam Levine and Arvin Shahbazi-Moghaddam for useful discussions.
The work of MM was supported by the US Department of Energy under grant number DE-SC0014123.
The work of PR and VPS was supported in part by the Berkeley Center for Theoretical Physics, by the National Science Foundation (award number PHY-1521446), and by the U.S. Department of Energy under contract DE-AC02-05CH11231 and award DE-SC0019380. This research used the Savio computational cluster resource provided by the Berkeley Research Computing program at the University of California, Berkeley (supported by the UC Berkeley Chancellor, Vice Chancellor for Research, and Chief Information Officer). VPS gratefully acknowledges support by the NSF GRFP under Grant No. DGE 1752814.

\appendix

\section{Finiteness of SRD for \texorpdfstring{$1/2 \, \le \, n \, < \, 1$}{1/2 < n < 1}}
\label{app-finite}

In Sec.~(\ref{sec-SRDQFT}), we stated that the SRD between any two cyclic and separating states is finite for $1/2 \le  n  <  1$.
In this appendix, we provide a justification for this statement.
To do this, we first introduce another Renyi divergence, called the Petz divergence, which for cyclic and separating states $\ket{\Psi}$ and $\ket{\Phi}$
is defined as \cite{Petz,Lashkari:2018nsl}
\begin{align}
    D^{\mathcal{M}}_{n}(\Psi||\Phi) \, \equiv \, \frac{1}{n-1} \, \log \, \bra{\Psi} \, \left( \Delta_{\Psi|\Phi}  \right)^{1-n} \, \ket{\Psi} \, ,
\end{align}
where $n \in [0,1) \, \cup \, (1,2]$.
Just like the SRD, it is non-negative and it approaches the relative entropy $S^{\mathcal{M}}_{\text{rel}}(\Psi||\Phi)$ in the limit $n \, \to \, 1$ \cite{Petz}. 


It is known that for $1/2 \le n  \le 2$, the Petz divergence provides a upper bound on the SRD \cite{Berta:2018ecp,Lashkari:2018nsl}.
In particular, we have
\begin{align}
    D^{\mathcal{M}}_{n}(\Psi||\Phi) \, \ge \, S^{\mathcal{M}}_{n}(\Psi||\Phi) \, .\label{eq-app-ALT}
\end{align}
This follows from a generalization of the Araki-Lieb-Thirring inequality \cite{Lieb1991,ALT} to arbitrary von Neumann algebras which was recently proven in \cite{Berta:2018ecp}.

Now, $\Delta_{\Psi|\Phi}$ is a positive-definite operator when the states $\ket{\Psi}$ and $\ket{\Phi}$ are cyclic and separating \cite{Witten:2018zxz}.
This implies that
\begin{align}
    \bra{\Psi} \, \left( \Delta_{\Psi|\Phi} \right)^{1-n} \, \ket{\Psi} \, > \, 0 \, .
\end{align}
As a result, we find that the Petz divergence between cyclic and separating states is finite for $n<1$, i.e.,
\begin{align}
    D^{\mathcal{M}}_{n}(\Psi||\Phi) \, < \, \infty \, .
\end{align}
Combining this result with Eq.~\eqref{eq-app-ALT}, we deduce that the SRD between two cyclic and separating states is finite for $1/2 \le n < 1$.

\section{Details of Calculation for Free Field Theory}
\subsection{Perturbative expansion of SRD} \label{sec-app-iden}

In Sec.~(\ref{sec-null-zn-corr}), we claimed that the formula for $Z_{n}^{(1,1)}$ in Eq.~\eqref{eq-null-z11-int-2}, which we derived by assuming $n$ to be an integer, is valid for arbitrary $n$.
In this appendix, we present the derivation of Eq.~\eqref{eq-null-z11-int-2} for arbitrary non-integer $n$, thus proving our claim.

To derive Eq.~\eqref{eq-null-z11-int-2}, we start with the identity in Eq.~\eqref{eq-null-zn-eqv} which relates the calculation of $Z_{n}$ for arbitrary states to that for two `nearby' states.
That is, we have
\begin{align}
    Z_{n} \, &= \, \frac{1}{n-1} \, \tr \widehat{\rho}^{\, n} \, ,
\end{align}
where $\widehat{\rho}$ is defined as
\begin{align}
    \widehat{\rho} \, &= \, \sigma^{1/n} \, + \, \sigma^{\frac{1-n}{2n}} \, \delta\rho \, \sigma^{\frac{1-n}{2n}} \, . \label{eq-app-zn-sp}
\end{align}
To keep our notation simpler, we are replacing $\trho^{(0)}$ with $\sigma$ and $\mathcal{A}^{1/2} \, \trho^{(1)}$ with $\delta\rho$.
The modular Hamiltonian of $\widehat{\rho}$ is then defined as
\begin{align}
    \widehat{K} \, \equiv \, - \log \, \widehat{\rho} \, .
\end{align}
Now using the resolvent identity
\begin{align}
    \widehat{K} \, = \, \int_{0}^{\infty} \, d\lambda \, \left( \frac{1}{\widehat{\rho}+\lambda} - \frac{1}{1+\lambda} \right) \, ,
\end{align}
we can express $\widehat{K}$ as
\begin{align}
    \widehat{K} \, = \, \frac{1}{n} \, K_{\sigma} \, + \, \delta\widehat{K} \, + \, \delta^{2}\widehat{K} \, + \, O\big(\delta\rho^{3}\big) \, , \label{eq-app-k-exp}
\end{align}
where $K_{\sigma} \, = \, - \log \sigma$, and
\begin{align}
    \delta\widehat{K} \, =&\, \, - \, \int_{0}^{\infty} \, d\lambda \,\, \frac{\sigma^{\frac{1-n}{2n}}}{\sigma^{1/n}+\lambda} \, \delta\rho \, \frac{\sigma^{\frac{1-n}{2n}}}{\sigma^{1/n}+\lambda} \, ,\label{eq-del-K-1}\\
    \delta^{2}\widehat{K} \, =&\, \, \int_{0}^{\infty} \, d\lambda \,\, \frac{\sigma^{\frac{1-n}{2n}}}{\sigma^{1/n}+\lambda} \, \delta\rho \, \frac{\sigma^{\frac{1-n}{n}}}{\sigma^{1/n}+\lambda} \, \delta\rho \, \frac{\sigma^{\frac{1-n}{2n}}}{\sigma^{1/n}+\lambda} \, .\label{eq-del-K-2}
\end{align}

Now, we note that
\begin{align}
    \frac{d}{d\alpha} \, \left( \widehat{\rho}^{\, \alpha} \, \sigma^{-\alpha/n} \right) \, = \, - \, \widehat{\rho}^{\, \alpha} \, \left( \widehat{K} \, - \, \frac{1}{n} K_{\sigma} \right) \, \sigma^{-\alpha/n} \, .
\end{align}
Integrating this equation yields
\begin{align}
    \widehat{\rho}^{\, \alpha} \, \sigma^{-\alpha/n} \, - \, 1 \, = \, - \int_{0}^{\alpha} d\alpha' \,\, \widehat{\rho}^{\, \alpha'} \, \left( \widehat{K} \, - \, \frac{1}{n} K_{\sigma} \right) \, \sigma^{-\alpha'/n} \, ,
\end{align}
or equivalently
\begin{align}
    \widehat{\rho}^{\, \alpha} \, = \,  \sigma^{\alpha/n} \, - \, \int_{0}^{\alpha} d\alpha' \,\, \widehat{\rho}^{\, \alpha'} \, \left( \widehat{K} \, - \, \frac{1}{n} K_{\sigma} \right) \, \sigma^{(\alpha-\alpha')/n} \, . \label{eq-rho-trans}
\end{align}
This is a transcendental equation which, in principle, can be solved by iteration.
Performing the first iteration leads us to
\begin{align}
    \widehat{\rho}^{\, \alpha} \, =&\, \,  \sigma^{\alpha/n} \, - \, \int_{0}^{\alpha} d\alpha' \,\, \sigma^{\alpha'/n} \, \left( \widehat{K} \, - \, \frac{1}{n} K_{\sigma} \right) \, \sigma^{(\alpha-\alpha')/n} \label{eq-rho-trans-2}\\ +&\,  \int_{0}^{\alpha} d\alpha' \, \int_{0}^{\alpha'} d\alpha'' \,\, \widehat{\rho}^{\, \alpha''} \, \left( \widehat{K} \, - \, \frac{1}{n} K_{\sigma} \right) \, \sigma^{(\alpha'-\alpha'')/n} \,  \left( \widehat{K} \, - \, \frac{1}{n} K_{\sigma} \right) \, \sigma^{(\alpha-\alpha')/n} \,  . \nonumber
\end{align}
This equation allows us to expand $\widehat{\rho}^{\, n}$ to second-order in $\delta\rho$.
More precisely, by using Eq.~\eqref{eq-app-k-exp} and by taking $\alpha \, = \, n$, we get
\begin{align}
    \widehat{\rho}^{\, n} \, =&\, \,  \sigma \, - \, \int_{0}^{n} d\alpha' \,\, \sigma^{\alpha'/n} \, \left( \delta\widehat{K} \, + \, \delta^{2}\widehat{K} \right) \, \sigma^{1 - \alpha'/n} \, \label{eq-rho-trans-3}\\ \, +&\,  \int_{0}^{n} d\alpha' \, \int_{0}^{\alpha'} d\alpha'' \,\, \sigma^{ \alpha''/n} \, \delta\widehat{K}  \, \sigma^{(\alpha'-\alpha'')/n} \,  \delta\widehat{K}  \, \sigma^{1 - \alpha'/n} \, + \, O(\delta\rho^{3}) . \nonumber
\end{align}
Taking the trace of this equation, we get
\begin{align}
    \tr \, \widehat{\rho}^{\, n} \, = \,  \tr \, \sigma \, - \, n \, \tr \left( \sigma  \left( \delta\widehat{K} \, + \, \delta^{2}\widehat{K} \right) \right) \, + \, \frac{n^{2}}{2} \, \int_{0}^{1} d\alpha \,  \tr \, \left( \sigma^{1 -  \alpha} \, \delta\widehat{K}  \, \sigma^{\alpha} \,  \delta\widehat{K}  \, \right) \, + \, O(\delta\rho^{3}) . \label{eq-app-tr-rho-n-1}
\end{align}

Now, combining Eqs.~\eqref{eq-del-K-1}-\eqref{eq-del-K-2} with Eq.~\eqref{eq-app-tr-rho-n-1}, we deduce that the second-order `correction' to $Z_{n}$ is given by
\begin{align}
    \delta^{2}Z_{n} \, =&\, \, - \frac{n}{n-1} \, \int_{0}^{\infty} d\lambda \, \tr \, \left[ \frac{\sigma^{1/n}}{(\sigma^{1/n}+\lambda)^{2}} \, \delta\rho \, \frac{\sigma^{(1-n)/n}}{(\sigma^{1/n}+\lambda)} \, \delta\rho \right] \, \label{eq-app-zn-2-monster} \\ +&\, \frac{n^{2}}{2(n-1)} \, \int_{0}^{1} d\alpha  \int_{0}^{\infty} d\lambda d\lambda' \, \tr \, \left[ \frac{\sigma^{1/n - \alpha}}{(\sigma^{1/n}+\lambda)(\sigma^{1/n}+\lambda')} \, \delta\rho \, \frac{\sigma^{1/n - 1 + \alpha}}{(\sigma^{1/n}+\lambda)(\sigma^{1/n}+\lambda')} \, \delta\rho \, \right] \, . \nonumber
\end{align}
To compare this expression with Eq.~\eqref{eq-null-z11-int-2}, we evaluate each of the traces in a basis in which $\sigma$ is diagonal, i.e.,
\begin{align}
    \sigma \, \ket{\omega} \, = \, e^{-2\pi \omega} \, \ket{\omega} \, .
\end{align}
In this basis, Eq.~\eqref{eq-app-zn-2-monster} becomes
\begin{align}
    \delta^{2}Z_{n} \, = \, \frac{1}{2} \, \int d\omega \, \int d\omega' \, e^{2\pi\omega'} \, F_{n}(\omega,\omega') \, \left|\bra{\omega} \delta\rho \ket{\omega'}\right|^{2} \, ,
\end{align}
where
\begin{align}
    F_{n}(\omega,\omega') \, = \, &\, \frac{n^{2}}{n-1} \, \int_{0}^{1} d\alpha \, e^{2\pi\alpha(\omega-\omega')} \, \left[ \int_{0}^{\infty} d\lambda \, \frac{e^{-\pi(\omega+\omega')/n}}{\left(e^{-2\pi\omega/n} + \lambda\right) \left(e^{-2\pi\omega'/n} + \lambda\right)} \right]^{2}\\
    -&\, \, \frac{n}{n-1} \, \int_{0}^{\infty} d\lambda \, \left[ \frac{e^{-2\pi(\omega+\omega')/n}}{\left(e^{-2\pi\omega/n} + \lambda\right)^{2} \left(e^{-2\pi\omega'/n} + \lambda\right)} \, + \, \frac{e^{2\pi(\omega-\omega')} \,  e^{-2\pi(\omega+\omega')/n}}{\left(e^{-2\pi\omega/n} + \lambda\right) \left(e^{-2\pi\omega'/n} + \lambda\right)^{2}} \right] \nonumber \, .
\end{align}
Finally, by performing these integrals we get
\begin{align}
    F_{n}(\omega,\omega') \, = \, \frac{n}{1-n} \, \frac{e^{2\pi\left(\frac{n-1}{n}\right)(\omega-\omega')} \, - \, 1}{e^{-2\pi(\omega-\omega')/n} \, - \, 1}  \, ,
\end{align}
which is the same as Eq.~\eqref{eq-null-Fn-def}.
This completes our derivation of Eq.~\eqref{eq-null-z11-int-2} for arbitrary $n$.

\subsection{Calculation of Correlation Functions} \label{app-int-Gw}

In this appendix, we present the calculation of the correlation functions $\mathcal{G}(s)$ and $G(\omega)$ defined in Eq.~\eqref{eq-null-def-Gs} and in Eq.~\eqref{eq-null-def-Gw} respectively.

\subsection*{Calculation of $\mathcal{G}(s)$}

Recall from Eq.~\eqref{eq-null-def-Gs} that
\begin{align}
    \mathcal{G}(s) \, \equiv \, &\left\langle \big( \trho^{(0)} \big)^{-is/2\pi} \, {\OO}^{\alpha\beta}(r_{1},\theta_{1}) \, \big( \trho^{(0)} \big)^{is/2\pi} \,   \ddot{\OO}^{\alpha'\beta'}(r_{2},\theta_{2}) \right\rangle \,\nonumber\\ + \, &\Big\langle \big( \trho^{(0)} \big)^{-is/2\pi} \dot{\OO}^{\alpha\beta}(r_{1},\theta_{1}) \, \big( \trho^{(0)} \big)^{is/2\pi} \, \dot{\OO}^{\alpha'\beta'}(r_{2},\theta_{2}) \Big\rangle \, . 
\end{align}
Since $\trho^{(0)}$ factorizes between the pencil and auxiliary system, $\mathcal{G}(s)$ also factorizes.
Namely, by using the definition of $\trho^{(0)}$, Eq.~\eqref{eq-null-sand-0}, and the definition of ${\OO}^{\alpha\beta}(r,\theta)$, Eqs.~\eqref{eq-OO-0}-\eqref{eq-ddOO-0}, we get
\begin{align}
    \mathcal{G}(s) \, = \, \mathcal{G}_{p}(s) \, \cdot \, \mathcal{G}_{\aux}(s) \, , \label{eq-Gs-prod}
\end{align}
where
\begin{align}
    \mathcal{G}_{p}(s) \, \equiv \, &\left\langle \sigma_{p}^{-is/2\pi} \, \partial\Phi(r_{1} e^{i\theta_{1}}) \, \sigma_{p}^{is/2\pi} \,   \partial^{3}\Phi(r_{2} e^{i\theta_{2}}) \right\rangle_{p} \, \nonumber\\ + \, &\left\langle \sigma_{p}^{-is/2\pi} \, \partial^{2}\Phi(r_{1} e^{i\theta_{1}}) \, \sigma_{p}^{is/2\pi} \,   \partial^{2}\Phi(r_{2} e^{i\theta_{2}}) \right\rangle_{p} \, , \label{eq-null-def-Gs-p}
\end{align}
and
\begin{align}
    \mathcal{G}_{\aux}(s) \, \equiv \, \left\langle \big( \trho_{\aux}^{(0)} \big)^{-is/2\pi} \, E_{\alpha\beta}(\theta_{1}) \, \big( \trho_{\aux}^{(0)} \big)^{is/2\pi} \,   E_{\alpha'\beta'}(\theta_{2}) \right\rangle_{\aux} \, .\label{eq-null-def-Gs-aux}
\end{align}
We first consider $\mathcal{G}_{\aux}(s)$.
For $\theta_{1}>\theta_{2}$, we have
\begin{align}
    \mathcal{G}_{\aux}(s) \, =& \,\, \tr_{\aux} \left( \, \trho^{(0)} \, \big( \trho_{\aux}^{(0)} \big)^{-is/2\pi} \, E_{\alpha\beta}(\theta_{1}) \, \big( \trho_{\aux}^{(0)} \big)^{is/2\pi} \,   E_{\alpha'\beta'}(\theta_{2}) \, \right) \, ,\\
    =& \,\, e^{-2\pi K_{\alpha}} \, e^{(\theta_{1}-\theta_{2}) v_{\alpha\beta} } \, e^{is v_{\alpha\beta}} \, \delta_{\alpha\beta'} \, \delta_{\beta\alpha'} \, , \label{eq-null-gs-aux-int-1}
\end{align}
where we have used Eq.~\eqref{eq-null-trho-eigen} and Eq.~\eqref{eq-null-def-Eab}.
We have also defined $v_{\alpha \beta}$ as
\begin{align}
    v_{\alpha\beta} \, \equiv \, K_{\alpha} \, - \, K_{\beta} \, .
\end{align}
For $\theta_{2}>\theta_{1}$, we similarly obtain
\begin{align}
    \mathcal{G}_{\aux}(s) \, =& \,\, \tr_{\aux} \left( \, \trho^{(0)} \,  E_{\alpha'\beta'}(\theta_{2}) \, \big( \trho_{\aux}^{(0)} \big)^{-is/2\pi} \, E_{\alpha\beta}(\theta_{1}) \, \big( \trho_{\aux}^{(0)} \big)^{is/2\pi} \, \right) \, ,\\
    =& \,\, e^{-2\pi K_{\beta}} \, e^{(\theta_{1}-\theta_{2}) v_{\alpha\beta} } \, e^{is v_{\alpha\beta}} \, \delta_{\alpha\beta'} \, \delta_{\beta\alpha'} \, . \label{eq-null-gs-aux-int-2}
\end{align}
By combining Eq.~\eqref{eq-null-gs-aux-int-1} and Eq.~\eqref{eq-null-gs-aux-int-2}, we get
\begin{align}
   \mathcal{G}_{\aux}(s) \, =\, e^{-\pi(K_{\alpha}+K_{\beta})} \, e^{(\theta_{1}-\theta_{2}+is) v_{\alpha\beta} } \, e^{- \pi \, \text{sgn}(\theta_{1}-\theta_{2}) v_{\alpha\beta} } \,   \delta_{\alpha\beta'} \, \delta_{\beta\alpha'} \, . \label{eq-null-gs-aux}
\end{align}
Now, we consider $\mathcal{G}_{p}(s)$.
Using the known correlation functions for a free field theory of a chiral boson, we have
\begin{align}
\left\langle \partial\Phi(z) \, \partial\Phi(w) \right\rangle_{p} \, = \, - \, \frac{1}{(z-w)^{2}} \, .
\end{align}
Using the transformation property of operators under vacuum modular flow, which acts as a Lorentz boost, we have
\begin{align}
    \sigma_{p}^{-is/2\pi} \, \partial\Phi(r e^{i\theta}) \, \sigma_{p}^{is/2\pi} \, = \, e^{-s} \, \partial\Phi(r e^{i\theta -s}) \, , \label{eq-null-boost-1}
\end{align}
and
\begin{align}
    \sigma_{p}^{-is/2\pi} \, \partial^{2}\Phi(r e^{i\theta}) \, \sigma_{p}^{is/2\pi} \, = \, e^{-2s} \, \partial^{2}\Phi(r e^{i\theta -s}) \, .\label{eq-null-boost-2}
\end{align}
With these results, we can compute $\mathcal{G}_{p}(s)$ in Eq.~\eqref{eq-null-def-Gs-p} to obtain
\begin{align}
    \mathcal{G}_{p}(s) \, = \, - \, \frac{6}{\big(r_{1} e^{i\theta_{1}-s} - r_{2} e^{i\theta_{2}}\big)^{4}} \, \left(e^{-s} - e^{-2s}\right) \,  .\label{eq-null-Gsp-fin}
\end{align}
Finally, combining Eq.~\eqref{eq-null-gs-aux} and Eq.~\eqref{eq-null-Gsp-fin},  we find that $G(s)$ in Eq.~\eqref{eq-Gs-prod} is given by
\begin{align}
    \mathcal{G}(s) \, = \, -6 \, e^{-\pi(K_{\alpha}+K_{\beta})} \, e^{(\theta_{1}-\theta_{2}) v_{\alpha\beta} } \, e^{- \pi \, \text{sgn}(\theta_{1}-\theta_{2}) v_{\alpha\beta} } \, \, \delta_{\alpha\beta'} \, \delta_{\beta\alpha'} \,  \frac{\left(e^{-s} - e^{-2s}\right) \, e^{is v_{\alpha\beta}}}{\big( r_{2} e^{i\theta_{2}} \, - \, r_{1} e^{i\theta_{1}-s}\big)^{4}} \, . \label{eq-null-Gs-final}
\end{align}

\subsection*{Calculation of $G(\omega)$}

Recall from Eq.~\eqref{eq-null-def-Gw} that $G(\omega)$ is defined by a Fourier transform, i.e.,
\begin{align}
    G(\omega) \, = \, \frac{1}{2\pi} \, \int_{-\infty}^{\infty} ds \, e^{-is\omega} \, \mathcal{G}(s) \, . \label{eq-app-def-Gw}
\end{align}
From Eq.~\eqref{eq-null-Gs-final}, we see that $\mathcal{G}(s) \, \to \, 0 $ for $\text{Re}(s) \, \to \, \pm \infty$.
Moreover, note that
\begin{align}
    \mathcal{G}(s+2\pi i) \, = \, e^{-2\pi v_{\alpha\beta}} \, \mathcal{G}(s) \, .
\end{align}
With these observations, we can write $G(\omega)$ in Eq.~\eqref{eq-app-def-Gw} as
\begin{align}
    G(\omega) \, = \, \frac{1}{2\pi} \, \frac{1}{1 \, - \, e^{2\pi(\omega-v_{\alpha\beta})}} \, \oint_{C} ds \, e^{-is\omega} \, \mathcal{G}(s) \, , \label{eq-app-Gw-cont}
\end{align}
where $C$ is a closed rectangular contour given by
\begin{align}
    C \, : \, (-\infty,\infty) \, \cup \, (\infty,\infty+2\pi i) \, \cup \, (\infty+2\pi i,-\infty+2\pi i) \, \cup \,  (-\infty+2\pi i,-\infty) \, .
\end{align}

The Fourier transform now expressed as the contour integral in Eq.~\eqref{eq-app-Gw-cont}, can now be computed using the residue theorem.
There is only one pole of $\mathcal{G}(s)$ inside the contour $C$, located at $s=s_{*}$ given by
\begin{align}
    s_{*} \, = \, \log \left(\frac{r_{1}}{r_{2}}\right) \, + \, i(\theta_{1}-\theta_{2} + \pi) \, - \, i \pi\,  \text{sgn}(\theta_{1}-\theta_{2}) \, .
\end{align}
By the residue theorem, Eq.~\eqref{eq-app-Gw-cont} becomes
\begin{align}
    G(\omega) \, = \, \frac{i}{1 \, - \, e^{2\pi(\omega-v_{\alpha\beta})}} \,\, \text{Res}\Big[ e^{-is\omega} \, \mathcal{G}(s) \, ; \, s = s_{*} \Big] \,  . \label{eq-app-Gw-res}
\end{align}
This residue can be calculated using standard methods.
For any complex number $z$, we have the general result
\begin{align}
    \text{Res}\left[ \frac{e^{-2s} \, e^{is z}}{\big( 1 \, - \, r e^{i\theta-s}\big)^{4}} \, ; \, s = s_{*} \right] \, = \, -\frac{i \, e^{-\pi z}}{6} \, z(z^{2}+1) \, \left( r e^{i\theta} \right)^{-2+iz} \, e^{\pi z \, \text{sgn}(\theta_{1}-\theta_{2})} \, .
\end{align}
Finally, combining this general result with Eq.~\eqref{eq-null-Gs-final} and making some trivial simplifications, we find that $G(\omega)$ in Eq.~\eqref{eq-app-Gw-res} is given by
\begin{align}
    {G}(\omega) \, = \, &\, \, \frac{1}{2} \, \delta_{\alpha\beta'} \, \delta_{\beta\alpha'} \, e^{-\pi\left(K_{\alpha}+K_{\beta}\right)} \, \frac{1}{\big(r_{1} e^{i\theta_{1}}\big)^{2} \, \big(r_{2} e^{i\theta_{2}}\big)^{2}} \, \left(\frac{r_{1}}{r_{2}}\right)^{iv_{\alpha\beta}} \, e^{-\pi \, \text{sgn}(\theta_{1}-\theta_{2}) \, \omega} \, \label{eq-app-Gw-final}\\ &\quad\times \Bigg[ Q\big(v_{\alpha\beta}-\omega\big) \, \left(\frac{r_{1}e^{i\theta_{1}}}{r_{2}e^{i\theta_{2}}}\right)^{-i\omega} \, + \, Q\big(v_{\alpha\beta}-\omega-i\big) \, \left(\frac{r_{1}e^{i\theta_{1}}}{r_{2}e^{i\theta_{2}}}\right)^{1-i\omega} \Bigg] \, ,\nonumber
\end{align}
where we have defined
\begin{align}
    Q(x) \, \equiv \, \frac{x(x^{2}+1)}{\sinh (\pi x)} \, . 
\end{align}
This completes our derivation of Eq.~\eqref{eq-null-Gw-final}.

\subsection{Details of the \texorpdfstring{$n \to 1$}{n to 1} limit} \label{app-n-1}

In this appendix, our goal is to show that the limits $n\to 1^{+}$ and $n\to 1^{-}$ of $\widehat{F}_{n}(\omega)$ are given by Eq.~\eqref{eq-null-fnw-lim-3} and Eq.~\eqref{eq-null-fnw-lim-below} respectively.
These limits ensure that $\ddot{Z}_{n}^{(1,1)}$ is continuous at $n=1$ and reduces to the result obtained in the proof of the QNEC \cite{Bousso:2015wca} when we take the limit $n\to 1$.
Using the expression of $\widehat{F}_{n}(\omega)$ for general $n$ in Eq.~\eqref{eq-null-fnw-final}, we get
\begin{align}
    \lim_{n\to 1^{\pm}} \, \widehat{F}_{n}\big(\omega\big) \,
    =& \,\, \frac{2 \, \sinh(\pi\omega)}{\tanh\big(\pi\omega\big)} \, \lim_{n \to 1^{\pm}} \, \left[ \frac{1}{n-1} \, \frac{\sin^{2}\big(\pi/n\big)}{\sinh^{2}\big(\pi\omega/n\big)+\sin^{2}\big(\pi/n\big)} \right] \, . \label{eq-null-fnw-lim-1}
\end{align}
Near $n = 1$ we have
\begin{align}
    \sin^{2}(\pi/n) \, = \, \pi^{2} \, (n-1)^{2} \, + O\Big( (n-1)^{3} \Big) \, ,
\end{align}
which implies that Eq.~\eqref{eq-null-fnw-lim-1} can be written as
\begin{align}
    \lim_{n\to 1^{\pm}} \, \widehat{F}_{n}\big(\omega\big) \, = \, \frac{2 \, \sinh(\pi\omega)}{\tanh\big(\pi\omega\big)} \, \lim_{n \to 1^{\pm}} \, \left[  \frac{(n-1)}{(n-1)^{2} \, + \, \pi^{-2} \, \sinh^{2}\big(\pi\omega\big)} \right] \, . \label{eq-null-fnw-lim-2}
\end{align}
Using the identity
\begin{align}
    \delta(x) \, = \, \lim_{\epsilon\to 0^{+}} \, \frac{1}{\pi} \, \frac{\epsilon}{\epsilon^{2}+x^{2}} \, ,
\end{align}
we can reduce Eq.~\eqref{eq-null-fnw-lim-2} to
\begin{align}
    \lim_{n\to 1^{\pm}} \, \widehat{F}_{n}\big(\omega\big) \, = \, \pm \, 2\pi \, \frac{ \sinh(\pi\omega)}{\tanh\big(\pi\omega\big)}  \, \delta\Big(\pi^{-1} \, \sinh\big(\pi\omega\big)\Big)  \,  = \, \pm \, 2\pi \, \delta\big(\omega\big)  \, . 
\end{align}
This completes our derivation of Eq.~\eqref{eq-null-fnw-lim-3} and Eq.~\eqref{eq-null-fnw-lim-below}.

\bibliographystyle{JHEP}
\bibliography{all}

\providecommand{\href}[2]{#2}\begingroup\raggedright\begin{thebibliography}{10}

\bibitem{Bousso:2015mna}
R.~Bousso, Z.~Fisher, S.~Leichenauer and A.~C. Wall, \emph{{Quantum focusing
  conjecture}}, \href{https://doi.org/10.1103/PhysRevD.93.064044}{\emph{Phys.
  Rev. D} {\bfseries 93} (2016) 064044}
  [\href{https://arxiv.org/abs/1506.02669}{{\ttfamily 1506.02669}}].

\bibitem{Bousso:2019dxk}
R.~Bousso, V.~Chandrasekaran and A.~Shahbazi-Moghaddam, \emph{{From black hole
  entropy to energy-minimizing states in QFT}},
  \href{https://doi.org/10.1103/PhysRevD.101.046001}{\emph{Phys. Rev. D}
  {\bfseries 101} (2020) 046001}
  [\href{https://arxiv.org/abs/1906.05299}{{\ttfamily 1906.05299}}].

\bibitem{Casini:2017roe}
H.~Casini, E.~Teste and G.~Torroba, \emph{{Modular Hamiltonians on the null
  plane and the Markov property of the vacuum state}},
  \href{https://doi.org/10.1088/1751-8121/aa7eaa}{\emph{J. Phys. A} {\bfseries
  50} (2017) 364001} [\href{https://arxiv.org/abs/1703.10656}{{\ttfamily
  1703.10656}}].

\bibitem{Leichenauer:2018obf}
S.~Leichenauer, A.~Levine and A.~Shahbazi-Moghaddam, \emph{{Energy density from
  second shape variations of the von Neumann entropy}},
  \href{https://doi.org/10.1103/PhysRevD.98.086013}{\emph{Phys. Rev. D}
  {\bfseries 98} (2018) 086013}
  [\href{https://arxiv.org/abs/1802.02584}{{\ttfamily 1802.02584}}].

\bibitem{Bousso:2015wca}
R.~Bousso, Z.~Fisher, J.~Koeller, S.~Leichenauer and A.~C. Wall, \emph{{Proof
  of the Quantum Null Energy Condition}},
  \href{https://doi.org/10.1103/PhysRevD.93.024017}{\emph{Phys. Rev.}
  {\bfseries D93} (2016) 024017}
  [\href{https://arxiv.org/abs/1509.02542}{{\ttfamily 1509.02542}}].

\bibitem{Koeller:2015qmn}
J.~Koeller and S.~Leichenauer, \emph{{Holographic Proof of the Quantum Null
  Energy Condition}},
  \href{https://doi.org/10.1103/PhysRevD.94.024026}{\emph{Phys. Rev. D}
  {\bfseries 94} (2016) 024026}
  [\href{https://arxiv.org/abs/1512.06109}{{\ttfamily 1512.06109}}].

\bibitem{Malik:2019dpg}
T.~A. Malik and R.~Lopez-Mobilia, \emph{{Proof of the quantum null energy
  condition for free fermionic field theories}},
  \href{https://doi.org/10.1103/PhysRevD.101.066028}{\emph{Phys. Rev. D}
  {\bfseries 101} (2020) 066028}
  [\href{https://arxiv.org/abs/1910.07594}{{\ttfamily 1910.07594}}].

\bibitem{Balakrishnan:2017bjg}
S.~Balakrishnan, T.~Faulkner, Z.~U. Khandker and H.~Wang, \emph{{A General
  Proof of the Quantum Null Energy Condition}},
  \href{https://doi.org/10.1007/JHEP09(2019)020}{\emph{JHEP} {\bfseries 09}
  (2019) 020} [\href{https://arxiv.org/abs/1706.09432}{{\ttfamily
  1706.09432}}].

\bibitem{Ceyhan:2018zfg}
F.~Ceyhan and T.~Faulkner, \emph{{Recovering the QNEC from the ANEC}},
  \href{https://doi.org/10.1007/s00220-020-03751-y}{\emph{Commun. Math. Phys.}
  {\bfseries 377} (2020) 999}
  [\href{https://arxiv.org/abs/1812.04683}{{\ttfamily 1812.04683}}].

\bibitem{Lashkari:2018nsl}
N.~Lashkari, \emph{{Constraining Quantum Fields using Modular Theory}},
  \href{https://doi.org/10.1007/JHEP01(2019)059}{\emph{JHEP} {\bfseries 01}
  (2019) 059} [\href{https://arxiv.org/abs/1810.09306}{{\ttfamily
  1810.09306}}].

\bibitem{Mezei:2019sla}
M.~Mezei and J.~Virrueta, \emph{{The Quantum Null Energy Condition and
  Entanglement Entropy in Quenches}},
  \href{https://arxiv.org/abs/1909.00919}{{\ttfamily 1909.00919}}.

\bibitem{Balakrishnan:2019gxl}
S.~Balakrishnan, V.~Chandrasekaran, T.~Faulkner, A.~Levine and
  A.~Shahbazi-Moghaddam, \emph{{Entropy Variations and Light Ray Operators from
  Replica Defects}},  \href{https://arxiv.org/abs/1906.08274}{{\ttfamily
  1906.08274}}.

\bibitem{Ecker:2019ocp}
C.~Ecker, D.~Grumiller, W.~van~der Schee, M.~Sheikh-Jabbari and P.~Stanzer,
  \emph{{Quantum Null Energy Condition and its (non)saturation in 2d CFTs}},
  \href{https://doi.org/10.21468/SciPostPhys.6.3.036}{\emph{SciPost Phys.}
  {\bfseries 6} (2019) 036} [\href{https://arxiv.org/abs/1901.04499}{{\ttfamily
  1901.04499}}].

\bibitem{Ecker:2020gnw}
C.~Ecker, D.~Grumiller, H.~Soltanpanahi and P.~Stanzer, \emph{{QNEC2 in
  deformed holographic CFTs}},
  \href{https://arxiv.org/abs/2007.10367}{{\ttfamily 2007.10367}}.

\bibitem{muller2013quantum}
M.~M{\"u}ller-Lennert, F.~Dupuis, O.~Szehr, S.~Fehr and M.~Tomamichel, \emph{On
  quantum r{\'e}nyi entropies: A new generalization and some properties},
  {\emph{Journal of Mathematical Physics} {\bfseries 54} (2013) 122203}.

\bibitem{Wilde:2014eda}
M.~M. Wilde, A.~Winter and D.~Yang, \emph{{Strong Converse for the Classical
  Capacity of Entanglement-Breaking and Hadamard Channels via a Sandwiched
  Renyi Relative Entropy}},
  \href{https://doi.org/10.1007/s00220-014-2122-x}{\emph{Commun. Math. Phys.}
  {\bfseries 331} (2014) 593}
  [\href{https://arxiv.org/abs/1306.1586}{{\ttfamily 1306.1586}}].

\bibitem{Berta:2018ecp}
M.~Berta, V.~B. Scholz and M.~Tomamichel, \emph{{R{\'e}nyi Divergences as
  Weighted Non-commutative Vector-Valued $L_p$-Spaces}},
  \href{https://doi.org/10.1007/s00023-018-0670-x}{\emph{Annales Henri
  Poincare} {\bfseries 19} (2018) 1843}
  [\href{https://arxiv.org/abs/1608.05317}{{\ttfamily 1608.05317}}].

\bibitem{Jencova-1}
A.~{Jen{\v{c}}ov{\'a}}, \emph{{R{\'e}nyi Relative Entropies and Noncommutative
  $L_p$-Spaces}},
  \href{https://doi.org/10.1007/s00023-018-0683-5}{\emph{Annales Henri
  Poincar{\'e};} {\bfseries 19} (2018) 2513}
  [\href{https://arxiv.org/abs/1609.08462}{{\ttfamily 1609.08462}}].

\bibitem{Jencova-2}
A.~{Jen{\v{c}}ov{\'a}}, \emph{{R{\'e}nyi relative entropies and noncommutative
  $L_p$-spaces II}}, {\emph{arXiv e-prints} (2017) arXiv:1707.00047}
  [\href{https://arxiv.org/abs/1707.00047}{{\ttfamily 1707.00047}}].

\bibitem{Harlow:2016vwg}
D.~Harlow, \emph{{The Ryu--Takayanagi Formula from Quantum Error Correction}},
  \href{https://doi.org/10.1007/s00220-017-2904-z}{\emph{Commun. Math. Phys.}
  {\bfseries 354} (2017) 865}
  [\href{https://arxiv.org/abs/1607.03901}{{\ttfamily 1607.03901}}].

\bibitem{Akers:2018fow}
C.~Akers and P.~Rath, \emph{{Holographic Renyi Entropy from Quantum Error
  Correction}}, \href{https://doi.org/10.1007/JHEP05(2019)052}{\emph{JHEP}
  {\bfseries 05} (2019) 052}
  [\href{https://arxiv.org/abs/1811.05171}{{\ttfamily 1811.05171}}].

\bibitem{Dong:2018seb}
X.~Dong, D.~Harlow and D.~Marolf, \emph{{Flat entanglement spectra in
  fixed-area states of quantum gravity}},
  \href{https://doi.org/10.1007/JHEP10(2019)240}{\emph{JHEP} {\bfseries 10}
  (2019) 240} [\href{https://arxiv.org/abs/1811.05382}{{\ttfamily
  1811.05382}}].

\bibitem{frank2013monotonicity}
R.~L. Frank and E.~H. Lieb, \emph{Monotonicity of a relative r{\'e}nyi
  entropy}, {\emph{Journal of Mathematical Physics} {\bfseries 54} (2013)
  122201}.

\bibitem{beigi2013sandwiched}
S.~Beigi, \emph{Sandwiched r{\'e}nyi divergence satisfies data processing
  inequality}, {\emph{Journal of Mathematical Physics} {\bfseries 54} (2013)
  122202}.

\bibitem{tomita1967quasi}
M.~Tomita, \emph{Quasi-standard von neumann algebras}, {\emph{preprint} (1967)
  }.

\bibitem{takesaki2006tomita}
M.~Takesaki, \emph{Tomita's theory of modular Hilbert algebras and its
  applications}, vol.~128. Springer, 2006.

\bibitem{haag2012local}
R.~Haag, \emph{Local quantum physics: Fields, particles, algebras}. Springer
  Science \& Business Media, 2012.

\bibitem{Borchers:2000pv}
H.~Borchers, \emph{{On revolutionizing quantum field theory with Tomita's
  modular theory}}, \href{https://doi.org/10.1063/1.533323}{\emph{J. Math.
  Phys.} {\bfseries 41} (2000) 3604}.

\bibitem{Hollands:2017dov}
S.~Hollands and K.~Sanders, \emph{{Entanglement measures and their properties
  in quantum field theory}},
  \href{https://arxiv.org/abs/1702.04924}{{\ttfamily 1702.04924}}.

\bibitem{Witten:2018zxz}
E.~Witten, \emph{{APS Medal for Exceptional Achievement in Research: Invited
  article on entanglement properties of quantum field theory}},
  \href{https://doi.org/10.1103/RevModPhys.90.045003}{\emph{Rev. Mod. Phys.}
  {\bfseries 90} (2018) 045003}
  [\href{https://arxiv.org/abs/1803.04993}{{\ttfamily 1803.04993}}].

\bibitem{Bisognano:1976za}
J.~Bisognano and E.~Wichmann, \emph{{On the Duality Condition for Quantum
  Fields}}, \href{https://doi.org/10.1063/1.522898}{\emph{J. Math. Phys.}
  {\bfseries 17} (1976) 303}.

\bibitem{Faulkner:2020iou}
T.~Faulkner, S.~Hollands, B.~Swingle and Y.~Wang, \emph{{Approximate recovery
  and relative entropy I. general von Neumann subalgebras}},
  \href{https://arxiv.org/abs/2006.08002}{{\ttfamily 2006.08002}}.

\bibitem{Araki:1976}
H.~Araki, \emph{{Relative Entropy of States of von Neumann Algebras}},
  \href{https://doi.org/10.2977/prims/1195191148}{\emph{Publications of the
  Research Institute for Mathematical Sciences} (1976) 809}.

\bibitem{Araki:1982}
H.~Araki and T.~Masuda, \emph{{Positive Cones and Lp-Spaces for von Neumann
  Algebras}},
  \href{https://doi.org/10.2977/prims/1195183577}{\emph{Publications of the
  Research Institute for Mathematical Sciences} (1982) 759}.

\bibitem{Wall:2011hj}
A.~C. Wall, \emph{{A proof of the generalized second law for rapidly changing
  fields and arbitrary horizon slices}},
  \href{https://doi.org/10.1103/PhysRevD.85.104049}{\emph{Phys. Rev. D}
  {\bfseries 85} (2012) 104049}
  [\href{https://arxiv.org/abs/1105.3445}{{\ttfamily 1105.3445}}].

\bibitem{Faulkner:2014jva}
T.~Faulkner, \emph{{Bulk Emergence and the RG Flow of Entanglement Entropy}},
  \href{https://doi.org/10.1007/JHEP05(2015)033}{\emph{JHEP} {\bfseries 05}
  (2015) 033} [\href{https://arxiv.org/abs/1412.5648}{{\ttfamily 1412.5648}}].

\bibitem{Lashkari:2015hha}
N.~Lashkari and M.~Van~Raamsdonk, \emph{{Canonical Energy is Quantum Fisher
  Information}}, \href{https://doi.org/10.1007/JHEP04(2016)153}{\emph{JHEP}
  {\bfseries 04} (2016) 153}
  [\href{https://arxiv.org/abs/1508.00897}{{\ttfamily 1508.00897}}].

\bibitem{Faulkner:2017tkh}
T.~Faulkner, F.~M. Haehl, E.~Hijano, O.~Parrikar, C.~Rabideau and
  M.~Van~Raamsdonk, \emph{{Nonlinear Gravity from Entanglement in Conformal
  Field Theories}}, \href{https://doi.org/10.1007/JHEP08(2017)057}{\emph{JHEP}
  {\bfseries 08} (2017) 057}
  [\href{https://arxiv.org/abs/1705.03026}{{\ttfamily 1705.03026}}].

\bibitem{May:2018tir}
A.~May and E.~Hijano, \emph{{The holographic entropy zoo}},
  \href{https://doi.org/10.1007/JHEP10(2018)036}{\emph{JHEP} {\bfseries 10}
  (2018) 036} [\href{https://arxiv.org/abs/1806.06077}{{\ttfamily
  1806.06077}}].

\bibitem{Ugajin:2018rwd}
T.~Ugajin, \emph{{Perturbative expansions of R{\'e}nyi relative divergences and
  holography}},  \href{https://arxiv.org/abs/1812.01135}{{\ttfamily
  1812.01135}}.

\bibitem{Bao:2019aol}
N.~Bao, M.~Moosa and I.~Shehzad, \emph{{The holographic dual of R{\'e}nyi
  relative entropy}},
  \href{https://doi.org/10.1007/JHEP08(2019)099}{\emph{JHEP} {\bfseries 08}
  (2019) 099} [\href{https://arxiv.org/abs/1904.08433}{{\ttfamily
  1904.08433}}].

\bibitem{Rosenhaus:2014woa}
V.~Rosenhaus and M.~Smolkin, \emph{{Entanglement Entropy: A Perturbative
  Calculation}}, \href{https://doi.org/10.1007/JHEP12(2014)179}{\emph{JHEP}
  {\bfseries 12} (2014) 179} [\href{https://arxiv.org/abs/1403.3733}{{\ttfamily
  1403.3733}}].

\bibitem{Rosenhaus:2014ula}
V.~Rosenhaus and M.~Smolkin, \emph{{Entanglement entropy, planar surfaces, and
  spectral functions}},
  \href{https://doi.org/10.1007/JHEP09(2014)119}{\emph{JHEP} {\bfseries 09}
  (2014) 119} [\href{https://arxiv.org/abs/1407.2891}{{\ttfamily 1407.2891}}].

\bibitem{Allais:2014ata}
A.~Allais and M.~Mezei, \emph{{Some results on the shape dependence of
  entanglement and R{\'e}nyi entropies}},
  \href{https://doi.org/10.1103/PhysRevD.91.046002}{\emph{Phys. Rev. D}
  {\bfseries 91} (2015) 046002}
  [\href{https://arxiv.org/abs/1407.7249}{{\ttfamily 1407.7249}}].

\bibitem{Lewkowycz:2014jia}
A.~Lewkowycz and E.~Perlmutter, \emph{{Universality in the geometric dependence
  of Renyi entropy}},
  \href{https://doi.org/10.1007/JHEP01(2015)080}{\emph{JHEP} {\bfseries 01}
  (2015) 080} [\href{https://arxiv.org/abs/1407.8171}{{\ttfamily 1407.8171}}].

\bibitem{Rosenhaus:2014zza}
V.~Rosenhaus and M.~Smolkin, \emph{{Entanglement Entropy for Relevant and
  Geometric Perturbations}},
  \href{https://doi.org/10.1007/JHEP02(2015)015}{\emph{JHEP} {\bfseries 02}
  (2015) 015} [\href{https://arxiv.org/abs/1410.6530}{{\ttfamily 1410.6530}}].

\bibitem{Mezei:2014zla}
M.~Mezei, \emph{{Entanglement entropy across a deformed sphere}},
  \href{https://doi.org/10.1103/PhysRevD.91.045038}{\emph{Phys. Rev. D}
  {\bfseries 91} (2015) 045038}
  [\href{https://arxiv.org/abs/1411.7011}{{\ttfamily 1411.7011}}].

\bibitem{Carmi:2015dla}
D.~Carmi, \emph{{On the Shape Dependence of Entanglement Entropy}},
  \href{https://doi.org/10.1007/JHEP12(2015)043}{\emph{JHEP} {\bfseries 12}
  (2015) 043} [\href{https://arxiv.org/abs/1506.07528}{{\ttfamily
  1506.07528}}].

\bibitem{Faulkner:2015csl}
T.~Faulkner, R.~G. Leigh and O.~Parrikar, \emph{{Shape Dependence of
  Entanglement Entropy in Conformal Field Theories}},
  \href{https://doi.org/10.1007/JHEP04(2016)088}{\emph{JHEP} {\bfseries 04}
  (2016) 088} [\href{https://arxiv.org/abs/1511.05179}{{\ttfamily
  1511.05179}}].

\bibitem{Leichenauer:2016rxw}
S.~Leichenauer, M.~Moosa and M.~Smolkin, \emph{{Dynamics of the Area Law of
  Entanglement Entropy}},
  \href{https://doi.org/10.1007/JHEP09(2016)035}{\emph{JHEP} {\bfseries 09}
  (2016) 035} [\href{https://arxiv.org/abs/1604.00388}{{\ttfamily
  1604.00388}}].

\bibitem{Belin:2018juv}
A.~Belin, N.~Iqbal and S.~F. Lokhande, \emph{{Bulk entanglement entropy in
  perturbative excited states}},
  \href{https://doi.org/10.21468/SciPostPhys.5.3.024}{\emph{SciPost Phys.}
  {\bfseries 5} (2018) 024} [\href{https://arxiv.org/abs/1805.08782}{{\ttfamily
  1805.08782}}].

\bibitem{Agon:2020fqs}
C.~A. Agón, S.~F. Lokhande and J.~F. Pedraza, \emph{{Local quenches, bulk
  entanglement entropy and a unitary Page curve}},
  \href{https://arxiv.org/abs/2004.15010}{{\ttfamily 2004.15010}}.

\bibitem{gordon10}
M.~Gordon and L.~Loura, \emph{Exponential generalized distributions},
  {\emph{Math.~J.~Okayama Univ.} {\bfseries 52} (2010) 159}.

\bibitem{Lashkari:2018oke}
N.~Lashkari, H.~Liu and S.~Rajagopal, \emph{{Modular Flow of Excited States}},
  \href{https://arxiv.org/abs/1811.05052}{{\ttfamily 1811.05052}}.

\bibitem{wiesbrock1993}
H.-W. Wiesbrock, \emph{Half-sided modular inclusions of von-neumann-algebras},
  {\emph{Comm. Math. Phys.} {\bfseries 157} (1993) 83}.

\bibitem{Hofman:2008ar}
D.~M. Hofman and J.~Maldacena, \emph{{Conformal collider physics: Energy and
  charge correlations}},
  \href{https://doi.org/10.1088/1126-6708/2008/05/012}{\emph{JHEP} {\bfseries
  05} (2008) 012} [\href{https://arxiv.org/abs/0803.1467}{{\ttfamily
  0803.1467}}].

\bibitem{Kologlu:2019mfz}
M.~Kologlu, P.~Kravchuk, D.~Simmons-Duffin and A.~Zhiboedov, \emph{{The
  light-ray OPE and conformal colliders}},
  \href{https://arxiv.org/abs/1905.01311}{{\ttfamily 1905.01311}}.

\bibitem{Komargodski:2012ek}
Z.~Komargodski and A.~Zhiboedov, \emph{{Convexity and Liberation at Large
  Spin}}, \href{https://doi.org/10.1007/JHEP11(2013)140}{\emph{JHEP} {\bfseries
  11} (2013) 140} [\href{https://arxiv.org/abs/1212.4103}{{\ttfamily
  1212.4103}}].

\bibitem{Costa:2017twz}
M.~S. Costa, T.~Hansen and J.~Penedones, \emph{{Bounds for OPE coefficients on
  the Regge trajectory}},
  \href{https://doi.org/10.1007/JHEP10(2017)197}{\emph{JHEP} {\bfseries 10}
  (2017) 197} [\href{https://arxiv.org/abs/1707.07689}{{\ttfamily
  1707.07689}}].

\bibitem{Balakrishnan:2020lbp}
S.~Balakrishnan and O.~Parrikar, \emph{{Modular Hamiltonians for Euclidean Path
  Integral States}},  \href{https://arxiv.org/abs/2002.00018}{{\ttfamily
  2002.00018}}.

\bibitem{Ryu:2006bv}
S.~Ryu and T.~Takayanagi, \emph{{Holographic derivation of entanglement entropy
  from AdS/CFT}},
  \href{https://doi.org/10.1103/PhysRevLett.96.181602}{\emph{Phys. Rev. Lett.}
  {\bfseries 96} (2006) 181602}
  [\href{https://arxiv.org/abs/hep-th/0603001}{{\ttfamily hep-th/0603001}}].

\bibitem{Hubeny:2007xt}
V.~E. Hubeny, M.~Rangamani and T.~Takayanagi, \emph{{A Covariant holographic
  entanglement entropy proposal}},
  \href{https://doi.org/10.1088/1126-6708/2007/07/062}{\emph{JHEP} {\bfseries
  07} (2007) 062} [\href{https://arxiv.org/abs/0705.0016}{{\ttfamily
  0705.0016}}].

\bibitem{Wall:2017blw}
A.~C. Wall, \emph{{Lower Bound on the Energy Density in Classical and Quantum
  Field Theories}},
  \href{https://doi.org/10.1103/PhysRevLett.118.151601}{\emph{Phys. Rev. Lett.}
  {\bfseries 118} (2017) 151601}
  [\href{https://arxiv.org/abs/1701.03196}{{\ttfamily 1701.03196}}].

\bibitem{Bousso:2020yxi}
R.~Bousso, V.~Chandrasekaran, P.~Rath and A.~Shahbazi-Moghaddam, \emph{{Gravity
  Dual of Connes Cocycle Flow}},
  \href{https://arxiv.org/abs/2007.00230}{{\ttfamily 2007.00230}}.

\bibitem{Calabrese:2009qy}
P.~Calabrese and J.~Cardy, \emph{{Entanglement entropy and conformal field
  theory}}, \href{https://doi.org/10.1088/1751-8113/42/50/504005}{\emph{J.
  Phys. A} {\bfseries 42} (2009) 504005}
  [\href{https://arxiv.org/abs/0905.4013}{{\ttfamily 0905.4013}}].

\bibitem{Nakagawa:2017fzo}
Y.~O. Nakagawa and T.~Ugajin, \emph{{Numerical calculations on the relative
  entanglement entropy in critical spin chains}},
  \href{https://doi.org/10.1088/1742-5468/aa85c1}{\emph{J. Stat. Mech.}
  {\bfseries 1709} (2017) 093104}
  [\href{https://arxiv.org/abs/1705.07899}{{\ttfamily 1705.07899}}].

\bibitem{White:1992zz}
S.~R. White, \emph{{Density matrix formulation for quantum renormalization
  groups}}, \href{https://doi.org/10.1103/PhysRevLett.69.2863}{\emph{Phys. Rev.
  Lett.} {\bfseries 69} (1992) 2863}.

\bibitem{Lashkari:2014yva}
N.~Lashkari, \emph{{Relative Entropies in Conformal Field Theory}},
  \href{https://doi.org/10.1103/PhysRevLett.113.051602}{\emph{Phys. Rev. Lett.}
  {\bfseries 113} (2014) 051602}
  [\href{https://arxiv.org/abs/1404.3216}{{\ttfamily 1404.3216}}].

\bibitem{Headrick:2012fk}
M.~Headrick, A.~Lawrence and M.~Roberts, \emph{{Bose-Fermi duality and
  entanglement entropies}},
  \href{https://doi.org/10.1088/1742-5468/2013/02/P02022}{\emph{J. Stat. Mech.}
  {\bfseries 1302} (2013) P02022}
  [\href{https://arxiv.org/abs/1209.2428}{{\ttfamily 1209.2428}}].

\bibitem{Radicevic:2016tlt}
D.~Radicevic, \emph{{Entanglement Entropy and Duality}},
  \href{https://doi.org/10.1007/JHEP11(2016)130}{\emph{JHEP} {\bfseries 11}
  (2016) 130} [\href{https://arxiv.org/abs/1605.09396}{{\ttfamily
  1605.09396}}].

\bibitem{F_hringer_2008}
M.~F{\"u}hringer, S.~Rachel, R.~Thomale, M.~Greiter and P.~Schmitteckert,
  \emph{Dmrg studies of critical su(n) spin chains},
  \href{https://doi.org/10.1002/andp.200810326}{\emph{Annalen der Physik}
  {\bfseries 17} (2008) 922–936}.

\bibitem{alcaraz1994critical}
F.~C. Alcaraz and A.~L. Malvezzi, \emph{Critical and off-critical properties of
  the $xxz$ chain in external homogeneous and staggered magnetic fields},
  1994.

\bibitem{DiFrancesco:1997nk}
P.~Di~Francesco, P.~Mathieu and D.~Senechal, \emph{{Conformal Field Theory}},
  Graduate Texts in Contemporary Physics. Springer-Verlag, New York, 1997,
  \href{https://doi.org/10.1007/978-1-4612-2256-9}{10.1007/978-1-4612-2256-9}.

\bibitem{petz1986quasi}
D.~Petz, \emph{Quasi-entropies for finite quantum systems}, {\emph{Reports on
  mathematical physics} {\bfseries 23} (1986) 57}.

\bibitem{Petz}
D.~Petz, \emph{{Quasi-entropies for states of a von neumann algebra}},
  {\emph{Publications of the Research Institute for Mathematical Sciences}
  {\bfseries 21} (1985) 787}.

\bibitem{audenaert2013alpha}
K.~M. Audenaert and N.~Datta, \emph{alpha-z-relative renyi entropies},
  {\emph{arXiv preprint arXiv:1310.7178} (2013) }.

\bibitem{Lieb1991}
E.~H. Lieb and W.~E. Thirring, \emph{Inequalities for the Moments of the
  Eigenvalues of the Schrodinger Hamiltonian and Their Relation to Sobolev
  Inequalities}, pp.~135--169.
\newblock Springer Berlin Heidelberg, Berlin, Heidelberg, 1991.

\bibitem{ALT}
H.~Araki, \emph{{On an inequality of Lieb and Thirring}}, {\emph{Lett Math
  Phys} {\bfseries 19} (1990) 167}.

\end{thebibliography}\endgroup

\end{document}